\documentclass[12pt]{book}

\pdfoutput=1
\usepackage{braket}
\usepackage{amsmath}
\usepackage{graphicx}
\usepackage{mathtools}
\usepackage{amssymb}
\usepackage[utf8]{inputenc}
\usepackage[english]{babel}
\usepackage{amsthm}
\usepackage{bbm}
\usepackage{braket}
\usepackage{subcaption} 
\usepackage[titletoc,title]{appendix}
\usepackage{url}
\usepackage{lmodern}
\theoremstyle{plain}
\newtheorem{theorem}{Theorem}[section]

\newtheorem{lemma}[theorem]{Lemma}
\newtheorem*{lemma*}{Lemma}

\theoremstyle{definition}
\newtheorem{definition}[theorem]{Definition} 
 
\newtheorem{example}[theorem]{Example} 

\DeclareMathOperator{\Tr}{Tr} 
\DeclareMathOperator{\Real}{Re}
\DeclareMathOperator{\Imag}{Im}

\begin{document}
\begin{titlepage}
    \begin{center}
        %\vspace*{0.8cm}
        
        \Huge
        \textbf{On the Practical Applications of Information Field Dynamics}
        
        \LARGE
        
        \vspace{1.5cm}
        \textbf{Martin Dupont}
        
        \vspace{1cm}
        
        %Master Thesis
        
        %\vspace{0.8cm}
        
        \includegraphics[width=0.4\textwidth]{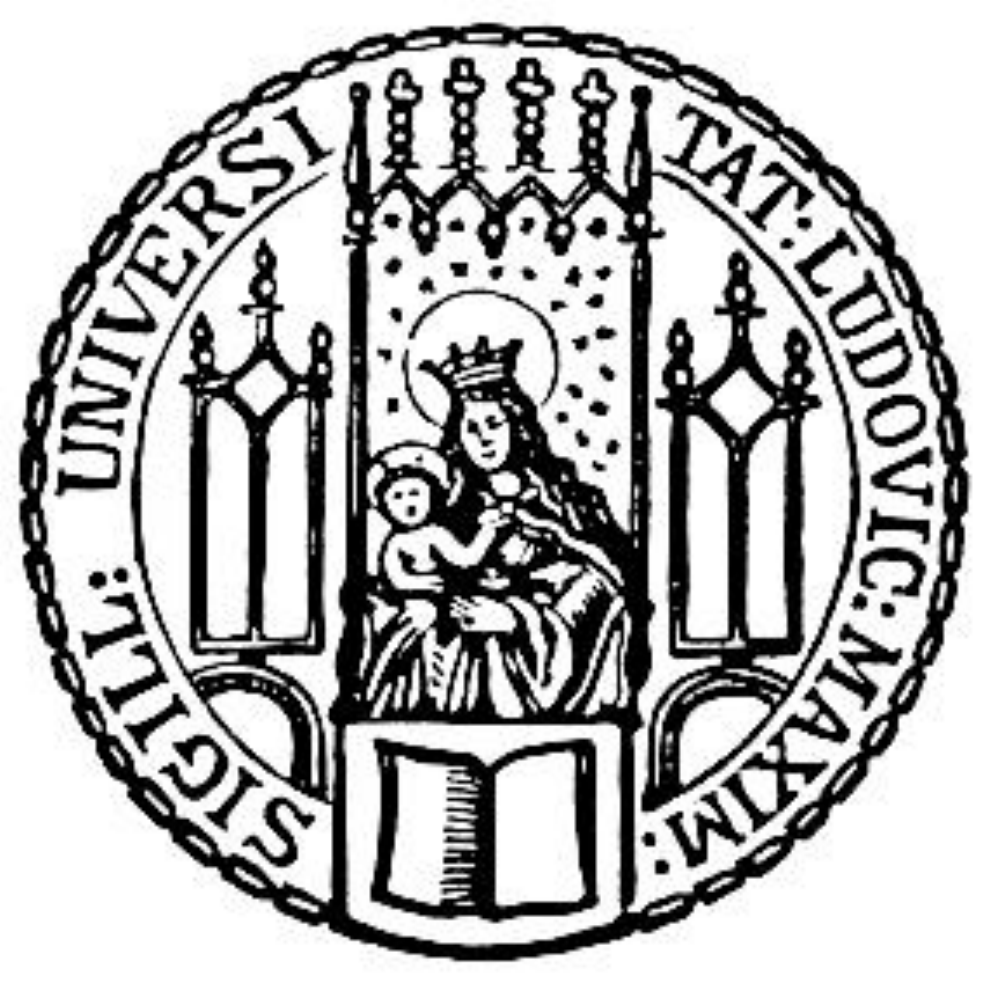}
        \vspace{0.8cm}
				
        \large
        Submitted for the master course in 
			 Theoretical and Mathematical Physics at the 
        Ludwig--Maximilians--Universit\"at 
        Munich\\

    \end{center}
  
	\vspace{3.2cm}
	\begin{flushleft}
    \large Supervisor:  Prof. Dr. Torsten En\ss{}lin \\[1mm]
    \large Second referee: Prof. Dr. Ewald Müller \\[1mm]
    \large Defence date: 1st August 2017 \\
  \end{flushleft}

\end{titlepage}

%\title{On the practical applications of Information Field Dynamics}
%\author{Martin Dupont}

%\maketitl
\section*{Acknowledgements}
First and foremost I am grateful to my friends Anthony and Max for sacrificing their weekend to proofread my thesis cover to cover. Thanks to Anthony, whose helpful commentary and LaTeX skills were invaluable to my thesis. Thanks to Max, whose unhelpful commentary and sarcasm were at least entertaining. Special thanks needs to go to my supervisor Torsten En\ss{}lin for tolerating my pessimism, and my colleague Reimar Leike, for reading my numerous rambling manifestos.

\tableofcontents

\chapter{Introduction}

As physicists, we use mathematical equations to describe the world around us. However, the unfortunate state of the world is that most physically interesting processes have equations of motion described by partial differential equations (PDEs) which have no analytical solutions. This necessitates the use of simulation schemes for differential equations, which can only approximate the behaviour of the true solution. Given a PDE, the solutions are functions, which contain an infinite number of degrees of freedom. Whereas for any implementable approximation, the number of degrees of freedom must be finite. Thus there is always a gap between any simulation and reality. \newline

There is already a vast and well-known literature base of numerical methods for differential equations, with each method having its own advantages and drawbacks. 
One often-used approach is that of subgrid models. A field will typically be represented by a series of discrete samples of the field value at certain points, and the simulation scheme will generally assume that the field has some structure between those points. For example, subgrid models may often assume a linear interpolation of the field between the data points. This assumption is then applied somewhere in the scheme in the hope of obtaining more accurate results \cite{Hydro}. \newline

Information field dynamics (IFD)\cite{IFD} is a new framework for developing numerical schemes, and can be thought of as an improvement on, or generalization of subgrid models. The main idea of IFD is that rather than making any concrete assumptions about the nature of the field, we use Bayesian statistics to infer the most likely continuous field configuration given some data in computer memory, and this continuous reconstruction is used to inform the numerical simulation scheme. The continuous field reconstruction is achieved using a framework already developed by T. En{\ss}lin known as Information Field Theory (IFT) \cite{IFT}.
The general mathematical framework of this scheme has indeed already been laid out in \cite{IFDmath}, but has yet to be practically implemented.\newline

The original goal of this project was to develop the first real-world application of the IFD framework. The problem chosen for this was the simulation of cosmic ray transport in space. It was believed that IFD would be well-suited for this. However, throughout the course of this project, it became apparent that there were many unsolved problems on the general level which needed to be ironed out before a practical implementation could be carried out. Specifically speaking, an analysis of the stability, errors and convergence of codes in the IFD framework had not yet been performed. The early implementations of the cosmic ray codes were plagued by instabilities and numerical artefacts, which necessitated a theoretical analysis before moving forward. \newline

The content of this work is as follows: after an introduction to the general framework of IFT and Bayesian statistics in chapter \ref{nummer2}, we will present the construction of the IFD framework in chapter \ref{EE}. In this chapter we will also present a number of small results and improvements to the framework, before moving on to discuss errors, stability and convergence. We will also introduce the trial problem for this project: cosmic ray simulations. Chapter \ref{Tinvariant} will then explore a broad class of IFD models whose stability and convergence properties can be analytically examined. In tandem, we will develop a toy model which serves as an illustrative example of the strengths and weaknesses of this general class of models. This example will then be extended to something which at least superficially resembles a cosmic ray evolution simulation. \newline

A second class of models will also be presented in chapter \ref{ChapSPH}, which is also based on IFD, and draws inspiration from so-called \textit{Smooth Particle Hydrodynamics} algorithms. This model is unsuccessful, but draws into sharp focus some of the weaknesses presented by the previous class of models. While any numerical scheme has advantages and disadvantages, it is believed that the weaknesses uncovered during the course of this project will need to be addressed before attempting to simulate a truly nontrivial and scientifically interesting system. This general weakness, and possible solutions for a way forward will be presented last. This will be followed in chapter \ref{summary} by a summary of all the results presented in this work.

\chapter{The IFT framework}\label{nummer2}

\section{Priors, posterior and Bayes Theorem}

To understand IFT, one first needs a quick introduction to Bayesian statistics. In Bayesian probability theory, probabilities are assigned to events, and take the form of real numbers. These real numbers express a subjective belief about how likely a given even is to happen. The assigned probabilities for any event range between zero and one, with one implying absolute confidence in a result. The sum of probabilities (or integral) over all possible events within some set must equal one, i.e. \textit{some} event must occur. In Bayesian statistics, a rational observers degree of belief about an event may change in response to new information. This updating of beliefs is the foundation of Bayesian inference, typically one has some prior belief about a system, which is updated to a new belief about the system after performing an experiment and obtaining new information. \newline

The probability of an event $a$ occurring is denoted by $\mathcal{P}(a)$. Typically, probabilities of events are dependent on some background condition. We write 
$\mathcal{P}(a|b)$ to denote the probability of $a$ occurring given that we know $b$ to be true. The product rule of probabilities states that given two events $a$ and $b$, the joint probability of them both occurring is given by $\mathcal{P}(ab)= \mathcal{P}(a|b) \mathcal{P}(b)$. Two events are said to be mutually exclusive if they never occur in unison, and a set of events is said to be exhaustive if one element of the set must always occur. A further piece of terminology used in this thesis is that of marginalizing over probabilities; given a probability distribution $\mathcal{P}(a_i,b_j)$ in two sets of variables $\{a_i\}$ and $\{b_j\}$ and the set of $b$'s are mutually exclusive and exhaustive, then the probability distribution for just the first variable is given by

\begin{equation}
\mathcal{P}(a_i) = \sum_j \mathcal{P}(a_i,b_j)
\end{equation}

In the case where the $b$'s form a continuous variable, the sum will be replaced by an integral. \newline

Some more terms need to be defined as well. Suppose one had a system whose state is governed by some variable $\theta$, and one performs an experiment on the system yielding some data $d$, from which we want to infer the state of the system. In Bayesian statistics, we will have some preconceived probability on the state of the system $\mathcal{P}(\theta)$ before any measurement is carried out. This is known as the \textit{prior distribution}. This prior may come from a variety of sources, such as past experiments, or symmetry principles (all states are equally likely, etc.) or even expert intuition. The \textit{likelihood} is the probability of obtaining some data given a fixed state of the system, i.e. $\mathcal{P}(d|\theta)$. The \textit{posterior} distribution is the probability of a given field configuration given some data, $\mathcal{P}(\theta | d)$. In this language, the posterior distribution is what we wish to obtain from an experiment. The posterior can be obtained from the prior and the likelihood by using Bayes theorem:

%numbers between one and zero expressing how likely an event is, and the sum/integral over all possible events must equal one. In Bayesian statistics, one often has beliefs which are updated upon performing some measurement. Probabilites are always dependent on some background state or criteria. We denote this by writing $\mathcal{P}(a|b)$, as expressing the probability of an event $a$ occurring given that we know $b$. In Bayesian statistics there are some important distributions. Suppose we have some parameters, which for us are fields $\phi$ that we want to infer, and some data $d$. The prior distribution is the probability distribution over the fields before we have made any measurement $\mathcal{P}(\phi)$, given that this is in general unknowable, this probability will typically be an approximation based on past experience, assumptions etc. The \textit{likelihood} is the probability of obtaining some data given that the field has some value, i.e. $\mathcal{P}(d|\phi)$. The \textit{posterior} distribution is the probability of a given field configuration given some data, $\mathcal{P}(\phi | d)$. (CITE ALL OF THIS). The posterior is typically the quantity we want to obtain when doing statistics. The posterior can be obtained from the prior and the likelihood by using Bayes theorem:

\begin{equation}
\mathcal{P}(\theta|d)=\frac{\mathcal{P}(d|\theta)\mathcal{P}(\theta)}{\mathcal{P}(d)}=\frac{\mathcal{P}(d,\theta)}{\mathcal{P}(d)}
\end{equation}

which is a simple corollary of the product rule for probabilities. While this formula does depend on $\mathcal{P}(d)$, which is in general unknown, for a given trial it is constant, and so can be thrown out as an irrelevant normalization constant. The information in this section is well known was taken from a variety of sources, but may be found in any good probability textbook, like \cite{Jaynes} for example. \newline

\section{Field inference}

Now that the basic language of inference problems has been established, the goal of Information Field Theory can be stated. Suppose that the system one was investigating was a continuous field $\phi(x)$ which is a function of the variable $x$ in some set $\Omega$. Suppose one also had experimental apparatus that measured that field, the goal is then to infer the value of the continuous field given a prior distribution and some data. To do this, we need to be able to write down probability distributions over fields, which means $\mathcal{P}(\phi)$ and $\mathcal{P}(\phi|d)$ become functionals.\newline

Given that many operations on probability distributions such as normalization, expectation values etc. occur under an integral, we immediately see that we will have to commit the minor sin of resorting to the functional integral. If the probability distribution in question is everywhere greater than zero, then it can be rewritten in the form: 

%The framework is as follows. Suppose we are experimentalists or otherwise, and we have somehow obtained through inference a probability distribution over a field on %some space at a given time. We know the dynamics of said field, so by all rights we should be able to model how our probability distribution evolves too. Before we %arrive at this we must address the idea of a probability distribution over a field. The language to do this was also developed by Ensslin (CITE), and is known as% information field theory, on which IFD is built. For the purposes of IFD, a valid probability distribution over a field is any functional $\mathcal{P}(\phi)$ acting %on a field $\phi$ that is everywhere greater or equal to zero, and integrates to one under the functional integral

%\begin{equation}
%\int \mathcal{D}\phi \mathcal{P}(\phi)
%\end{equation}

%We acknowledge that using the functional integral is indeed a minor sin. Given the obvious link to quantum field theory/condensed matter theory, we will often rewrite the probability distribution as ($\pm$ HERE?):

\begin{align}
\mathcal{P}(\phi|d) = \exp(-H(\phi|d)) \quad \text{where} \quad H(\phi|d)= -\ln (\mathcal{P}(\phi|d))
\end{align}

which, in our applications, we will always be able to do. We refer to $H(\phi|d)$ as the \textit{information Hamiltonian} as a deliberate analogy to the physical Hamiltonians occurring in quantum/condensed matter field theories \cite{IFT}. This direct analogy allows techniques developed in QFT to be applied to extracting relevant information such as expectation values and correlation functions from these formally ill-defined objects. \newline

The goal is now to construct a posterior distribution given some prior $\mathcal{P}(\phi)$ on the field. In our framework we will generally assume that the measured data is a function of two things:
a deterministic function of the field, $R$, known as the response function, plus some random measurement noise $n$ which is independent of the field. We write this as  $d=R(\phi)+n$. The data and noise will be assumed to live in $\mathbb{R}^m$ for some dimension $m$, although we will often extend to $\mathbb{C}^m$ in certain cases where it is mathematically convenient. The fields $\phi(x)$ defined on some set $\Omega$ must live in some subspace of $\mathcal{L}^2(\Omega)$, as we will shortly see that the IFT formalism requires the use of inner-products. These two spaces will be referred to as data space and signal space, respectively.\newline

 The full generality of information field theory can handle nonlinear responses, as well as prior distributions which are non-Gaussian in the fields, 
by expanding out the desired expectation values etc. in terms of Feynmann diagrams. For the purposes of a short masters project however, we specialize to what we refer to as the linear case, analogous to the free theory in QFT. This means that we assume that the noise and the prior can be expressed as independent Gaussians:

\begin{align}
\mathcal{P}(\phi) \propto \exp(-\frac{1}{2}\bra{\phi-\psi}\Phi^{-1} \ket{\phi-\psi}_{s}) && \mathcal{P}(n) \propto \exp(-\frac{1}{2}\bra{n}N^{-1} \ket{n}_{d}) 
\end{align}

where $N$ and $\Phi$ are some positive, self-adjoint operators on the data and field spaces respectively, and $\psi$ is some assumed mean value of the field. The linear case also assumes that the response $R$ is a linear map from signal space to data space. The spaces in which the inner-products are taken are denoted as subscripts $s$ and $d$ on the brackets for signal and data respectively. \newline

The above Gaussians have the properties that $\mathbb{E}[\ket{\phi-\psi} \bra{\phi-\psi}]_{\mathcal{P}(\phi)} = \Phi$ and 
$\mathbb{E}[\ket{n} \bra{n}]_{\mathcal{P}(n)}=N$, these are often referred to as the correlation structures of the fields, or the covariance matrices, or simply covariances. $\mathbb{E}$ is used to denote an expectation value, with a subscript denoting which random variable the expectation is taken. 
The likelihood of the data given the signal is found by marginalizing over all possible values of the noise:

\begin{equation}
\mathcal{P}(d|\phi) \propto \int \delta(d - R\phi -n) \exp \big{(}-\frac{1}{2}\bra{n}N^{-1} \ket{n}\big{)} dn \propto \mathcal{G}(d-R\phi,N)
\end{equation}

where from now on we will use the notation $\mathcal{G}(a,A)$ to denote a zero-centred Gaussian in $a$ with covariance $A$. To obtain the posterior from the likelihood, we use Bayes' theorem and multiply by the signal prior, i.e. we calculate $\mathcal{P}(d|\phi)\mathcal{P}(\phi)$. For the meantime, we assume that there is no prior mean field, $\psi =0$. Given that all the involved terms are Gaussians, we can omit the
 $\exp$ terms and focus on the additive terms in the exponent:

%\begin{equation}
\begin{align}
&-2\ln (\mathcal{P}(\phi|d))=\bra{\phi}\Phi^{-1} \ket{\phi} + \bra{d-R\phi} N^{-1} \ket{d-R\phi} +C_0 \nonumber \\
&=\bra{\phi}\underbrace{(\Phi^{-1} + R^\dagger N^{-1} R)}_{D^{-1}}\ket{\phi} - \braket{R^\dagger N^{-1} d |\phi} - \braket{\phi | R^\dagger N^{-1} d} + \bra{d} N^{-1} \ket{d} +C_0 \nonumber \\
&=\bra{\phi}D^{-1}\ket{\phi} - 
\braket{D^{-1} D R^\dagger N^{-1} d |\phi} - \braket{\phi | D^{-1} D R^\dagger N^{-1} d} +C_1 \nonumber \\
&= \bra{\phi- \underbrace{DR^\dagger N^{-1} d}_{m(d)}}D^{-1}\ket{\phi-DR^\dagger N^{-1} d} + C_2 \nonumber \\
&= \bra{\phi- m(d)}D^{-1}\ket{\phi-m(d)} + C_2
\end{align}

%&-2\ln (\mathcal{P}(\phi|d))=\bra{\phi}\Phi^{-1} \ket{\phi} + \bra{d-R\phi} N^{-1} \ket{d-R\phi} \\
%&=\bra{\phi}\underbrace{(\Phi^{-1} + R^\dagger N^{-1} R)}_{D^{-1}}\ket{\phi} - \braket{\underbrace{R^\dagger N^{-1} d}_{j(d)} |\phi} - \braket{\phi | R^\dagger N^{-1} %d} + \bra{d} N^{-1} \ket{d} \\
%&=\bra{\phi}(\Phi^{-1} + R^\dagger N^{-1} R)\ket{\phi} - 
%\braket{D^{-1} D j(d) |\phi} - \braket{\phi | D^{-1} D j(d)} \\
%&= \bra{\phi- Dj(d)}D^{-1}\ket{\phi-Dj(d)}

%\end{equation}
Note that we completed the square, and that any terms independent of  $\phi$ (such as $\bra{d} N^{-1} \ket{d}$) have been absorbed into the constants $C_0$, $C_1$ and $C_2$, which drop out as irrelevant normalization factors. One can immediately see that the result is again a Gaussian, so the expected mean field and variance can be simply read off: $\mathbb{E}[\phi]_{\mathcal{P}(\phi|d)}=m(d)$, $\mathbb{E}[(\phi-m(d))(\phi-m(d))^\dagger]_{\mathcal{P}(\phi|d)}=D$. We refer to this $D$ as the \textit{uncertainty variance}. The expected mean field $m(d)$ is a linear function of the data, and as such can be expressed by the action of a single linear operator, referred to as the \textit{Wiener filter} \cite{Wiener}. We denote this object by $W$, such that $m(d)=Wd$,          

%It is however convenient to rewrite the 
%expected mean field $Dj(d)$ in a more convenient form better suited to the linear case. The $j(d)$ and $D$ are standard notations in IFD which are useful for dealing %with nonlinearitied. For the linear case, $Dj(d)$ is a linear map on the data $d$, so we express it as a single operator known as the \textit{Wiener filter}(CITE):

\begin{equation}
W= (\Phi^{-1} + R^\dagger N^{-1} R)^{-1} R^\dagger N^{-1} \\
=\Phi R^\dagger (R\Phi R^\dagger + N)^{-1}
\end{equation}

The first equation is known as the signal space representation \cite{IFDmath}, and the second as the data space representation, after which space the inversion
takes place in.  We will rely on the second representation quite heavily throughout this text, as the inversion may be performed much more easily in data space, as well as being analytically easier to deal with. It is also stable in the limit of low-noise, i.e. $N \to 0$, which will often be assumed throughout the course of this thesis. \newline

The derivation for the case of a nonzero prior mean $\psi$ can be easily deduced, and is detailed in \cite{IFDmath}. We will simply state that the posterior mean field $m(d)$ in the presence of a prior mean field is given by:

\begin{equation}
m(d) = \psi + W(d-R\psi) = D(R^\dagger N^{-1} d + \Phi^{-1}\psi)
\end{equation}
with $W$ and $D$ the same as before.

\chapter{Dynamics}\label{EE}

\noindent With the machinery for field inference now in-place, we can begin discussing dynamics. We assume that the field under consideration has an equation of motion of the form 

\begin{equation}
\partial_t \phi(x,t) =   f(\phi(x,t))
\end{equation}

for some function $f$.
Many partial differential equations, including ones that are higher order in time, can be brought into this form. \newline

Given this knowledge of the time evolution of the field, it should be possible to evolve the posterior probability distribution self-consistently. We take the view that if we assign a given field configuration a certain probability initially, if our views are consistent, then in the absence of outside factors the time-evolved field should have that same probability. We express this idea rigorously by defining $\phi_0$ to be a value of the field at an initial time $t_0$, which undergoes evolution to a new configuration $\phi(t)$ at time $t$. We denote the operator taking $\phi_0$ to $\phi(t)$ by $U(t)$ such that $\phi(t)=U(t)(\phi_0)$.
Under these conditions, the time-evolved posterior distribution should be of the form:

\begin{equation}
\exp \bigg(-\frac{1}{2} \bra{U^{-1}(\phi(t))-Wd} D^{-1} \ket{U^{-1}(\phi(t))-Wd} \bigg) |\det(J(U^{-1}))|
\end{equation}

where we need some Jacobian volume factor $|\det(J(U^{-1}))|$ to account for the changing probability mass\footnote{This Jacobian may formally be infinite-dimensional, however considering that we are already working with the functional integral, it doesn't really matter}, which in general will not be a constant if the time evolution is nonlinear in the fields. However, if the time evolution is nonlinear, then the original Gaussian posterior will evolve into a distribution which is non-Gaussian, which will in general be hard to deal with. In order to manage the scope of this project, we specialize to the case where the time evolution is linear, so that the equations of motion are given by $\partial_t \phi =L \phi$ for some linear operator $L(t)$. This ensures that the time evolution operator $U(t)$ will also be linear. Despite being denoted by $U$, this operator is not necessarily unitary. The evolved probability distribution can then be rewritten as:

\begin{align}
&\exp \bigg(-\frac{1}{2} \bra{U^{-1}(\phi(t)-UWd)} D^{-1} \ket{U^{-1}(\phi(t)-UWd)} \bigg) |\det(J(U^{-1}))| \nonumber\\
&\propto \exp \bigg(-\frac{1}{2} \bra{(\phi(t)-UWd)}  {U^{-1}}^\dagger D^{-1} U^{-1}\ket{(\phi(t)-UWd)} \bigg )
\end{align}
Where we note that the Jacobian dropped out because $|\det(J(U^{-1}))|$ is now a constant independent of the field values, and can be dropped as an irrelevant normalization factor. For this derivation to be correct however, it must still preserve information, i.e. $|\det(J(U))| \ne 0$. \newline

Gaussians can be described by only two quantities, the mean field and the covariance. For the linear case, the mean field unsurprisingly evolves according to the equations of motion, and the uncertainty variance $D$ evolves as  $U D U^\dagger$. We are at the stage now where we at least have a self-consistent equation of motion for the posterior given some known time evolution of the fields. However, we have not yet arrived at a simulation scheme. Since, for any system that we need to simulate, we will not be able to compute the time evolution analytically, the previous equation will remain only a formal solution. Any construction of a practical simulation scheme will need to make a finite-dimensional approximation to this infinite-dimensional object.

\section{Constructing the simulation scheme}

To construct the actual simulation scheme, we break the time evolution up into a finite number of smaller timesteps, which we index by the variable $i$, $\{t_i\}$ for $i$ going from 1 to $N_{step}$. Over these timesteps $[t_i,t_{i+1}]$ the field evolves according to some linear operator $U((t_{i+1},t_i))$ which we assume can be represented as a matrix-valued Taylor series in 
$\Delta t = t_{i+1}-t_i$, i.e. $U(\Delta t)= \mathbbm{1} +\Delta t L(t_i) + ...$. For notational convenience we set $L$ to be constant in time (i.e. the system is linear \textit{and} time-invariant), so $L(t)=L$. This assumption can be dropped at any time and does not affect the derivations. The formal solution for the time evolution is then $U(t)=\exp(t L)$. \newline

We assume that we also have data $d_i$ at timestep $t_i$ that comes from some linear measurement of the field, as described in the previous sections. This data can result from a real-world experiment, or it can be a hypothetical measurement. In the latter case (which we will typically use), the response and prior simply define a rule for reconstructing the field given the data. This means that the response, prior and noise covariances are simply parameters which describe the nature of the ``subgrid model'' that we are using.
We also declare that at the next timestep $t_{i+1}$, we will have some new response, prior and noise, which are considered fixed, and may in general be 
different\footnote{Of course not all choices of new responses and priors will give decent results, for example we will see later that the prior should be chosen so that it is consistent with the time evolution of the system.}
from those at $t_i$ , and some yet-to-be determined data $d_{i+1}$. We distinguish the new and old responses etc. with subscripts $i$ and $i+1$. \newline

Given the initial reconstruction at time $t_i$, the posterior probability distribution can be evolved to time $t_{i+1}$, where we also have a second posterior distribution from the hypothetical measurement at time $t_{i+1}$. The goal is now to select the data $d_1$ so that these two distributions match as well as possible. \newline

%The goal is: given an evolved reconstruction from timestep $t_i$, we want to find some new data at a new timestep $t_{i+1}$ which captures this information effectively. 
%To this end, we suppose that at timestep $t_{i+1}$, we will also be performing a reconstruction of the field given some data $d_{i+1}$, with some new $R_{i+1}$, $\Phi{i+1}$ and $N_{i+1}$. These can in general be different\footnote{Of course not all choices of new responses and priors will give decent results, for example we will see later that the prior should be chosen so that .}

%As an example, take a linear time invariant (LTI) system, which has equation of motion $\partial_t \phi = L \phi$ for some linear operator $L$. Then the formal solution is $U(t)=\exp(t L)$. We would like a more finite representation of the time evolved Gaussian. To this end, imagine there was an observer at time $t_{i+1}$, and they were hypothetically to make their own measurement of the field with a (in general, different \footnote{Later we will see that to even get decent results for our mean field updating, we must also apply some strict criteria on how the priors and responses change at each timestep.} ) response $R_{i+1}$, prior $\Phi_{i+1}$, and noise $N_{i+1}$. 

%From this measurement they will generate their own posterior distribution for the field given their measured data. The goal is to now find the best way of approximating the evolved distribution with the posterior distribution of this new observer \newline 

\noindent This is done by picking the new data $d_{i+1}$ such that the \textit{Kullback-Leibler divergence} (a.k.a relative entropy)
between the true evolved probability distribution from time $t_i$ and the approximation at timestep $t_{i+1}$ is minimized. %The KL divergence is an \textit{asymmetric} functional of two probability distributions, which measures the information loss when one probability distribution is used to approximate another. 
For probability distributions $P$ and $Q$ over an arbitrary random variable $x$, it is given by:

\begin{equation}
\mathrm{KL}(P||Q)= \int P(x)\ln \bigg( \frac{P(x)}{Q(x)} \bigg) dx
\end{equation}

It has the property of being always greater than zero, and has a minimum if and only if $P=Q$. The relative entropy measures the amount of information lost when $Q$ is used to approximate $P$ \cite[p.~51]{KLref}, and crucially, this measure is asymmetric. Considering this asymmetry, it matters in which direction we take the divergence. It was pointed out by a member of the research group \cite{Reimar}, that the previous publications on the subject of IFD \cite{IFD} \cite{IFDmath} have both taken the KL divergence in the wrong direction, so the IFD update equations will need to be rederived for this report. We have a true evolved probability distribution, which we are approximating by a Gaussian at the new timestep, so the Gaussian at timestep $t_{i+1}$ will play the role of $Q$. For clarity and conceptual understanding, it is actually convenient to present the KL divergence for two arbitrary multivariate Gaussians, and then later insert the relevant terms. The proof is instructive, but tedious, and is not an original result \cite{MultiKL}. It has therefore been relegated to the appendix.

\begin{lemma} \label{lemmaKL}
For two Gaussians $\mathcal{G}(\phi-a,A)$ and $\mathcal{G}(\phi-b,B)$, the KL divergence between the two $\mathrm{KL}(\mathcal{G}(\phi-a,A)||\mathcal{G}(\phi-b,B))$is given by:
\begin{equation}
\frac{1}{2}\bigg{(} \Tr [\ln(BA^{-1}) -\mathbbm{1} + B^{-1}A ]+\bra{b-a} B^{-1} \ket{b-a} \bigg{)}
%\frac{1}{2} \big[ \Tr(BA^{-1}) + (b-a)^\dagger B (b-a) - \Tr{\mathbbm{1}} - \Tr\ln(BA^{-1}) \big]
\end{equation}
\end{lemma}

For our case, $\mathcal{G}(\phi-a,A)$ is the evolved probability distribution from $t_i$ and $\mathcal{G}(\phi-b,B)$ is the new probability distribution at time $t_{i+1}$. We momentarily specialize to the case of no prior mean field, i.e. $\psi_i = \psi_{i+1} =0$. This then allows us to set
$B=D_{i+1}$, $A= U D_i U^\dagger$, $b= W_{i+1}d_{i+1}$, and $a=U W_i d_i$, giving the full KL divergence:

%\begin{align*} \label{KL}
%&KL =\frac{1}{2}\bigg{(} \Tr [-\ln \big{(}(U D_i U^\dagger) D_{i+1}^{-1}\big{)} -\mathbbm{1} + D_{i+1}^{-1}(U D_i U^\dagger) ] \\
%&+(W_{i+1}d_{i+1}-U W_i d_i)^\dagger D_{i+1}^{-1} (W_{i+1}d_{i+1}-U W_i d_i) \bigg{)}
%\end{align*}

%Where the term inside the $\ln$ was switched with it's inverse and picked up a minus sign. This is for convenience of minimizing, we will i.g. minimize over the new %value $D_{i+1}$, and it is most convenient to work with when we deal with its inverse, because it's a simple sum $\Phi^{-1} + R^\dagger N^{-1} R$. \newline

\begin{align*} \label{KL}
&\mathrm{KL} =\frac{1}{2}\bigg{(} \Tr [\ln \big{(}D_{i+1}(U D_i U^\dagger)^{-1}\big{)} -\mathbbm{1} + D_{i+1}^{-1}(U D_i U^\dagger) ] \\
&+\bra{W_{i+1}d_{i+1}-U W_i d_i} D_{i+1}^{-1} \ket{W_{i+1}d_{i+1}-U W_i d_i} \bigg{)}
\end{align*}

\begin{figure}[h]\centering
\includegraphics[width=0.8\textwidth]{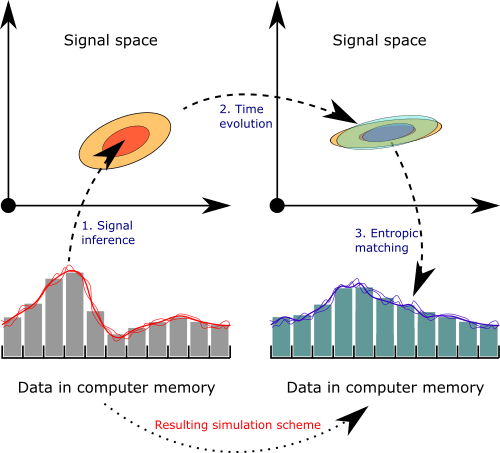}
\caption{Schematic representation of the IFD approach. The ovals in the above picture represent level sets of multivariate Gaussians. Picture courtesy of 
Torsten En\ss{}lin. }
\label{method}
\end{figure}

\noindent We now pick the new data $d_{i+1}$ such that the KL divergence is minimized. This is equivalent to saying \textit{we pick the new data so that we retain the maximum possible amount of information}. A visual representation of this process is shown in figure \ref{method}. To do this, first notice that there is only one term in the KL divergence which is dependent on the data, the term containing the inner product:

\begin{equation}\label{innerprod}
\braket{UW_i d_i -W_{i+1} d_{i+1}|D_{i+1}^{-1}|UW_i d_i -W_{i+1} d_{i+1} }
\end{equation}
we take the derivative with respect to $d_{i+1}$ and set it to zero, which yields

\begin{align*}
0= & W_{i+1}^\dagger D_{i+1}^{-1}W_{i+1} d_{i+1} -W_{i+1}^\dagger D_{i+1}^{-1} UW_i d_i \\
d_{i+1}= &(W_{i+1}^\dagger D_{i+1}^{-1}W_{i+1})^{-1}W_{i+1}^\dagger D_{i+1}^{-1} UW_i d_i
\end{align*}

The $t_{i+1}$ terms in the above equation can be simplified using the the Wiener filter formula:

\begin{align*}
&\big{(}W_{i+1}^\dagger D_{i+1}^{-1}W_{i+1}\big{)}^{-1}W_{i+1}^\dagger D_{i+1}^{-1}\\
&=\big{(}N_{i+1}^{-1}R_{i+1}D_{i+1}\underbrace{D_{i+1}^{-1}D_{i+1}}_{\mathbbm{1}}R_{i+1}^\dagger N_{i+1}^{-1}\big{)}^{-1} N_{i+1}^{-1} R_{i+1}
\underbrace{D_{i+1} D_{i+1}^{-1}}_{\mathbbm{1}} \\
&=\big{(}R_{i+1}\underbrace{D_{i+1}R_{i+1}^\dagger N_{i+1}^{-1}}_{=W_{i+1}}\big{)}^{-1}N_{i+1} N_{i+1}^{-1} R_{i+1} =(R_{i+1}W_{i+1})^{-1}R_{i+1}
\end{align*}
Thus giving a  full update equation of:

\begin{equation}\label{dataupdate}
\boxed{d_{i+1}=(R_{i+1}W_{i+1})^{-1}R_{i+1}UW_i d_i}
\end{equation}

Because it will often be referred to, $(R_{i+1}W_{i+1})^{-1}R_{i+1}UW_i$ will be called the update operator, or transport operator (following the terminology of \cite{Hydro}), and we denote it by $T_i$.\newline

To achieve a practical simulation scheme, the expansion of the operator $U$ has to be truncated to some desired order in $\Delta t$. This is because finding the time evolution operator is equivalent to solving the system. So if $U$ was known to arbitrary order, we would not need to do the simulation. From here on we denote the
truncated expansion by $\bar{U}=\sum_{k=0}^{\alpha} (\Delta t L)^k/k!$ for some order $\alpha$. \newline

This update operator is actually computable, despite the fact that many of the involved matrices are infinite-dimensional, because the update operator as a whole is still finite. Given a prior, response and the equations of motion for the system, this operator can often be computed algebraically. When not, one can simply take a very-high resolution approximation to signal space, whose resolution is much higher than that of data space, precompute the update operator numerically at the start of the simulation, and save it in memory. In light of these considerations, it becomes apparent that the current IFD framework is best suited to time-invariant systems. If 
the update operator needs to be continually recomputed, then the time spent computing the operators may eclipse the time spent actually updating the data, because the approximation to signal space must always be of much higher resolution than that of data space. \newline

To compute the data update equations for the case of a nonzero prior mean field, observe the general form for the KL in lemma \ref{lemmaKL}. If the roles of the two Gaussians are swapped, the operator in the middle of the inner product changes from $B^{-1} \to A^{-1}$. For our case this would correspond to taking the divergence in the ``wrong'' direction, and thus switching the roles of the evolved posterior from $t_i$ and the new posterior at $t_{i+1}$. Using this knowledge, for any incorrect formula in \cite{IFDmath}, we can fix it by sending $U D_i U^\dagger \to D_{i+1}$. For the case with the prior mean field, we copy the result presented in \cite[p.~55]{IFDmath} and fix the $D$ terms, which yields:

%\begin{equation}
\begin{align} \label{meanfield}
&d_{i+1}=(R_{i+1}W_{i+1})^{-1}R_{i+1}\big{(} U m(d_i) - \psi_{i+1} \big{)} + R_{i+1}\psi_{i+1} = \\
&(R_{i+1}W_{i+1})^{-1}R_{i+1}\big{(}U\big{[} \psi_i + W_i(d_i-R_i\psi_i) \big{]} - \psi_{i+1} \big{)} + R_{i+1}\psi_{i+1}
\end{align}
%\end{equation}

\section{Redundant parameters and simplifications} \label{redundancy}
\subsection{Prior mean field}
The IFD formalism, as presented so far, has placed absolutely no restrictions on the form of the response, prior, noise and mean field at each timestep. This was indeed a deliberate choice, intended to maximize the generality in the derivations of \cite{IFDmath}. It turns out however that the great freedom of choice for these parameters means that many of them are redundant, i.e. changing one is equivalent to changing another. Take for example the prior mean field, $\psi$. An assumed mean field makes sense in a pure inference problem, but it's role in a simulation scheme is not so clear. We expand eqn.~\ref{meanfield} and rearrange it into a term dependent on the data, and one dependent on the prior means:

\begin{align}
d_{i+1}= &(R_{i+1}W_{i+1})^{-1}R_{i+1} Ud_i \nonumber \\
&+ (R_{i+1}W_{i+1})^{-1}R_{i+1} (U(\mathbbm{1}-W_iR_i)\psi_i -\psi_{i+1}) +R_{i+1}\psi_{i+1} 
\end{align}

The first term is the one we want to keep, it is a linear update operator representing a linear equation, whereas the additional mean field terms introduce a drift at every timestep. The origin of this drift is easy to see: given the supposition of a prior mean field, the reconstruction of the data is an affine transformation, not a linear one, which moves the reconstruction towards the supposed mean. Repeatedly applying this operation will introduce a persistent artificial shift in the equations. It could be asked if a consistency condition is missing; if one believes something about the mean field at a certain time, then they should update their beliefs self-consistently, i.e. one should set $\psi_{i+1} = U \psi_i$, or perhaps $\psi_{i+1} = \psi_i$. Neither of these identities however, when substituted into the previous equation, eliminate the drift. \newline

%\begin{equation}
%(R_{i+1}W_{i+1})^{-1}R_{i+1} (U(W_iR_i)\psi_i) +R_{i+1}U \psi_{i} 
%\end{equation}

%The origin of this drift is easy to see. Given the supposition of a prior mean field, the reconstruction of the data is an affine transformation, not a linear one, %which moves the reconstruction towards the supposed mean. Repeatedly applying this operation will introduce a persistent shift in the equations. \newline
To see that the parameter is also redundant, we write down the inner-product term from the KL divergence for the case with nonzero prior means, from \cite{IFDmath}, with the appropriate correction:
\begin{equation}
\braket{Um(d_i) -m (d_{i+1})|D_{i+1}^{-1}|Um(d_i) -m (d_{i+1}) }
\end{equation}

with $m(d_i) = \psi_i + W_i(d_i-R_i\psi_i)$ and likewise for $m(d_{i+1})$. This term is minimized by making $m(d_{i+1})$ as close as possible to $Um(d_i)$. The former depends lineary on the data and the prior mean $\psi_{i+1}$, thus shifting one simply introduces a compensatory shift in the data to attempt to return to the minimum KL. Given the redundancy, and the persistent drift for which we have not yet found a use, we discard it as a simulation parameter in this project. \newline

%This shift is not perfect however, due to the mismatch between the reconstructions as mentioned earlier. This redundancy can be more easily seen, when the observer at time $t_{i+1}$ takes his assumption of the prior mean field to be $U(m(d_i))$, i.e. their prior mean is the evolved posterior mean from the previous timestep. This yields:

%\begin{align}
%&d_{i+1}=(R_{i+1}W_{i+1})^{-1}R_{i+1}\big{(} U m(d_i) - U m(d_i) \big{)} + R_{i+1}U m(d_i) \\
%&d_{i+1}=R_{i+1}U m(d_i) = R_{i+1}U W_i d_i
%\end{align}

%This is exactly equal to the no-noise version of the update equations. (in the limit of no noise, from the data space version of the wiener filter  definition, one can see that it has the property $RW =\mathbbm{1}$, so the $(R_{i+1}W_{i+1})^{-1}$ term disappears. So, the mean field is left arbitrary, but we see that there are choices of the mean field that offload the data processing onto that of the mean field. This redundancy is expected to hold into the nonlinear case as well. FIX THIS WHOLE SECTION\newline
\subsection{Noise}
We now focus on the next redundant parameter, which is the noise. The noise is a parameter which typically adds uncertainty to a measurement. However, at every timestep except the first (which may be the result of a real measurement), this noise is completely fictitious, as the simulation scheme is performing hypothetical measurements. One naturally asks why we should have update equations which adjust for a nonexistent uncertainty. The answer is that the noise term was initially kept in the equations in the hope that it may be a useful tunable parameter for the reconstructions, or it may help to describe the uncertainty in the simulation coming from numerical error. It turns out however, that the noise can simply be discarded as a parameter in IFD:

\begin{lemma}
The equations of motion for linear IFD are independent of the noise up to a simple equivalence.
\begin{proof}
For a simulation scheme with timesteps ${t_i}$ for $i \in \{1,...n\}$, responses $\{R_i\}$, priors $\{ \Phi_i\}$, noises $\{ N_i\} $, Wiener filters $\{ W_i=\Phi_i R_i^\dagger (R_i \Phi_i R_i^\dagger +N_i)^{-1}\}$, and linear time evolution operators $\bar{U}_i= 1+\Delta t L_i+...$, the data update equations are given by:
\begin{multline}
d_{i+1}=(R_{i+1}W_{i+1})^{-1}R_{i+1}\bar{U}_i W_i d_i \\
=\big{[}R_{i+1}\Phi_{i+1} R_{i+1}^\dagger (R_{i+1}\Phi_{i+1} R_{i+1}^\dagger +N_{i+1})^{-1}\big{]}^{-1}R_{i+1}\bar{U}_i \Phi_{i} R_{i}^\dagger (R_{i}\Phi_{i} R_{i}^\dagger +N_i)^{-1}
\end{multline}
The second line is obtained by inserting the definition of the Wiener filter. We rename the terms: $(R_i \Phi_i R_i^\dagger +N_i)=C_i$ , $(R_i \Phi_i R_i^\dagger)=B_i$ and $R_{i+1} \bar{U}_i \Phi_i R_i^\dagger=A_i$,
yielding:
\begin{equation}
d_{i+1}= (B_{i+1} C_{i+1}^{-1})^{-1} A_i C_{i}^{-1} d_i =  C_{i+1} B_{i+1}^{-1} A_i C_{i}^{-1}d_i
\end{equation}
The update equations are then iterated $n$ times, yielding:

\begin{equation}
C_n B_n^{-1} A_{n-1} \underbrace{C_{n-1}^{-1} C_{n-1}}_{=\mathbbm{1}} ... A_0 C_0^{-1} d_0= C_n \big ( \prod_{i=0}^{n} B_{i+1}^{-1} A_i \big ) C_0^{-1}d_0 
\end{equation}

The only terms with any dependence on the noise were the $C$ terms and therefore, up to a change of basis at the beginning and end of the simulation, the equations of motion are independent of the noise. This holds even if the noise, response and prior change at every timestep. In the infinite-noise limit, $C \to N$, and in the zero noise limit $C \to B$.

\end{proof}
\end{lemma}

\noindent Given the equivalence, from here on in we will 
always work in the no-noise limit, and the data update procedure becomes: 
\begin{equation}
\prod_{i=0}^{n} A_i B_{i}^{-1}= \prod_{i=0}^{n}R_{i+1} \bar{U}_i \Phi_i R_i (R_i \Phi_i R_i^\dagger)^{-1}= \prod_{i=0}^{n}R_{i+1} \bar{U}_i W_i
\end{equation}
with $W_i$ now being the no-noise version of the Wiener filter. This means we can always write the transport operator as:  $T_i =R_{i+1} \bar{U}_i W_i$. Furthermore, many times throughout this project, an unchanging response will often be used. We can then exploit the fact that in the no noise case, the Wiener filter has the property $R_iW_i =\mathbbm{1}$, to rewrite the transport operator in the following useful form:

\begin{equation}
T_i = R (\mathbbm{1}+\Delta t L + \cdots) W_i = \mathbbm{1}+ \Delta t  R  L W_i + \Delta t^2  R  L^2 W_i+ \cdots
\label{constantR}
\end{equation}
This form is valid even if the prior changes at every timestep. The no-noise assumption allows us to free up the symbols $N$ and $n$, which will now denote integers etc. \newline

Worth mentioning is that the unsimplified transport operator found in \ref{dataupdate}, was of a similar form, but had a $(R_i W_i)^{-1}$ term out the front, whose purpose was unclear. Some readers may have also found the derivation of the update equations rather odd, since the entropic matching of two probability clouds, at first sight, has little to do with numerical simulation schemes. However the new form of the transport operator,  $R_{i+1} \bar{U}_i W_i$, has a particularly simple interpretation: we guess the true field using the Wiener filter, evolve it, then measure the field again at the new timestep. All of this is without any reference to KL divergences.\newline

%Given the time evolved posterior $\mathcal{P}(\phi(t_{i+1})|d_i)$, the update equations also now have the interpretation:

%\begin{equation}
%d_{i+1}= \mathbb{E}[R \phi(t_{i+1})]_{\mathcal{P}(\phi(t_{i+1})|d_i)} =R \mathbb{E}[\phi(t_{i+1})]_{\mathcal{P}(\phi(t_{i+1})|d_i)} =
% R\bar{U} \underbrace{\mathbb{E}[\phi(t_{i})]_{\mathcal{P}(\phi(t_{i})|d_i)}}_{W_i}
%\end{equation}

%where the second and third equalities come from the linearity of expectation values. This has a very simple interpretation: the new data is the expected value of a field measurement given the evolved posterior from the old data. All of this is without any reference to KL divergences.\newline
\subsection{Data/response equivalence}\label{DRequiv}

The next redundancy is the responses. The ability to select new responses at any time gives us the freedom to change to a more convenient coordinate system whenever we require. The responses can however be dynamically updated in a way that is equivalent to updating the data; by picking $R_{i+1}=R_i U^{-1}$. A time-evolved response should give an unchanging data output when acting on a time-evolved field: $d_{i+1}=R_{i+1} \phi(t_{i+1})= R_i U^{-1} U \phi(t_i) = R_i\phi(t_i)=d_i$. 
This holds at least for a hypothetical measurement of the field. We should check that this behaviour is also reflected in the update equations:

\begin{equation}
d_{i+1}=R_{i+1}U \Phi_i R_i^\dagger (R_i \Phi_i R_i^\dagger)^{-1}d_i = R_i U^{-1}U \Phi_i R_i^\dagger (R_i \Phi_i R_i^\dagger)^{-1}d_i =d_i
\end{equation}

which is consistent with our expectations. If the responses are updated according to the field evolution, then the data is static. Given a nontrivial time evolution, $R U^{-1}$ will of course need to be simulated. This however, would mean that we have gone from attempting to model a field, to attempting to model an operator on the space of fields, and thus we have gone up a whole level of complexity. It will therefore in general be hard to exploit this equivalence. \newline

%The IFD formalism requires that any approximation to signal space should be of much higher resolution than that of data space. Given that it is computationally much harder to evolve objects in signal space, it will in general be hard to exploit this equivalence. \newline %Furthermore, if we did have the resources to simulate effectively in signal space, then it would be more efficient to simulate the mean field $Um(d_0)$ directly than to simulate a whole matrix $RU^{-1}$, and not bother with repeatedly passing through data space. \newline

As we will see later though, a code is developed in which the responses are partially updated using this equivalence, and in that specific case some data processing is offloaded onto the response, where the computation is more convenient. It must be pointed out that in this process, we have swapped out an information-theoretic update on the data side (one that is constructed to minimize information loss) for a non information-theoretic update on the response side. This is because we have not yet prescribed \textit{how} $RU^{-1}$ is approximated. An information-theoretic response update would involve taking equation 
\ref{KL}, fixing $R_i$ and minimizing w.r.t $R_{i+1}$, which should yield a set of responses which best capture the time evolved reconstruction and lose the least information.\newline

However a short look at eqn.~\ref{KL}, shows that inserting ${D=(\Phi + R^\dagger N^{-1}R)^{-1}}$, ${W= \Phi R^\dagger (R \Phi R^\dagger +N)^{-1}}$ etc. and attempting to minimize the KL will result in a highly nonlinear equation in the response matrices, objects which are themselves very high dimensional. For the purposes of this project, we discard the possibility of updating the responses information-theoretically; 
trying to model the linear evolution of a field by going through a nonlinear equation in operators \textit{on} fields is probably suboptimal\footnote{The response/data equivalence was already discovered in \cite{IFD}, where it was shown that for a prior which is invariant under time evolution ($ U \Phi U ^\dagger = \Phi$), evolving the response as $RU^{-1}$ perfectly preserves information. However if $U$ is not analytically known, we are back at square one.}. \newline

There was a point to this detour: the previously mentioned code in which we partially update the responses is not 100\% information-theoretic. This should be kept in mind.

\section{Errors, stability and convergence}\label{ESC}

Until now, it has not been proven that the IFD update equations actually converge to the true time evolution of the field.  The previous publication \cite{IFDmath} proved that the objects involved are mathematically well-defined, but it was not proved that schemes produced using this framework
actually represent the equations of motion of the system. However, given the extremely general nature of the IFD framework, such a discussion is difficult. The IFD equations can represent sensible schemes, and they can also deliver rubbish. The reader will unfortunately have to wait a while before an example of a sensible scheme is presented, but we can present an illustrative example of the latter case.

\begin{example}
In the worst case scenario, suppose one has a time-invariant system, and that $U$ is expanded to first order $\bar{U}=\mathbbm{1}+ \Delta t L$. Suppose further that the responses and prior are static. It is possible to choose a response and prior that are so poorly designed, that for every vector $v$ in the image of the Wiener filter, $Lv$ lies in the kernel of the response. This would mean that

\begin{equation}
d_{i+1} = R (\mathbbm{1} + \Delta t L) W d_i = d_i + \Delta t RLW d_i = d_i
\end{equation}

Thus the data does not evolve in time, and the reconstructions do not evolve either, and the model achieves nothing. The kernel of the response is infinite dimensional, so we should expect the useless schemes to outnumber the useful ones.

%Suppose now one has a collection of very narrow boxes spaced $\Delta x$ apart, and each having width less than $\Delta x/2$. Now suppose we have a correlation function (prior is convolution over this correlation function) which is compactly supported on a domain of width also less than $\Delta x/2$, then $R \Phi R^\dagger$ is proportional to the identity, because the convolution over a box still does not intersect with any other box. Now suppose our time evolution was simple advection to the right. (NEED A PICTURE), our smoothed boxes get shifted, but not enough so that they cross over into any neighbouring box. Thus when we act on this with the response, there is no off-diagonal terms. Furthermore, because some mass has been shifted outside the box, the full update equations are simply multiplication with a multiple of the identity matrix whose value is less than one. Our equations now describe an exponentially falling field. When this is repeated multiple times, the reconstruction has the exact same form, just smaller. So it is not that we initially had an undersampled field, and the update equations represent an accurate simulation of our (bad) knowledge state. No, the equations do not even reflect the evolution of our knowledge state either. NOT EVEN GUARANTEED TO CONVERGE IN THE LIMIT OF DELTA T TO ZERO.
\end{example}

\subsection{Error}

Given some intuition about the problem, we are now ready to derive some general formulas for the error. The previous example showed that we cannot expect to prove that in IFD the error is always less than some bound, or that the codes produced always converge. Therefore, errors/convergence must be checked specifically for each new model that we develop.\newline

The error is defined as the difference between the true solution and the simulated solution. For a simulation on a discretized grid in time and space with locations $\{ x_i \}$ and $\{t_j \}$, if the simulated solution is denoted by $\phi^j_i$ and the true solution is $\phi(t_j, x_i)$, then the global error at timestep $j$ is typically defined as:

\begin{equation}
E_j = \phi^j_i -\phi(t_j, x_i)
\end{equation}

The true analytic solution is in general unknowable, and the error is often estimated by looking at the \textit{one-step error} (OSE) or \textit{local error}, which is the error accumulated in a single timestep. We represent one step of the numerical simulation by the operator $T$. We then assume that at timestep $t_j$, there is no error: $\phi^j_i =\phi(t_j, x_i)$, and then define the one-step error by $\mathrm{OSE}=T[\phi^j_i] -\phi(t_{j+1}, x_i)$. The global error is then bounded by summing the absolute values of the local error at every timestep (\cite{Numerics} p.593). \newline

IFD was constructed with the explicit goal of minimizing the information lost at each timestep via the KL divergence. Thus in IFD, the criteria for success is to minimize the information theoretic error rather than the traditional error. Given that the KL divergence is an abstract distance between probability distributions, these two notions may have little to do with one-another. However, it turns out that in IFD there are three valid ways of interpreting error, and they are all roughly equivalent\footnote{It must be noted that this argumentation applies when the data is the only parameter being updated information-theoretically. If, for example, the responses are being updated information theoretically as discussed in subsection \ref{DRequiv}, then the full KL formula will need to be considered.}. Observe the inner-product term in the KL divergence formula:

\begin{equation}
\bra{W_{i+1}d_{i+1}-U W_i d_i)} D_{i+1}^{-1} \ket{W_{i+1}d_{i+1}-U W_i d_i}
\end{equation}

this represents the information lost when passing from one timestep to the next, and is thus the local information-theoretic error, which will now be denoted by $E_{KL}$. One sees from inspection that $E_{KL}$ goes to zero if and only if the local signal-space error $E_s$: 
\begin{equation}
E_s \equiv ||W_{i+1}d_{i+1}-U W_i d_i|| \quad \text{($\mathcal{L}^2$ norm)} 
\end{equation} 
goes to zero. The global signal space error is naturally $||W_{i}d_{i}-\phi(t_i)||$. We have the liberty of measuring the error in signal space because we have a formula for reconstructing the field given the data, in contrast to other numerical schemes. For the KL and signal space errors, if the error approaches zero in one norm, then it approaches zero in the other. % The terms only really differ by a factor of  $D_{i+1}^{-1}$, and a square. \newline

%Now, we note that there is another notion of error for the data, which is the actual error in data space: $||d_{i}-R_{i}\phi(t_i)||$ with $\phi(t)$ being the actual real field. %=||R_{i+1}(W_{i+1}d_{i+1}-U W_i d_i)|| \le ||R||\cdot||W_{i+1}d_{i+1}-U W_i d_i||$. 
%This corresponds to the typical notion of error for finite-difference codes. 
There is also a correspondence between the signal-space error and the typical notion of error, which we will refer to as data-space error, and denote by $E_d$. This notion must be slightly generalized, because IFD allows us to take arbitrary measurements of the data. We set $E_d = d_i - R_i \phi(t_i)$, and for a response which is a series of point-measurements of the field (i.e. a grid of delta-functions), this agrees with the old definition. We use the no-noise Wiener filter identity $R_i W_i =\mathbbm{1}$ to write:

\begin{equation}
|E_d| = ||d_i -R_i \phi(t_i)||=||R_i (W_i d_{i}-\phi(t_i))|| \le ||R_i||\cdot|| (W_i d_{i}-\phi(t_i))||
\end{equation}
%We set the global error in data space to be $d_i - R_i \phi(t_i)$, and for a response which is a series of point-measurements of the field (i.e. a grid of delta-%functions), this agrees with the old definition. We find the local error by assuming $d_i -R_i \phi(t_i) =0$ at timestep $t_i$. We then use the no-noise Wiener filter %identity $R_i W_i =\mathbbm{1}$ to write:
%
%\begin{equation}
%||E_d|| = ||d_{i+1} -R_{i+1} U \phi(t_i)||=||R_{i+1} (W_{i+1} d_{i+1}-U\phi(t_i))|| \le ||R_{i+1}||\cdot|| (W_{i+1} d_{i+1}-U\phi(t_i))||
%\end{equation}
where $||R_i||$ denotes the operator norm.
This equation shows that convergence in signal space implies convergence in data space, but not vice versa. As long as the Wiener finter reconstructions are not perfect, there will be some signal-space error. When analysing error, since the true behaviour of the field is unknown, the error can typically only be expressed by an upper bound that has some dependence on the time and space resolutions $\Delta t$ and $\Delta x$.\footnote{Astute readers will note that a notion of $\Delta x$ for IFD has not been constructed yet, this will be done in the next section.} One then asks how the error scales in the limit $\Delta x, \Delta t \to 0$. In the limit of high resolutions, we expect the Wiener filter reconstructions to become perfect, i.e. $W_iR_i \to \mathbbm{1}$, and in this limit, signal space and data space approach one another and the notions of error become equivalent\footnote{Note that $W_iR_i$ is an operation on signal space describing the act of measuring then reconstructing. This is different to the operator on data space, $R_iW_i$, which describes reconstructing then measuring, and is always $\mathbbm{1}$ for our purposes.}. \newline

%Because the image of the Wiener filter can only be of the same dimension as data space, the reconstruction will never achieve perfect results. This means that for the error in signal space to go to zero, we must take some sort of limit of high spatial resolution such that the reconstruction converges. In this limit, convergence in signal space and data space become equivalent. \newline

%or this reason, the error calculations will be carried out in signal space, because the scaling is the same as that of the KL, up to a square. This simplifies calculations by avoiding the $D_{i+1}^{-1}$ term in the KL error. Note however that the square in the scaling of the KL divergence is purely cosmetic factor that has to do with units, and does not represent an underlying higher accuracy in the IFD framework. \newline
% Furthermore, it will often suffice just to look at the data-space error, because at each timestep we project back down to data space, it doesn't matter how good our reconstruction is, if none of this information is retained in the next timestep. So the error that actually accumulates is the error which shows up in the data at the next timestep. I.e. if $d_i = R\phi(t_i) +E_i$ with some acculmulated error $E$, then the reconstruction is off by at least $||W_i E_i||$ in addition to the reconstruction error. \newline

Thus, there are three ways to analyse the error in IFD, all of which are roughly equivalent. Note however that the scaling of data/signal space errors differs from the KL error by a factor of a square. This is however a purely cosmetic factor that has to do with units, and does not represent an underlying higher accuracy in the IFD framework. \newline

We now seek a general formula for the local data space error. We start by assuming zero accumulated error, i.e. $d_i =R_i \phi(t_i)$.
We must then pick an expansion of $U$ to some finite order $\alpha$, $\bar{U}=\sum_{k=0}^{\alpha} (\Delta t L)^k/k!$. This gives the data update equation
 $d_{i+1}=R_{i+1} \bar{U} W_i d_i$, which can then be used to bring the local error into the following form:

\begin{equation}
\begin{aligned} \label{error}
&E_d=||d_{i+1}-R_{i+1}\phi(t_{i+1})||=||R_{i+1}\bar{U} W_i d_i-R_{i+1}U \phi(t_i)|| \\%= ||(W_{i+1}R_{i+1}-\mathbbm{1})U W_i d_i||
& \le ||R_{i+1}\bar{U} (W_iR_i -\mathbbm{1}) \phi(t_i)||+ ||R_{i+1} \sum_{k=\alpha+1}^{\infty} \frac{(\Delta t L)^k}{k!} \phi(t_i)||
%=||R_{i+1}\bar{U} (W_iR_i -\mathbbm{1}) \phi(t_i))|| \\
%&\le ||(W_{i+1}R_{i+1}-\mathbbm{1})\bar{U} W_i d_i|| + || \sum_{k=\alpha+1}^{\infty} \frac{(\Delta t L)^k}{k!} W_i d_i||
\end{aligned}
\end{equation}
%\begin{equation}
%E_s=||W_{i+1}R_{i+1}\hat{U} W_i d_i-U W_i d_i|| \le ||(W_{i+1}R_{i+1}-\mathbbm{1})\hat{U} W_i d_i|| + || \sum_{k=\alpha+1}^{\infty} \frac{(\Delta t L)^k}{k!} W_i d_i||
%\end{equation}

The second term is the expected truncation error from an order $\alpha$ time expansion. All other higher order terms in time are still present, but are modified by the term $(W_{i+1}R_{i+1}-\mathbbm{1})$, which measures the accuracy of the Wiener filter reconstruction. Thus to bound the error to any order, a bound needs to be placed on the spatial part of the reconstruction.\newline 

This formula shows that for any IFD code, the one-step error is determined by only three factors: how much is lost by measuring then reconstructing $(W_{i+1}R_{i+1})$, the order of the expansion of $U$, and how much of the time-evolved reconstruction is captured by the new response $R_{i+1}$. Error will be analysed for only one class of models in this report, where we only observe the scaling behaviour, and thus are working in the limit where data and signal spaces approach each other. Nonetheless, it is useful to have a general formula for the error in IFD. \newline

\subsection{Stability}

Stability is easy to describe intuitively. If there are certain solutions to a finite-difference scheme that grow without bound, when the actual physical solutions do not, then the code is unstable. The unbounded solutions will grow to dominate any simulation, no matter how small they are initially. To define stability rigorously, in the form that we need it, we first need to define a notion of spatial resolution in IFD, i.e. we need a $\Delta x$ to match our $\Delta t$. \newline

Now, the IFD framework technically doesn't need a notion of spatial resolution. After all, the responses can just be any arbitrary linear functions of the field, and don't need to be localized anywhere. However, there is already a significant wealth of theorems on stability, convergence and errors that exist in the literature, that rely on some notion of a spatial grid. Constructing a notion of $\Delta x$ will give us access to these. Because we are also not extraordinarily creative, every system of responses used in this report will correspond to some sort of spatial grid anyway. For this project we can simply state that for a 1D simulation domain of length $l$, and an $n$ dimensional data space, then $\Delta x = l/n$ as per usual. Higher dimensional domains are defined analogously. \newline

For more creative systems of responses, a concept of resolution is still easy to define. Any simulation scheme for PDE's is a finite approximation to an infinite-dimensional object, so there is always a notion of resolution. Therefore it will always make sense to ask what happens in the limit of high resolutions. If in doubt, for any IFD scheme one can simply take the dimension $n$ of data space, and set $\Delta x = C/n$ for some arbitrary constant $C$. \newline

%However, there is already a significant wealth of theorems on stability, convergence and errors that exist in the literature, which rely on some notion of a spatial grid. Constructing a notion of $\Delta x$ will give us access to these. \newline

%Because we are also not extraordinarily creative, most of the systems of responses we choose will correspond to a localized measurement of the field values (a.k.a some sort of spatial grid) anyway. But there may be supplementary coordinates, for example, say we were simulating some density field $f(x)$, then the total mass of the field $R f =\int f(x) dx$ could be a useful supplementary coordinate in addition to our grid coordinates, which would help enforce global conservation of mass. Despite this, it is very easy to write down a generalized notion of $\Delta x$: suppose $n$ is the dimension of data space, then pick an arbitrary positive constant $C$ and set $\Delta x = C/n$. The constant $C$ is arbitrary because we only care about the scaling of the error w.r.t $\Delta x$ anyway. For the typical notion of spatial grids over some domain of length $l$, 
%this returns the usual $\Delta x = l/n$.\newline

The \textit{Lax-Richtmyer} definition of stability \cite{Lax}, is that given a transport operator $T$ for the simulation, which defines $d_{i+1}=Td_i$, which is defined for some $\Delta x$ and $\Delta t$ which both go to zero, and a total simulation time $\tau$ such that $\Delta t =\tau /N$ for a number of timesteps $N$, then the simulation is defined to be stable if the set:
\begin{align}
& \{ T^n(\Delta t,\Delta x) | n \in \mathbb{Z}, 0\le n \le N \}
\end{align}

is uniformly bounded in the operator norm in the limit $\Delta x, \Delta t \to 0$. If the set is not uniformly bounded, then that means there is a solution whose magnitude grows without bound. \newline

There is a another notion of stability known as \textit{Von Neumann stability}, which is far easier to compute. Von Neumann stability analysis assumes that the coefficients of the PDE do not change in space, the grid spacing is regular, and that the boundary conditions are nice enough such that the transport operator will be translation-invariant and will thus have a diagonal representation in Fourier space. The eigenvalues $\{ \lambda_i \}$ of the transport matrix are then analysed, if the magnitude $|\lambda_i |$ of any of them is greater than one, then there is an exponentially growing mode and the simulation is unstable. Instability in the Von Neumann sense implies instability in the Lax-Richtmyer sense, but is only equivalent in certain cases \cite{Lax}. Given that stability is hard to check analytically for numerical solvers with nonconstant spatial coefficients, grids etc., it is common practice to analyse a new numerical scheme on a constant-coefficient problem, and use it as an indicator of stability for the nontrivial problem \cite[ch.~7]{Godunov}. 

\subsection{Convergence}

A notion closely related to error is that of convergence, which asks if the simulation scheme actually approaches the true behaviour of the field in the limit of high resolutions $\Delta x$ and $\Delta t \to 0$. This is equivalent to asking if the global error goes to zero.

%Until now, it has not been proven that the IFD codes in general converge to anything, or even represent the equations of motion. What is convergence? 
%That the error goes to zero in the limit of high accuracy. The previous document \cite{IFDmath} proved that the IFD scheme minimized the information theoretic error at each timestep, but that isn't the same as convergence. 
%Now, typically, to discuss convergence, people study the behaviour of codes in the limit of $\Delta t$ and $\Delta x$ going to zero. But IFD hasn't yet developed a notion of $\Delta x$, and in general, the responses don't need to necessarily correspond to spatially-localized measurements. 
%Or, we may have bins for example, supplemented by an extra piece of information that is a linear function of the field in question. Say we were simulating a CR spectrum, we could also measure the total energy of the system $\int \epsilon(p) f(p) dp$, and that would be a supplementary coordinate in addition to our grid coordinates. 
%But, we can define a generalized notion of $\Delta x$: suppose $n$ is the dimension of data space, then pick an arbitrary positive constant $C$ and set $\Delta x = C/n$.The constant $C$ is arbitrary cause we only care about scaling w.r.t $\Delta x$ anyway. For the typical notion of spatial grids over some domain of length $l$, we get the typical $\Delta x = l/n$ back out.\newline

Of the many theorems on convergence that a notion of $\Delta x$ gives access to, the absolute most important is the Lax-Richtmyer equivalence theorem \cite{Lax}. For a differential equation of the form $\partial_t \phi(t,x) = L(t)\phi(t,x)$, and a set of well-posed boundary conditions, the theorem states:

\begin{theorem}
Given a properly posed initial value problem, and a finite difference
approximation $T(\Delta t)$ to it that satisfies the consistency condition,
stability is a necessary and sufficient condition that $T(\Delta t)$ be a convergent
approximation.
\end{theorem}

What is consistency? Consistency essentially states that in the limit of high resolutions, the approximated operator actually approaches the true differential operator. This may seem obvious, but often one can write down sensible-looking schemes which are in fact not consistent. In the notation of Lax and Richtmyer, they assume that the spatial and time resolutions are coupled through some function $\Delta x =g(\Delta t)$ which ensures that the simultaneous limit of $\Delta x, \Delta t \to 0$ is always taken. We follow their notation, but note that it is mostly a formality.

\begin{definition}[Consistency] \label{cons}
For an operator $T(\Delta t)$ which approximates $U(t)$, with $U(t)$ being the analytic time evolution operator corresponding to $L(t)$, 
the approximation is said to be consistent, if for some set of solutions, $\Omega$, to the differential equation, then for any $\phi \in \Omega$, 
\begin{equation}
\lim_{\Delta t \to 0} \bigg{\|} \big{(} T(\Delta t) - U(\Delta t) \big{)} \phi(t,x) \bigg{\|} =0
\end{equation}
uniformly in $t$.
\end{definition}

The definition has been (harmlessly) paraphrased, with some technical details left out. It needs to be pointed out that the definition of consistency involves comparing operators which live in different spaces: $T(\Delta t)$ acts on a discrete space, yet $U(\Delta t)$ acts on a continuous space. The original paper by Lax et al. assumes that there is some sufficient level of smoothness such that Taylor series expansions or smooth interpolation etc. may be used to bridge the gap between spaces. We will not use the exact details here, and allow ourselves to freely make such statements as $(f_{i+1}-f_{i-1})/2\Delta x \to \partial_x f(x)$ as $\Delta x \to 0$.\newline

Thus, to analyse consistency for any general model, we will need to specify a rule for comparing these operators on different spaces, which will depend on our choice of responses etc. So a general formula for consistency of IFD schemes is not presented here. The Lax equivalence theorem is extremely useful because convergence is often hard to check, as it is a global property. Stability and consistency however, are both easy-to-check local properties. \newline

%We in fact already have such a rule, the Wiener filter. We place this in the definition by setting $T(\Delta t) = W_i R_i \hat{U}$.??!?!??!?!?!?!?!???! When we concatenate multiple such operators together, we get the data update equation back out. Putting this in the definition yields

%\begin{equation}
%|| \big{(}W_iR_i \hat{U}(\Delta t)- U(\Delta t)\big{)}\phi(t) ||
%\end{equation}
%
%Now, we may want to keep higher orders of $\Delta t$ in our equations for accuracy, but for consistency, we only care about whether or not we converge to the %solution, regardless of how fast. Keeping only first order in $\Delta t$, we get 
%
%\begin{equation}
%|| \big{(} (W_iR_i(\Delta t)- \mathbbm{1} \big{)} ( \mathbbm{1}+ {\Delta t} L) \phi(t) ||
%\end{equation}
%
%Where the $W_iR_i(\Delta t) $ are written through some coupling of $\Delta t$ and $\Delta x$. From this, it is exceedingly obvious that we need $W_iR_i(\Delta x) \to %\mathbbm{1}$, and then we will have consistency. Now, as previously stated, given the extremely general nature of the IFD equations, we cannot go any further with %this on a general level. Consistency will have to be analysed for specific cases. 

\section{Implicit methods}

IFD, has so far been constructed as an \textit{explicit scheme}, the data at a new timestep is solved as an explicit function of the data at the previous timestep. The most basic example of a forward difference scheme is the Euler method. Given the DE $\partial_t \phi(x,t) =f(\phi(x,t))$, the update scheme is then ${(\phi_{i+1}-\phi_i)/\Delta t} = f(\phi(t_i)) \Rightarrow {\phi_{i+1} =\phi_i +\Delta t f(\phi(t_i))}$. \newline 

Forward Euler schemes can have drawbacks, for example, they are often unstable. One common remedy for this is to use an \textit{implicit scheme}, such as the backward Euler method, which solves an implicit equation expressing the data at $t_i$ in terms of $t_{i+1}$, i.e. ${\phi_{i+1} =\phi_i +\Delta t f(\phi(t_{i+1}))}$. For nonlinear equations, the implicit schemes are in general harder to solve \cite[sec.~13]{Numerics}. We did attempt the use of a backward scheme during this project, but was not included in this report. \newline

If one chooses constant responses and prior, and a first order time expansion, then eqn.~\ref{dataupdate} is brought into the form $d_{i+1} = d_i + \Delta t RLW d_i$, which is cosmetically identical to the forward Euler scheme. One is then tempted to interpret the above as a DE in continuous time: $d'(t) =RLW d(t)$, from which a backward Euler method is defined. However the data is not a continuous variable, and this construction doesn't generalize to nonconstant responses, which are only defined at discrete timesteps. Even if one could construct an $R(t)$, it wouldn't account for the possibility of the dimension of data space changing between timesteps, which is not a continuous transformation. \newline

%\begin{align*}
%&R\sum_{k=0}^{\alpha} \frac{(\Delta t L)^k}{k!} \Phi R^\dagger (R \Phi R^\dagger)^{-1}& \text{instead of    } & 
%\sum_{k=0}^{\alpha} (\Delta t )^k \frac{(R L\Phi R^\dagger (R \Phi R^\dagger)^{-1})^{k}}{k!}
%\end{align*}

Implicit methods may still be constructed in IFD, they just need to be defined correctly. Start with a reconstructed posterior at a time $t_{i+1}$ given some data $d_{i+1}$, and time evolve this distribution backwards, and then minimize the KL divergence w.r.t. the data at $t_i$, this gives:

\begin{align}
&d_i = R_i \bar{U}^{-1} W_{i+1} d_{i+1} = R_i (\mathbbm{1} - \Delta t L +\ldots) W_{i+1} d_{i+1} \nonumber \\
&\Rightarrow d_{i+1} = [R_i (\mathbbm{1} - \Delta t L +\ldots) W_{i+1}]^{-1} d_i
\end{align}

The fix applied to the definition was scarcely worth mentioning, except for the fact there are a variety of creative timestepping schemes in existence such as Runge-Kutta methods etc. that future studies may want to apply to IFD. The moral here is that often such schemes assume that one is solving an underlying continuous equation, which eqn.~\ref{dataupdate} is not. So, one should check the validity of such schemes in IFD before proceeding.

\section{Boundary conditions}
The astute reader will have noticed that boundary conditions have not been mentioned yet. This is because they aren't baked into the IFD formalism. To be able to compute the necessary Gaussian functional integrals, we needed to exploit the fact that we are working in a vector space of functions. Now for general boundary conditions, this isn't true. As a trivial example, take two functions $f$ and $g$ on an interval $[0,L]$, given that $f(0)=g(0)=a$ and $f(L)=g(L)=b$, but $f+g$ does not equal $a$ or $b$ on the boundaries. The addition of boundary conditions typically restricts us to an affine space. It would probably be easy to extend the IFD formalism to handle boundary conditions and affine spaces, but that is beyond the scope of this project. \newline

The complicating factor with IFD is that the reconstructions of the field given the data are typically nonlocal. This is the desired behaviour, in that the spatial correlation structure of the field is used to get better inferences of the field at a location than one would get from a local reconstruction. The nonlocality comes from the $(R\Phi R^\dagger)^{-1}$ term in the Wiener filter. With local spatial correlations, the $R \Phi R^\dagger$ matrix is indeed local (i.e. almost diagonal in the spatial indices), however the matrix inversion depends on global properties of the system. Thus throughout this thesis, boundary conditions, when implemented at all, will be done in an ad-hoc manner, and only for some models.

%Well, the IFD equations are in general nonlocal, even for local time evolution equations. Mathematically, this comes from the $(R\Phi R^\dagger)^{-1}$ term. While the non-inverted matrix is typically local (i.e. vanishes away from the main diagonal), the inverse of such a matrix in general is not. Intuitively, this non-locality comes from the fact that we used the spatial correlations of the field to guide the reconstructions, so a reconstructed field value depends on data points that may be very far away. However, if the operators we are modelling are local, experience shows that we generally get out an operator that is quasi-local. i.e. an updated data point typically only has meaningful contributions from points in some small neighbourhood around it, with contributions dropping to zero with distance. We can typically implement boundary conditions by not just specifying some value at the boundary, but on some region extending outside the boundary as well. Or, one can have a 'buffer zone' around the boundary of normal, non-IFD finite-difference solvers, such that the nonlocal part of the IFD equations doesn't really touch the boundary THIS WHOLE PARAGRAPH MAKES NO SENSE. We find a way of falling back on implementing normal finite-diff boundary conditions.

\section{Trial system: Cosmic Ray Simulations}

The original goal of this project was to develop a first real-world application for IFD, and the problem chosen for this was the simulation of cosmic ray transport. Cosmic rays (CR's) are very high energy particles originating in space, and consist mainly of charged protons. On galactic scales they can be described by the equations of Magnetohydrodynamics, which describes the behaviour of charged plasmas by combining Maxwell's equations and the Navier-stokes equations.  In \cite{Skilling}, the author derives an equation for cosmic ray propagation, in which the cosmic ray plasma is represented by a number density distribution $f(\vec{x}, \vec{p}, t)$ in the three-dimensional location $\vec{x}$ and momenta $\vec{p}$ and time $t$. For low energy cosmic rays, it is believed \cite{CRmain} that over galactic scales, the magnetic field of interstellar space has a far stronger effect on the movement of cosmic rays than the transport induced by their own momenta $\vec{p}$. In addition, the distribution of cosmic ray momenta is assumed to be nearly isotropic. This justifies replacing the vector momentum by a scalar, $p$. Under these assumptions, and some others, the cosmic ray transport equation in phase space is given by \cite{CRmain}:

\begin{equation}\label{CR}
\frac{\partial f}{\partial t} + \frac{\partial}{\partial x_i} (\nu_i f) + \frac{\partial}{\partial p} (\dot p f) =
\frac{\partial}{\partial x_i}(\kappa_{ij}\frac{\partial}{\partial x_j} f ) + q -\frac{f}{\tau_{loss}}
\end{equation}

where $\nu_i(\vec{x})$ is the cosmic ray transport velocity which advects the particles along magnetic field lines. This term is an a For simplicity it is assumed to be a vector quantity independent of the momentum.
$\kappa_{ij}(\vec{x})$, is the spatial diffusion tensor, which is highly anisotropic in directions parallel and perpendicular to the background magnetic field. $\dot p$ describes acceleration and deceleration of CR's in response to plasma waves and interaction processes. $q(\vec{x},p,t)$ is an injection term, which
corresponds to various processes in space producing cosmic rays. $\tau_{loss}(\vec{x},t,p)$ is the catastrophic loss timescale, which describes CR's being removed from the population due to collisions with interstellar gas etc. As it is not relevant to this thesis, we neglected a momentum space diffusion term describing second order Fermi acceleration.\newline

The typical regions of interest for simulating CR populations are on the scales of galaxies and galaxy clusters \cite{CRsecond}, the latter of which are the largest bound objects in the universe. A complete treatment of the CR evolution equation would require simulating the momentum component with equal resolution to that of the spatial component. This promotes the $N^3$ scaling of the memory requirements for the spatial components to an $N^4$ scaling, which becomes unmanageable at the high spatial resolutions required for resolving cluster-scale structures \cite{Miniati}. Thus practical simulations are often limited to momentum resolutions of the order of only a handful of bins, $\approx$ 10, and assume a subgrid structure in order to get satisfactory results \cite{Miniati,Miniati-new}. This made the CR evolution equation a promising candidate for an application of IFD, as it was hoped that it would offer an improvement over current subgrid models. \newline

Developing a satisfactory simulation scheme was harder than initially expected, and there were many unexpected problems that needed to be solved, both theoretical and practical. Hence equation \ref{CR} wasn't solved in all its generality. It can be seen that if the injection and catastrophic loss terms are ignored, the equation is mostly just advection by a nonconstant velocity field and diffusion with a nonconstant diffusion tensor. The term $\frac{\partial}{\partial p} (\dot p f)$ is just advection in the momentum component by a velocity field $\dot p$. These difficulties required us to curb our expectations, and redefine
 ``success'' as simulating advection and diffusion in one dimension, with a nonconstant velocity field, and constant diffusion coefficient. The original problem was kept in mind, and thus a secondary goal for the project was developing a scheme which was viable at low resolutions.

\chapter{Translation-invariant schemes}\label{Tinvariant}
\section{Toy model: straight advection}

One-dimensional advection on a finite interval will form the first trial problem for which we can develop a toy model. The development of this toy model will be presented side-by-side with some analytical results on a broader class of models for systems where the time evolution is diagonal in Fourier space. These analytic results will be used to analyse the error and convergence properties of the toy model. This model will then be extended to nonperiodic boundary conditions and nonconstant velocity fields. The analytic results presented here are for the one-dimensional case, but they trivially generalize to higher dimensions. \newline

A disclaimer must be added at the start here, the error analysis for these models is carried out for the case of constant velocity and periodic boundary conditions. This case can already be solved analytically. However this practice is entirely normal in numerical methods, as many advanced schemes are too complicated to permit an analytic analysis. This is the case in IFD; as the codes are typically nonlocal, and the algebraic equations are always dependent on the geometry of the simulation domain. In this thesis, the best that we can do is prove convergence for the analytically solvable case, and then hope that these conclusions hold in the non-analytically solvable case. This should be kept in mind at all times. \newline

The equation to be solved is the one-dimensional advection equation:
\begin{equation}
\partial_t \phi(t,x) = \partial_x(v(x)\phi(t,x))
\end{equation}

for $v(x)$ some velocity field defined on $x \in [0,l]$, which will at first be set to a constant $v$ and brought in front of the partial derivative. \newline

We start by selecting the signal and data spaces. The simulated space inside the computer must always be of finite extent. For this reason, we choose the signal space to be $\mathcal{L}^2([0,l])$. This is despite the fact that the true field itself may live on all of $\mathbb{R}$. The justification for this is that $\mathcal{L}^2([0,l])$ should contain all the relevant information for the simulation, given appropriate boundary conditions. Furthermore, most results will be derived in Fourier space, where the restriction of domain converts contour integrals in momentum space to infinite sums, which are much more manageable. \newline

Now the prior must be selected. The system under consideration is fairly abstract, so there is a lot of room for choosing a prior which suits our goals. If one starts with the assumption that no point in the space is special\footnote{We will see later that for nonconstant velocity fields, this can cause problems, because some points \textit{are} special.}, then the prior will be translation-invariant. This property means that the prior will always have a diagonal representation in Fourier space. The positivity and self-adjointness conditions on the prior ensure that the eigenvalues in momentum space will be everywhere positive and greater than zero, and symmetric about the origin. Priors of this form are generally referred to as smoothness priors. Using $k$ to denote momentum, a prior $\Phi(k)$ whose values fall to zero as $k \to \infty$ essentially states that rapid oscillations in the signal are deemed unlikely; the field is smooth. \newline

Common examples of a prior include power laws in momentum, i.e. $|k|^\beta$ for some integer $\beta$, often supplemented by a regularizing mass term: $\Phi(k) = 1/(|k|^\beta + m^\beta)$. For our toy example, we choose $\Phi(k)= k^{-4}$, although any prior with $\Phi(k) \to 0$ as $k \to \infty$ sufficiently fast would also work.
%although we only really require that $\Phi(k) \to 0$ in $k$ fast enough such that $\Phi$ is defined on the functions we care about. 
The use of the Fourier transform tells us that we have already imposed periodic boundary conditions on the system. To extend to nonperiodic boundary conditions, we notice that in the case that $\Phi(k)$ has an inverse Fourier transform, then it's action on a function $\phi(x)$ takes the form of a convolution $\int \Phi(x-y) \phi(y) dy$. This helps extend to nonperiodic boundary conditions later. Throughout this report, we use the convention that the (signal space) Fourier modes are normalized as: $\frac{1}{\sqrt{L}}e^{ikx}$ with $k=2 \pi n /l$, for $n$ in the integers. \newline

An expansion order for $U$ must now be selected. For the toy model, we initially pick $\bar{U} = \mathbbm{1} + v\Delta t \partial_x$, but stability requirements will later require going to second order in time. \newline

Now to construct the responses. We assume that we have $N$ points which will be labelled with the index $j$. The responses are chosen to be constant in time, and the subscripts $R_j$ now denote spatial indices. The most natural and naive response is to choose that the index $j$ labels a regular grid of positions, and that the response is an average of the field value around that position.  We define $\Delta x=l/N$ and the box function:

\begin{equation}
B(x)=
\begin{cases} 
      1/{\Delta x} & 0\leq x\leq \Delta x \\
      0 & $otherwise$
   \end{cases}
\end{equation}

We then define the response operator from the signal space to the data space to be:

\begin{equation}
(R \phi)_j= \int_0^l dx B(x-x_j)\phi(x) \equiv \int_0^l dx B_j(x)\phi(x)
\end{equation}

where $x_j$ is the $x$-position of the $j$-th gridpoint, i.e. $x_j = \Delta x\cdot j$. This function simply averages the signal around a box centred at $\Delta x/2$.
The $B_j$'s form an orthogonal set, but they are not normalized since $\int B_j^2 dx =1/\Delta x$, nor are they a basis. \newline

It is quite easy to generalize this response to arbitrary functions $B(x)$ on $\mathcal{L}^2([0,l])$, simply set $(R \phi)_j= \int_0^l dx B_0(x-x_j)\phi(x)$. If the $x_j$'s are evenly spaced, we refer to any response of this form as a \textit{translation-invariant response}. The $B(x)$ functions will be referred to as the \textit{response bins} or just \textit{bins}. \newline

%We use the convention where our fourier modes are normalized as:
%%$\frac{1}{\sqrt{L}}e^{ikx}$ with $k=2 \pi n /L$.
%We consider the action of $R$ on the $k$-th Fourier mode, i.e we 
%%wish to find $R_{jk}$ where k indices will label fourier modes:
%
%\begin{equation}
%\begin{align*}
%& R_{jk}= \int_0^L dx B_j(x) \frac{1}{\sqrt{L}} e^{ikx} = \frac{1}{\sqrt{L}} \int_0^L dx B_j(x) e^{ikx} \\
%& = \frac{1}{\sqrt{L}} \int_{r_j}^{r_{j+1}} dx e^{ikx} = \frac{1}{\sqrt{L}} \bigg [ \frac{e^{ikx}}{ik} \bigg ]^{r_{j+1}}_{r_j} \\
%& \Rightarrow R_{jk} = \frac{1}{\sqrt{L}} \frac{e^{ikr_j}}{ik}\big( e^{ikl}-1 \big)
%\end{align*}
%\end{equation}
%
%And the adjoint of this is simply the complex conjugate.

We now want to begin to calculate the transport operator, starting with the computation of $(R \Phi R^\dagger)^{-1}$. Since both the responses and prior are invariant under translations of multiples of $\Delta x$, this allows us to actually make a very general statement. Given that a translation-invariant prior will be diagonal in momentum space, we can state:

\begin{lemma}\label{translation}
Given a signal space of the form $\mathcal{L}^2([0,l])$ with periodic boundary conditions, a translation invariant response $R_j$ whose bin function $B(x)$ has a Fourier series representation, as well as a prior $\Phi$ which is diagonal in momentum space, $(R \Phi R^\dagger)_{jl}$ will be of the form:
\begin{equation}
\sum_{k}^{} \Phi(k) |\widehat{B}(k)|^2 e^{ik(x_j-x_l)},
\end{equation}
where $\widehat{B}(k)$ is the Fourier coefficient of $B(x)$.
\begin{proof}
By the shift property of the Fourier transform, $\widehat{R}_{j,k}=e^{-ikx_j} \widehat{R}_{0,k} = \widehat{B}(k)$. Therefore $(R \Phi R^\dagger)_{jl}$ is
%\begin{equation}
\begin{equation}
(R\Phi R^\dagger)_{jl} = \sum_{k}^{} \sum_{q}^{} e^{ikx_j}\widehat{B}(k) \Phi(k) \delta_{kq} \widehat{B}^{*}(q)e^{-iqx_l}=\sum_{k}^{} \Phi(k) |\widehat{B}(k)|^2 e^{ik(x_j-x_l)}
\end{equation}
%\end{equation}
as desired. %WE GOTTA ADD IN THE conTiNUOus/ DISCRETE MOMENTUM REPRESENTATIONS.
\end{proof}

\end{lemma}

%\begin{corollary}
%If the signal space is $\mathcal{L}^2(\mathbbm{R})$, then the formula becomes:
%\begin{equation}
%\int \Phi(k) |\widehat{B}(k)|^2 e^{ik(x_j-x_l)} dk
%\end{equation}
%\begin{proof}
%Clear.
%\end{proof}
%\end{corollary}

\noindent The generalization to higher dimensions takes $x$ and $k$ to vectors. It must be stressed here that this formula holds regardless of how the update matrices are actually computed. We are not necessarily solving the equations in Fourier space, but we know that the operators always have such a representation. From now on, we will refer to any simulation scheme which satisfies the criteria of the previous lemma as a \textit{translation invariant scheme}.\newline

\noindent For the toy model, we compute the Fourier transform of $B(x)$:
%\begin{equation}
\begin{align*}
& \widehat{B}(k)= \int_0^l dx B(x) \frac{1}{\sqrt{l}} e^{ikx} = \frac{1}{\sqrt{l}} \int_0^l dx B(x) e^{ikx} \\
& = \frac{1}{\Delta x\sqrt{l}} \int_{0}^{\Delta x} dx e^{ikx} = \frac{1}{\Delta x\sqrt{l}} \bigg [ \frac{e^{ikx}}{ik} \bigg ]^{\Delta x}_{0} \\
&  = \frac{1}{\Delta x\sqrt{l}} \frac{\big( e^{ik\Delta x}-1 \big)}{ik}
\end{align*}
%\end{equation}

Thus, using the concrete form of $R$ and $\Phi(k) = 1/k^4$, we can immediately compute $(R\Phi R^\dagger)_{jl}$:
\begin{align}
(R\Phi R^\dagger)_{jl} = \frac{1}{\Delta x^2 l} \sum_{k}^{}\frac{e^{ik(x_j-x_l)}}{k^6} \big( e^{ik\Delta x}-1 \big) \big( e^{-ik\Delta x}-1 \big)& \nonumber \\
= \frac{2}{\Delta x^2 l} \sum_{k}^{} \frac{ 1-\cos(k\Delta x)}{k^6} e^{ik(x_j-x_l)}&
%& \Rightarrow R_{jk} = \frac{1}{\sqrt{L}} \frac{e^{ikr_j}}{ik}\big( e^{ikl}-1 \big)
\end{align}

This matrix now needs to be inverted, but it is actually trickier than it looks, and the inverse is not equal to the inverse of the Fourier coefficients. Why?
Observe that the spatial gridpoints are both finite \textit{and} discrete, which means 
that terms like $ \sum_{j}^{} e^{ix_j(k-q)}$ do not form kroenecker deltas $\delta_{kq}$. We need to study this behaviour more closely. Take our grid of
$N$ points and two momenta $k=2 \pi n /l$ and $q=2 \pi m /l$:

\begin{equation}
\sum_{j=0}^{N-1} e^{i(k-q)x_j} = \sum_{j=0}^{N-1} \exp \bigg[i(\frac{2\pi (n-m)}{l})\Delta xj \bigg ] 
\stackrel{\Delta x/l = 1/N }{=}
 \sum_{j=0}^{N-1} \exp \bigg[i(\frac{2\pi (n-m)}{N})j \bigg ]
\end{equation}

Now the sum is equal to N when $n = m$ as expected, but also when $(n-m)/N$ is an integer, i.e. $n=m \mod(N)$, making it hard to algebraically invert.\newline

What is going on here, is that data space is a discrete periodic interval, which has a discrete Fourier transform (DFT). For a DFT, the momentum values $k$ are the same as
those for the continuous interval, albeit with a highest uniquely resolvable frequency known as the \textit{Nyquist frequency}, which is equal to half of the sampling frequency. In this case, the Nyquist is $\frac{\pi} {\Delta x}$ and is denoted by $f_{N}$. Given that the matrix is indeed translation-invariant in data space, it must have \textit{some} diagonal representation in the discrete Fourier transform, i.e. some scalar function of $k$ for $k$ now less than the Nyquist frequency. 
It turns out this representation can be found by resumming over multiples of the Nyquist frequency.

\begin{lemma}\label{resum}
Given a regular, discrete grid of points $\{ x_j \}$ for $j \in \{ 1, \ldots, N \}$ on a periodic interval, and a matrix of the form:

\begin{equation}
A_{lj}=\sum^{\infty}_{k=-\infty} f(k)e^{ik(x_l-x_j)}
\end{equation}
for $f(k)$ some function of $k$, it has a diagonal representation in the DFT Fourier space, given by:

\begin{equation}
A_{lj}= \sum^{f_N}_{|k|} \bigg ( \sum^{}_{b \in 2f_N \mathbb{Z}}f(k+b) \bigg) e^{ik(x_l-x_j)} =\sum^{f_N}_{|k|} g(k) e^{ik(x_l-x_j)}
\end{equation}
where the new diagonal function $g(k)$ denotes the sum $\sum^{}_{b \in 2f_N \mathbb{Z}}f(k+b)$.

\begin{proof}
Observe what happens when the infinite sum over $k$ is partitioned into smaller sums shifted by multiples of the Nyquist frequency. For 
any $x_i$ and $x_j$ separated by a multiple of $\Delta x$ and $b = 2 \pi n/ \Delta x$, we have $(k+b)(x_i-x_j) = k(x_i-x_j) + 2\pi n$. This factor of $2 \pi$ then 
disappears in the complex exponential:

\begin{equation}
A_{lj}=\sum^{< f_N}_{|k|} \sum^{}_{b \in 2f_N \mathbb{Z}}f(k+b)e^{i(k+b)(x_l-x_j)} \stackrel{\text{Nyquist} }{=} 
\sum^{f_N}_{|k|} \bigg ( \sum^{}_{b \in 2f_N \mathbb{Z}}f(k+b) \bigg) e^{ik(x_l-x_j)}
\end{equation}

This resummed function is a diagonal function of the DFT frequencies $k < f_N$, and so must be the operator we were looking for.
\end{proof}
\end{lemma}

\noindent

Note that depending on whether the number of data points is even or odd, the domain of $|k|< f_N$ changes. For odd $N$ we use the convention that $k \in [-(N-1)/2, (N-1)/2]$ and if it's even we use $ k \in [-N/2, N/2 -1]$. Due to the physical analogy with Brillouin zones, we refer to the procedure of summing over multiples of the Nyquist frequency as \textit{the sum over Brillouin zones}.\newline

Now that we have obtained a representation of the operator which is diagonal in the DFT space, inverting becomes rather easy: 

\begin{equation}
(R \Phi R^\dagger)_{lj}^{-1}= \frac{1}{N} \sum^{f_N}_{|k|} \frac{1}{\sum^{}_{b \in 2f_N \mathbb{Z}}\Phi(k+b) |\widehat{B}(k+b)|^2} e^{ik(x_l-x_j)} 
\end{equation}

The factor of $N$ comes from the different normalizations of the DFT and the regular fourier transform. Fourier modes in the DFT are normalized as $\frac{1}{\sqrt{N} }e^{-ikx_j}$. For the example model, we can now write down the formula for  $(R\Phi R^\dagger)^{-1}_{lj}$:

\begin{align}\label{inverse}
%(R\Phi R^\dagger)^{-1}_{lj} = 
&\frac{\Delta x^2 l}{2N} \sum^{< f_N}_{|k|} \frac{e^{ik(x_l-x_j)} }{\sum_{b} [1-\cos((k+b)\Delta x)] (k+b)^6  } \nonumber \\
&\stackrel{\text{Nyquist} }{=} \frac{\Delta x^2 l}{2N} \sum^{< f_N}_{|k|} \frac{e^{ik(x_l-x_j)}}{ (1-\cos(k\Delta x)) \sum_{b} (k+b)^6  }
\end{align}

%Now, computing $RL\Phi R^\dagger(R\Phi R^\dagger)^{-1}$, we notice:

%\begin{equation}
%\frac{- \nabla \cdot \nu}{3} \bigg[ R(1+ik)\Phi R^\dagger (R\Phi R^\dagger)^{-1}]= 
%\frac{- \nabla \cdot \nu}{3} \bigg[ \mathbbm{1} +iRk\Phi R^\dagger (R\Phi R^\dagger)^{-1} \bigg]
%\end{equation}

Now it is time to compute the second part of the transport operator,  $R\bar{U}\Phi R^\dagger $. Given that $\bar{U}$ is assumed to be diagonal in Fourier space, the previous lemma \ref{translation} applies, and the operator will also be diagonal in the DFT space, with a sum over Brillouin zones. With this information, we may now write down the general form of the update operator $T= R\bar{U}\Phi R^\dagger (R\Phi R^\dagger)^{-1}$:

\begin{equation} \label{main}
T_{lj}= \sum^{f_N}_{|k|} 
\frac{\bigg ( \sum^{}_{b \in 2f_n \mathbb{Z}} \bar{U}(k+b) \Phi(k+b) |\widehat{B}(k+b)|^2 \bigg{)}}
{\sum^{}_{\hat b \in 2f_n \mathbb{Z}} \Phi(k+\hat b) |\widehat{B}(k+\hat b)|^2}e^{ik(x_l-x_j)}
\end{equation}

The factor of $1/N$ is cancelled by a factor of $N$ coming from the sum over spatial indices. For the toy model, we know the analytic form in Fourier space of all the objects involved. So, the transport operator can be computed by substituting into the above equation, giving an infinite algebraic series in Fourier space. This is summed numerically on a computer until it converges to within some degree of accuracy, yielding a function of $k$. The position space operator is then obtained by taking the inverse DFT of this function. Simple. For more complex models, computing the transport operator is typically done solely in position space. \newline

For the toy model, $L= ivk$, so the above formula applies. Since the responses are constant, we can use the expansion $T = \mathbbm{1} + \Delta t RL\Phi R^\dagger + \cdots$. For the meantime, we take the first order expansion and calculate $RL\Phi R^\dagger$:

\begin{equation}
(RL\Phi R^\dagger)_{lj}= \frac{2v \Delta x^2}{l}\sum^{f_N}_{|k|} (1-\cos(k\Delta x))\sum^{}_{b \in 2f_n \mathbb{Z}} \frac{ i(k+b)}{(k+b)^6}e^{ik(x_l-x_j)}
\end{equation}

This means that to first order, the $\Delta t$ term in the update operator is:

\begin{align}\label{kernel}
&v\frac{2}{\Delta x^2 l} \frac{\Delta x^2 l}{2}\sum^{f_N}_{|k|} (1-\cos(k\Delta x))\sum^{}_{\hat b} \frac{i}{(k+\hat b)^5}
  \bigg( \frac{1}{ (1-\cos(k\Delta x)) \sum_{b} (k+b)^6  } \bigg{)}  e^{iq(x_l-x_j)} \nonumber \\
&\Rightarrow T = \mathbbm{1} + \Delta t v\sum^{f_N}_{|k|} \bigg( \sum^{}_{\hat b} \frac{i}{(k+\hat b)^5} \frac{1}{\sum^{}_{ b} \frac{1}{(k+b)^6}} \bigg) e^{ik(x_l -x_j)}
\end{align}

\subsection{Results}

%This code turns out to perform far worse than the most basic finite-difference equation for advection. These codes exhibit a plethora of artefacts, which remain under changes in prior, grid sizes, and response functions. Namely, they develop extra maxima and minima, are often numerically unstable, and have large phase errors.An illustrative example is shown in figure \ref{trivial_v}. For this example, the time update operator $U$ was taken to second order in time, $U = 1 +i v\Delta t k - v^2\Delta t^2 k^2$. CHANGE FROM SHITTY THING TO $v$. The coefficient of the second term was modified, as the diffusion from the second order term is too strong. This can be equivalently be thought of as adding a small amount of numerical diffusion to the code. This is entirely standard practice (cite) \newline

When implementing the toy model, it becomes immediately apparent that a first order forward scheme is unstable. In fact, any first-order forward scheme for advection in this general class of models will be unstable. We show this via a Von Neumann stability analysis. We first insert $\bar{U} = \mathbbm{1} + v\Delta t \partial_x=1+ iv\Delta t k$ into eqn.~\ref{main} to give: \newline

\begin{equation}
T(k) = \mathbbm{1} + i v \Delta t  
\frac{ \sum^{}_{b \in 2f_N \mathbb{Z}} (k+b) \Phi(k+b) |\widehat{B}(k+b)|^2 }
{\sum^{}_{\hat b \in 2f_N \mathbb{Z}} \Phi(k+\hat b) |\widehat{B}(k+\hat b)|^2}
\end{equation}

Since the transport operator $T$ is diagonal in Fourier space, the magnitude of it's eigenvalues are simply $|T(k)|$. 
Noticing that the momentum-dependent term is purely imaginary, we get:
%$||T[e^{iqx_j}]|| = |T(k)|\cdot ||e^{iqx_j}||$. To calculate $|T(k)|$ we simply need to sum the squares of the real and imaginary parts:

\begin{equation}
|T(k)|^2 = 1 + (v \Delta t)^2 \bigg{[}  
\frac{ \sum^{}_{b } (k+b) \Phi(k+b) |\widehat{B}(k+b)|^2 }
{\sum^{}_{\hat{b} } \Phi(k+\hat b) |\widehat{B}(k+\hat b)|^2} \bigg{]}^2
\end{equation}

%\begin{multline}
%||d_{i+1}||^2= ||\bigg( \mathbbm{1} + \Delta t (RL\Phi R^\dagger(R\Phi R^\dagger)^{-1}) \bigg) \frac{1}{N}e^{iqx}||^2 \\
%=||1 -i \cdot \Delta t\frac{\nabla \cdot \nu}{3}\bigg( \sum^{}_{\hat b} \frac{1}{(q+\hat b)^5} \frac{1}{\sum^{}_{ b} \frac{1}{(q+b)^6}} \bigg)||^2 \\
%= Re^2 + Im^2 = 1 + \bigg[ \Delta t\frac{\nabla \cdot \nu}{3}\bigg( \sum^{}_{\hat b} \frac{1}{(q+\hat b)^5} \frac{1}{\sum^{}_{ b} \frac{1}{(q+b)^6}} \bigg) \bigg]^2
%\end{multline} 

which is everywhere greater than one, for all nonzero values of momentum. Thus all Fourier modes will undergo exponential growth, and the code is unstable. Thus, to stabilize the code, we go to second order in time.\newline

\begin{figure}
  \begin{minipage}[b]{0.5\linewidth}
    \includegraphics[width=\textwidth]{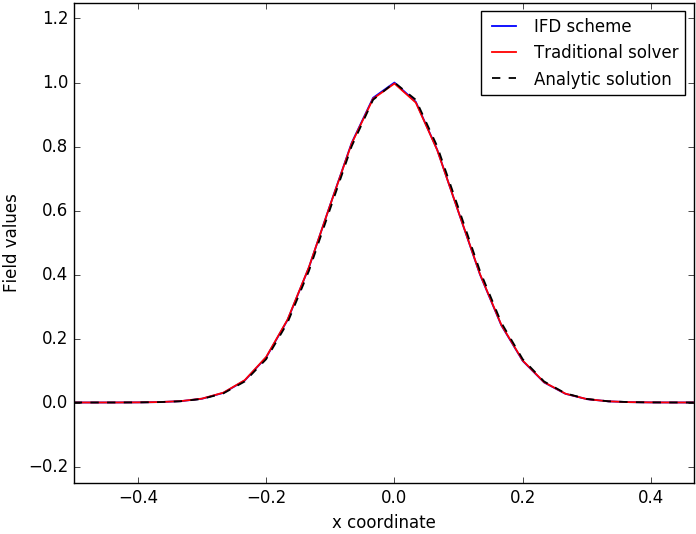}
    %\caption{Picture 1}
    %\label{fig:1}
  \end{minipage}
  \begin{minipage}[b]{0.5\linewidth}
    \includegraphics[width=\textwidth]{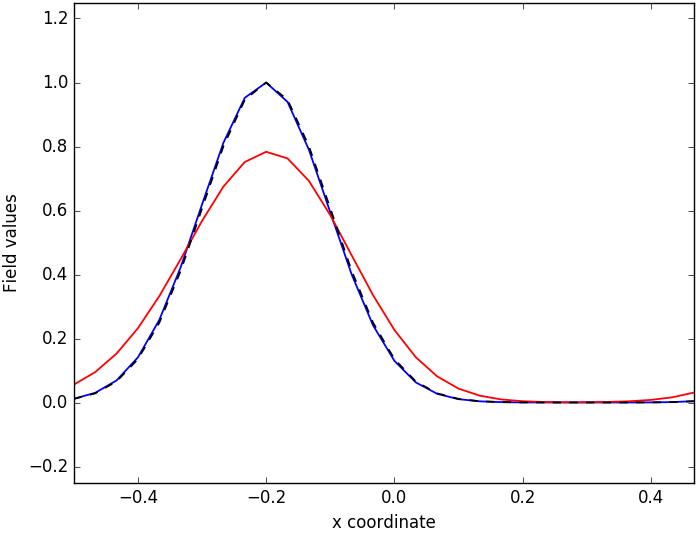}
    %\caption{Picture 2}
    %\label{fig:2}
  \end{minipage}
\begin{minipage}[b]{0.5\linewidth}
    \includegraphics[width=\textwidth]{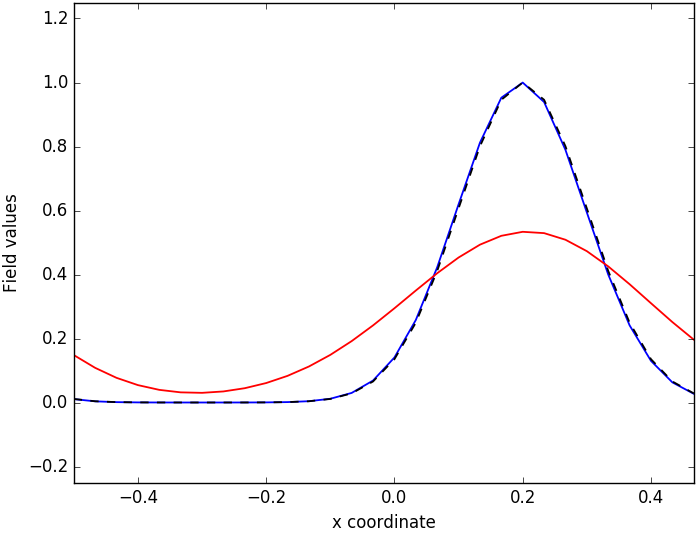}
    %\caption{Picture 2}
    %\label{fig:3}
  \end{minipage}
	\begin{minipage}[b]{0.5\linewidth}
    \includegraphics[width=\textwidth]{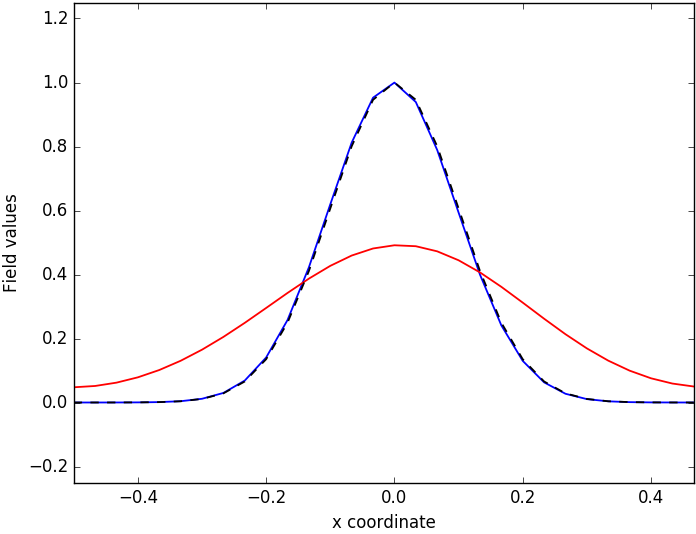}
    %\caption{Picture 2}
    %\label{fig:4}
  \end{minipage}
\caption{Simulated advection of an exponential pulse for a leftward directed velocity field, with periodic boundary conditions. The simulation space has 30 grid points and 500 timesteps, with a length, velocity and runtime of 1, in arbitrary units. The pictures detail one full cycle, at timesteps 0, 100, 400 and 499. The IFD code is rendered in blue, and the traditional first-order solver is shown in red. The analytic solution is rendered with a black dotted line. After one full cycle, the traditional solver has undergone significant broadening, whereas the IFD solution is indistinguishable from the analytic one.}
\label{trivial_v}
\end{figure}

The results for a second order advection code are shown in figure \ref{trivial_v}. The toy model was implemented in position space, in anticipation of the case of a nontrivial velocity field, where a Fourier space representation is not possible. The update operator was expanded to second order, $U = 1 +i v\Delta t k - v^2\Delta t^2 k^2$, so that $T(k)$ will have another term analogous to eqn.~\ref{kernel} with $iv\Delta t(k+b) \to -v^2 \Delta t^2 (k+b)^2$. $T(k)$ was then computed by summing over Brillouin zones numerically, until the sum converged. The resulting expression was then put through an inverse DFT to yield a position-space matrix representation of the transport operator.\newline

Even at relatively low resolutions, for a smooth initial pulse, the IFD scheme delivers results which are indistinguishable from the analytic solution by eye. This is in contrast to a basic first-order finite difference scheme, which suffers from the artefact of numerical diffusion; the simulated field spreads out despite the fact that there is no diffusion term in the equations.\newline

There are other ways of potentially stabilizing this toy code, like a Backward Euler scheme. Such a scheme was attempted and did indeed stabilize the code, however other undesireable properties remained. The backward Euler code developed extra maxima and minima which, thanks to the stabilizing property of the backward Euler, did not grow in size. But they did remain at a finite size and generally made things look bad. This is due to a general flaw in these codes. Namely, that they handle shocks quite poorly. Loosely defined, in a numerical simulation a \textit{shock} is a change in the field that is sharp relative to the grid spacing $\Delta x$. Given that our vague goal is to simulate systems at very low resolutions ($\approx$ 10 bins), at these resolutions, everything looks like a shock. To analyse the propagation of shocks, we need to not just compute the magnitude error of our code, but also the \textit{phase error}, which is the difference in propagation velocities for the true solution vs. the simulated solution, as a function of frequency. \newline

\subsection{Phase error}\label{phasesec}

%\begin{equation} \label{main}
%RU\Phi R^\dagger (R\Phi R^\dagger)^{-1}_{lj}= \sum^{f_N}_{|k|} 
%\frac{\bigg ( \sum^{}_{b \in 2f_n \mathbb{Z}} U(k+b) \Phi(k+b) |\widehat{B}(k+b)|^2 \bigg{)}}
%{\sum^{}_{\hat b \in 2f_n \mathbb{Z}} \Phi(k+\hat b) |\widehat{B}(k+\hat b)|^2}e^{ik(x_l-x_j)}
%\end{equation}
%
%Where the $B$'s are the integration kernels of the responses. We need a single name for this time update operator $RU\Phi R$... etc so i'll call it $T$ from now on. %\newline

Suppose we just consider straight advection, and observe the time evolution of a plane-wave $e^{-ikx}$ in this system. We know that the full analytic time evolution operator is given by $U(k)=e^{ivk\Delta t}$, which simply multiplies the plane wave by a phase factor. This phase, the term in the exponent, represents the velocity. We need to define an operation which extracts this phase, which we label $\omega$: 
\begin{equation} \label{planewave}
\tan(\omega) =\Imag(e^{ivk\Delta t})/\Real(e^{ivk\Delta t})= \sin(vk\Delta t)/\cos(vk\Delta t) \Rightarrow \omega = vk\Delta t
\end{equation}

gives the desired result. The phase error is defined to be the difference between the true phase ($\omega$) and the numerical phase ($\omega_n$) actually obtained in the simulation \cite{Godunov}. 
This can be calculated in the analytically solvable case by finding $\omega_n=\arctan(\Imag(T(k))/\Real(T(k))$. \newline

There is a mathematical subtlety that needs to be addressed here; in the generalized framework, the responses are simply integration over some regularly spaced grid of functions, whose form can be completely arbitrary. It has not yet been checked what the image in data space of a plane wave in signal space is. Pick a signal space plane wave $\phi(x) =e^{-ikx}$, and compute
%So, when we want to measure the propagation speeds in a numerical code, we want to calculate the phase that we actually get in our simulation. For pretty much all numerical schemes with constant velocites and constant grid spacings you can find a momentum-space representations, which gives a nice estimate of what will happen when you go to nonconstant grids/velocities. Doing a true phase error analysis for nonconstant velocities would require us to know the analytic solutions. 
%We simply define the numerical phase to be $\omega_n=\arctan [\Imag(T(k))/\Real(T(k))]$ and define the phase error to be the difference between that and the true phase. \newline

%So we need to look at what happens when we try to model a plane wave that is propagating in signal space. First, we need to find it's image in data space. 
%Define $\phi(x) =e^{ikx}$ 

\begin{equation}
R_j \phi =\int B(x-x_j)e^{-ikx} dx = e^{-ik x_j}\int B(x)e^{-ikx} dx = \widehat{B}(k)e^{-ik x_j}
\end{equation}

This is neat; regardless of the shape of the response bins, plane waves in signal space show up as plane waves in data space, as long as the grid is translation-invariant. The plane wave will get scaled by a constant factor $\widehat{B}(-k)$ which depends on its momentum, but it is still a plane wave (remember that we are in the position-space representation here). Note that if $k$ was above the Nyquist, the $e^{ik x_j}$ term with the discrete $x_j$ coordinates implies that the plane wave appears in the data with a frequency below the Nyquist, as expected. The above result allows us to perform a valid phase analysis by just looking at plane waves in data space.\newline

%In the discrete Fourier transform of data space, the update equations are diagonal, so the error analysis is easy. Simply take $\arctan(\Imag/\Real)$ of the kernel in equation \ref{main}. 
%As a side note: lately I've been revisiting my earlier codes, and when we go to higher orders in time, we've been doing it wrong. We've been exponentiating $T$ and truncating the exponential series. What we should be doing is taking the time evolution $U=\sum_{k=0}^{\alpha} (\Delta t L)^k/k!$ to some higher order $\alpha$ in signal space. I explain why later. 
We consider an expansion of $U$ to arbitrary order, $\alpha$, in time, because we can. %Of course for nontrivial velocities, this will become very hard. 
For this model, $L=ivk$, so every odd power in the Taylor series will have a factor of $i$ and every even power will not: this clearly forms truncated $\sin(v\Delta tk)$ and $\cos(v\Delta t k)$ series. The general phase error is then:

\begin{align}
\omega_n= 
&\arctan \bigg{(} \frac{ \sum^{}_{b} \sum^{\alpha}_{n=0} \frac{(-1)^n [v\Delta t(k+b)]^{2n+1}}{(2n+1)!}
 \Phi(k+b) |\widehat{B}(k+b)|^2 }
{\sum^{}_{\hat b} \Phi(k+\hat b) |\widehat{B}(k+\hat b)|^2} \nonumber \\
&\times \frac{\sum^{}_{\hat b} \Phi(k+\hat b) |\widehat{B}(k+\hat b)|^2}
{ \sum^{}_{b} \sum^{\alpha-1}_{n=0} \frac{(-1)^n [v\Delta t(k+b)]^{2n}}{(2n)!}
 \Phi(k+b) |\widehat{B}(k+b)|^2 }\bigg{)}
\end{align}
The denominator of the first term cancels with the numerator of the second, giving:
\begin{equation}\label{numphase}
\omega_n= 
\arctan \bigg{(}\frac{ \sum^{}_{b} \sum^{\alpha}_{n=0} \frac{(-1)^n [v\Delta t(k+b)]^{2n+1}}{(2n+1)!}
 \Phi(k+b) |\widehat{B}(k+b)|^2 }
{ \sum^{}_{\hat{b}} \sum^{\alpha-1}_{n=0} \frac{(-1)^n [v\Delta t(k+\hat{b})]^{2n}}{(2n)!}
 \Phi(k+\hat{b}) |\widehat{B}(k+\hat{b})|^2 } \bigg{)}
\end{equation}

\begin{figure}[h]%{0.4\textwidth}
\includegraphics[width=\textwidth]{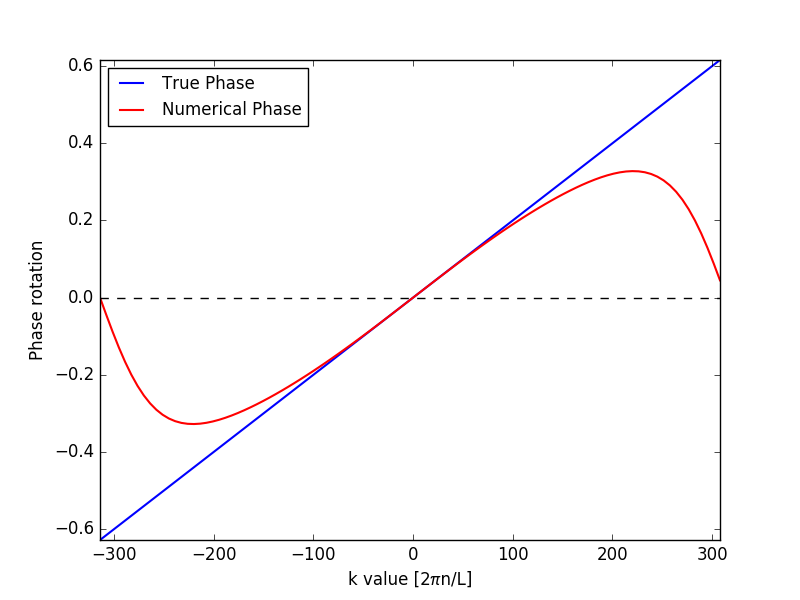}
\caption{Phase error of the second-order toy model over the domain $[-f_N, f_N]$, performed on a 100 point grid with length and velocity equal to 1 in arbitrary units. The numerical phase diverges quite strongly as the momentum approaches the Nyquist.}
\label{phase}
\end{figure}

Note that this equation is for odd powers in $\alpha$. For even powers, the term in the denominator has the higher power of $\alpha$. This equation may seem rather intimidating, and is somewhat hard to interpret at first. Begin by noticing that due to the sum over Brillouin zones, it must be periodic in $k \in [-f_N, f_N]$. This expression should attempt to approximate the graph of $vk$, i.e. a straight line. Unless the prior and responses were chosen terribly, the code should be reasonable enough that waves don't propagate backwards. i.e. to the right of the origin ($k=0$), the numerical phase will be everywhere positive, and to the left the phase will be everywhere negative. Periodicity implies that the phase must then drop to zero at $k = \pm f_N$, and the phase error approaches a maximum. An example of this behaviour is shown in the phase velocity plots for the toy model, figure \ref{phase}.

It is this feature which is responsible for the poor handling of shocks in this class of models. A feature that is sharp on the scale of the grid length must be resolved by high $k$ values, but at these high $k$ values, the structures do not propagate at all. This leads to the development of unwanted oscillations in the simulation. \newline

The argument that the phase error approaches a maximum at high momenta is just that: an argument. The $k$ values are discrete, rather than continuous, so there's no reason to say that the phase error cannot be reduced arbitrarily by making the phase plot approach a sawtooth function centred at zero.
 The bad phase error is rather a general and somewhat persistent feature of these models. Though this in itself is no great tragedy, many models such as the Lax-Wendroff scheme, which is a second order scheme, also suffer from this problem \cite{Hydro}. In fact, modelling shocks effectively is one of the most challenging and interesting areas of numerical hydrodynamics. \newline

There is a very relevant theorem here, called the \textit{Godunov Theorem} (\cite[p.~280]{Godunov}), which states that \textit{any linear algorithm for
solving partial differential equations, with the property of not producing new extrema, can be at
most first order.}\footnote{Note that we have not yet figured out to which spatial order these codes are accurate. This will be discussed later.} Given that we are exclusively working with linear codes, we cannot expect to escape these spurious oscillations, but hopefully by analysing eqn.~\ref{numphase}, they can be reduced as much as possible. \newline

Given the formula for the expansion to arbitrary order in time, it should be compared to the hypothetical ideal behaviour of an order $\alpha$ expansion, $U=\sum_{n=0}^{\alpha} (\Delta t L)^n/n!$:

\begin{equation}
\omega_{\text{ideal}}= 
\arctan \bigg{(}\frac{  \sum^{\alpha}_{n=0} \frac{(-1)^n [v\Delta t(k)]^{2n+1}}{(2n+1)!}}
{  \sum^{\alpha-1}_{n=0} \frac{(-1)^n [v\Delta t(k)]^{2n}}{(2n)!}} \bigg{)}
\end{equation}

%\begin{sloppypar}
Comparing the numerical phase and ideal phase, the discrepancy is clearly caused by the sum over Brillouin zones, if higher frequency modes could be damped or eliminated, then the code would approach the ideal behaviour.
The $B(x)$'s will very often be compactly supported, so by the uncertainty principle, their Fourier transforms will be nonzero all the way to infinity. Instead, imagine that the magnitude of the prior dropped off sharply for momenta above the Nyquist, this would cut out the influence of all the higher Brillouin zones. We conclude: \textit{the lower the probability the prior assigns to modes above the Nyquist, the better the phase error.}\footnote{So, we can minimize the phase error \textit{if} the prior kills frequencies above the Nyquist. Proving the converse statement, that the only way to take the phase error to zero is to assign zero probability to higher frequencies, will be very hard.} What is happening is that for any structure in the data below the Nyquist frequency, the prior associates a small but nonzero probability that it actually came from a structure above the Nyquist frequency, which has an entirely different propagation velocity, which then interferes with the simulation. \newline
%\end{sloppypar}

It is unfortunately time for a detour into matters of interpretation. We cannot just set the momentum-space representation of the transport operator to be $T(k)=\sum_{n=0}^{\alpha} (\Delta t L(k))^n/n!$ and get ideal results, because this will never generalize to nonconstant velocity fields. We remind the reader again that the simulations are not being carried out in Fourier space, it's just that for translation-invariant schemes they have such a representation, and an error analysis on this representation should give an idea as to what happens when nontrivial models are considered. The trickier problem however, is that of prior selection. \newline

There is an unresolved ambiguity in IFD as to whether the prior represents our honest beliefs about the behaviour of the system, or is simply a tunable parameter for constructing a subgrid model. If the former is true, then the prior cannot be adjusted so that it drops off sharply above the Nyquist, as that would mean we are deliberately selecting our beliefs about the system depending on the resolution at which we are observing it; which is rather illogical. Consider instead that the prior is fixed, and drops off to below some desired level after some frequency $k_0$, then for best results, we must \textit{pick} a resolution such that the Nyquist frequency is greater than $k_0$. Thus the phase error equation tells us something intuitively obvious: if we believe that the behaviour of a system is mostly described by structures above a certain length scale, then our simulations should have a resolution of at least that length scale. This assumption is implicit in normal finite-difference codes. The difference with IFD however, is that our assumptions are explicitly coded into the update equations, and we get poor results when the simulations violate these assumptions. This will turn out to be a recurring theme.\newline

If the prior is regarded as a tunable parameter, then all bets are off. It should be set to damp high-frequency modes as aggressively as possible, such that the reconstructions have little structure on scales above the below resolution. Intuition would say that a bit of higher structure should be kept when going to nonconstant velocity fields, as the advantage of subgrid models comes from the fact that they suppose the existence of extra structure between gridpoints. \newline

It appears that if we want good results, we need to discuss the limit of high resolutions, which means it's time to discuss consistency and convergence.\newline

%Notice that even if we take the hypothetical limit of $\alpha \to \infty$, the phase error still doesn't dissapear. Because
%
%\begin{equation}
%\sum^{}_{b} \sin(v\Delta t (k+b)) \Phi(k+b) |\widehat{B}(k+b)|^2 \ne \sin(v\Delta t k)\sum^{}_{b} \Phi(k+b) |\widehat{B}(k+b)|^2
%\end{equation}

%If only this were true, then the $\sin$ and $\cos$ terms would factor out of those sums over $b$ and $\hat{b}$ in the numerator and denominator, and those sums would then promptly cancel, giving the zero phase error. But there is in general no special relation between the $\Delta t$, $v$ and $\Delta x$ terms which would allow us to factor the $\sin$ and $\cos$ terms out\footnote{Note, there is a curious special condition, if $\Delta t$ \textit{perfectly} obeys the Courant condition, then the $\sin$ and $\cos$ terms factor out, giving perfect accuracy. Anything else will not.}. This essentially states that the spatial discretization is a limiting factor. So for good results, the limit of $\Delta x \to 0$ needs to be considered. Which means it's time to discuss consistency and convergence.\newline

\subsection{Consistency}

We suspect from equation \ref{numphase}, that in the limit of high resolutions, our simulations will approach reality. This is the same as proving that the transport operator is consistent. This can be shown in the translation-invariant case by only adding two extra assumptions: that the response bins $B(x)$ are compactly supported and have bounded Fourier transform, and that $U(k)\Phi(k) \to 0$ as $k \to \pm \infty$. \newline

The bounded Fourier transform requirement will almost always be true for any reasonable response. It holds for all smooth, compactly-supported functions, by the 
\textit{Paley-Wiener theorem} \cite{Reedsimon}.

\begin{theorem}[Paley-Wiener (weakened)]
If $f$ is a smooth, compactly supported function on $\mathbb{R}$, then it's Fourier transform $\hat{f}(k)$ can be bounded by
\begin{equation}
|\hat{f}(k)| \le C_n(1+|k|)^{-n}
\end{equation}
 for all $n \in \mathbb{Z}^+$, and some positive constant $C_n$
\end{theorem}

The box responses, despite not being smooth, also have bounded Fourier transform.\newline

To prove consistency, we ask if $T(k) \to U(k)$ in the limit of high resolution. In the Fourier representation, comparing the action of $T$ and $U$ is easy, despite the fact that they technically act on different spaces. The definition of consistency (def.~\ref{cons}) requires that the operators converge for any function in signal space that we pick, but not that they converge at the same rate for all functions\footnote{This would be impossible, for \textit{any} numerical scheme on a discretized grid, the Nyquist frequency dictates there is a function which the grid cannot resolve}. Pick a basis of signal space consisting of Fourier modes, then pick out a single mode of frequency $k$. As the resolution increases, eventually the Nyquist frequency will be greater than $k$ ($f_N = \pi /\Delta x$). Past this resolution, $T(k)$ and $U(k)$ can both be thought of as acting on the same space. $T(k)$ contains a time-order approximation $\bar{U}(k)$ to $U(k)$. If we can show that as $\Delta x \to 0$, $T(k) \to \bar{U}(k)$, then in the joint limit of time and space resolution going to infinity, then $T$ approaches $U$.\newline

Hence we need that for each fixed $k$, $T(k) \to \bar{U}(k)$ in $\Delta x$, but the convergence doesn't need to be uniform in $k$. 
 For the translation-invariant response, we want to increase the number of bins while simultaneously decreasing their width. 
Given some initial resolution $\Delta x_0$ for which all the bins fit evenly inside the interval, we pick an integer $\lambda$ that goes from 1 to infinity, then we set $\Delta x = \Delta x_0 / \lambda$. This guarantees that the new set of scaled bins $B(\lambda x) = B_\lambda(x)$ fits evenly inside the interval. \newline

The compact support property of the bins allows us to exploit the  
 fact that up to a  normalization constant, the coefficients $B(k)$ of the discrete values of $k$ in the Fourier series of the bins are the same as the values at $k$
in the continuous Fourier transform of $B$. This is because if a function is compact, it doesn't matter if it is integrated over a finite or infinite interval.
This allows us to exploit the scaling property of the Fourier transform $\hat{B_\lambda} (k) = \frac{1}{\lambda} B(k/\lambda)$. The normalization constant $\lambda$ and the constant from the differing normalizations of the Fourier transform cancel due to the division in eqn.~\ref{main}. \newline

Now observe the sum over the Brillouin zones. We sum over  $b \in 2\mathbb{Z}f_N$ where $f_N = \pi/\Delta x$ and thus $f_N^\lambda= \pi \lambda/\Delta x_0$  and $b^\lambda = 2 \pi n \lambda /\Delta x_0$ for $n \in \mathbb{Z}$. We don't scale the prior with $\lambda$ (our beliefs about the system shouldn't change depending on the resolution of our equipment) and look at the upper term of eqn.~\ref{main}:

\begin{equation}
\sum_{n \in \mathbb{Z}} \bar{U}(k + \frac{2\pi n \lambda}{\Delta x_0}) \Phi(k + \frac{2\pi n \lambda}{\Delta x_0}) | B(\frac{1}{\lambda}(k + \frac{2\pi n \lambda}{\Delta x_0}))|^2
\end{equation}
The $\lambda$ term inside $B$ can be absorbed to give:

\begin{equation}
| B(\frac{1}{\lambda}(k + \frac{2\pi n \lambda}{\Delta x_0}))|^2= | B(\frac{k}{\lambda} + \frac{2\pi n }{\Delta x_0}))|^2
\end{equation}
 
We want that in the limit of $\Delta x \to \infty$, the higher terms in the sum vanish, leaving only terms in the first Brillouin zone. The prior and bin terms in the numerator and denominator of \ref{main} would then cancel, leaving just $\bar{U}$, i.e. we want 

\begin{align} \label{convergence}
 \frac{\lim_{\lambda \to \infty} \sum_{n \in \mathbb{Z}} \bar{U}(k + \frac{2\pi n \lambda}{\Delta x_0}) \Phi(k + \frac{2\pi n \lambda}{\Delta x_0}) 
| B(\frac{k}{\lambda} + \frac{2\pi n }{\Delta x_0})|^2}
{\lim_{\lambda \to \infty}\sum_{m \in \mathbb{Z}} \Phi(k + \frac{2\pi m \lambda}{\Delta x_0}) | B(\frac{k}{\lambda} + \frac{2\pi m }{\Delta x_0})|^2} \nonumber &\\
= \frac{\bar{U}(k) \Phi(k) | B(\frac{k}{\lambda} )|^2}{ \Phi(k) | B(\frac{k}{\lambda} )|^2} = \bar{U}(k) &
\end{align}

We can expect that for all terms with $n \ne 0$, in the limit of $\lambda \to \infty$ each term goes to zero because $\bar{U}(k)\Phi(k)$ goes to zero at large $|k|$. Therefore we want to swap the limit and the infinite sum. \newline

Given  a sequence of functions $f_n(\lambda)$, swapping the limits $\lim_{\lambda \to \infty} \sum_{n=0}^\infty f_n(\lambda) = \sum_{n=0}^\infty \lim_{\lambda \to \infty} f_n(\lambda)$ is possible if and only if the sequence of functions converges uniformly. In our case, we throw out the $n=0$ term, and consider the positive and negative $n$ halves of the sum separately, but present only the case of positive $n$, as the working for negative $n$ is nearly identical. Our goal is that the functions converge to zero,  so we state: a function converges uniformly to zero if for any positive $\epsilon$, 
there is an $N$ such that $\forall n \ge N$, $|f_n(\lambda)| < \epsilon$ for all values of $\lambda$. The last requirement is the crucial part. \newline

For our purposes,  $f_n(\lambda) =\bar{U}(k + \frac{2\pi n \lambda}{\Delta x_0}) \Phi(k + \frac{2\pi n \lambda}{\Delta x_0}) 
| B(\frac{k}{\lambda} + \frac{2\pi n }{\Delta x_0})|^2$. We bound the whole function $|B(\frac{k}{\lambda} + \frac{2\pi n }{\Delta x_0})|^2 < C$
for some constant $C$, which we may do by assumption. The bin terms do not vanish in the limit of large $\lambda$, because as the bins become narrower, their Fourier transforms widen out, at the exact same rate as the Nyquist frequency is increasing. Hence we will need the property  $\Phi(k)\bar{U}(k)\to 0$ to do all of the work. \newline

This requirement means that for $|k|$ large enough $\bar{U}(k)\Phi(k)$ can be bounded by some monotonically decreasing function of $|k|$, call it $g(|k|)$. We start by finding a bound for 
 $\lambda=1$, and then show that this bound holds for all $\lambda$. For $\lambda=1$, and the desired $\epsilon$ bound, we can pick some $n$ large enough such that we are in this decreasing regime, hence $|\bar{U}(k + \frac{2\pi n \lambda}{\Delta x_0}) \Phi(k + \frac{2\pi n \lambda}{\Delta x_0})|C < g(k + \frac{2\pi n \lambda}{\Delta x_0}) < \epsilon$. For higher $\lambda$ and large $n$,  $|k + \frac{2\pi n}{\Delta x_0}| < |k + \frac{2\pi n \lambda}{\Delta x_0}|$, and since we have taken $n$ to be large enough that we are in the decreasing regime, the $g(k)$ bound also holds. Thus the bound holds for all lambda. \newline

Thus the sequence of functions is uniformly convergent, and equation \ref{convergence} holds. We can now say

\begin{theorem}\label{IFDconsistency}
For a translationally-invariant scheme, whose response bins are compactly supported with bounded Fourier transform, and some time-order approximation $\bar{U}(k)$ to $U(k)$, where $k$ denotes momentum, then the scheme is consistent provided $\lim_{k \to \infty} \bar{U}(k)\Phi(k) =0$. 
\end{theorem} 

%The over-aggressive bound on the boxes is justified in hindsight. In the limit of high resolutions, the boxes become arbitrarily narrow and approach delta functions, and so if $U(k)$ is some derivative, then $\partial_x (\Phi * B(x)) \to \partial_x \Phi(x)$. In the limit of high resolutions the shape of the boxes doesn't matter
%so momentum-space operators $U(k)$ which may be unbounded in $k$ acting on them will be undefined. Essentially, if the boxes become arbitrarily narrow, you end up taking the derivative of just the prior. In the limit of high resolutions, the shape of the boxes doesn't even matter, and one is essentially just taking the derivative of the prior. 
Important to note is that we only need $\bar{U}(k)\Phi(k) \to 0$, not $U(k)\Phi(k)$. For derivative operators s.t. $U = \exp(\Delta t \partial_x) =\exp(i\Delta t k)$ or similar, this would require that the prior is smooth. Using the approximated time expansion, the prior, $\Phi(x)$, only needs to be as many-times differentiable as the order of the expansion dictates.

\subsection{Error scaling}

Using our formula for the transport operator, we can estimate the data space error. We stay in the exact same limit we were before, scaling with $\lambda$, and picking a fixed $\phi(x)$ that is some plane wave, and exploiting the fact that below the Nyquist, signal space and data space are comparable. This means that the one-step error in data space $E_d$ is given by:

\begin{align} \label{fourier_error}
 &E_d = \bigg{|}\frac{ \sum_{n \in \mathbb{Z}} \bar{U}(k + \frac{2\pi n \lambda}{\Delta x_0}) \Phi(k + \frac{2\pi n \lambda}{\Delta x_0}) 
| B(\frac{k}{\lambda} + \frac{2\pi n }{\Delta x_0})|^2}
{\sum_{m \in \mathbb{Z}} \Phi(k + \frac{2\pi m \lambda}{\Delta x_0}) | B(\frac{k}{\lambda} + \frac{2\pi m }{\Delta x_0})|^2}- U(k)\bigg{|}
\end{align}

We pick some order in our expansion $\bar{U}=\sum_{p=0}^{N}(\Delta t L)^p/p!$, and expand out the error in terms of powers of $L$, as in equation \ref{error}.

\begin{align}\label{errorscaling}
 &E_d \le \sum_{p=0}^{N}\frac{\Delta t^p}{p!}\bigg{|}\frac{ \sum_{n \in \mathbb{Z}} L^p(k + \frac{2\pi n \lambda}{\Delta x_0}) \Phi(k + \frac{2\pi n \lambda}{\Delta x_0}) 
| B(\frac{k}{\lambda} + \frac{2\pi n }{\Delta x_0})|^2}
{\sum_{m \in \mathbb{Z}} \Phi(k + \frac{2\pi m \lambda}{\Delta x_0}) | B(\frac{k}{\lambda} + \frac{2\pi m }{\Delta x_0})|^2}- L^p(k)\bigg{|}
\end{align}

We analyse the scaling of each term individually, but start with the $\Delta t$ term, and then generalize.
In the limit of high resolutions, 
we expect the sum over the higher Brillouin zones to become small, so 
\begin{equation}
{\sum_{b} L(k+b)\Phi(k+b) |B(k+b)|^2 \approx L(k)\Phi(k) |B(0)|^2 + \epsilon (k)}
\end{equation} 
The denominator should also admit such an expansion. We then seek to bound the whole fraction. The scaling of the error can only be estimated if we know the scaling behaviour of the prior and $L(k)$. So, suppose that as $|k|$ becomes large, $\Phi(k)$ can be bounded by some decreasing power law in $k$, $|k|^{-\alpha}$ for $\alpha$ positive. We also assume that $L(k)$ can be bounded by some $|k|^\beta$ for $\beta$ positive, as $L$ will typically be a derivative operator, with $\partial_x^n = (ik)^n$. There will be constants of proportionality, but they don't matter. We factorize the numerator as:

\begin{equation}
L(k)\Phi(k)| B(\frac{k}{\lambda})|^2 + \sum_{n \ne 0} L(k + \frac{2\pi n \lambda}{\Delta x_0}) \Phi(k + \frac{2\pi n \lambda}{\Delta x_0}) | B(\frac{k}{\lambda} + \frac{2\pi n }{\Delta x_0})|^2
\end{equation}

Once again, in the limit $\lambda \to \infty$, the bin terms just converge to ${B(0 + \frac{2\pi n }{\Delta x_0})}$ which isn't useful for bounding anything so we replace it with the uniform bound $C$ from before. We then bound
\begin{align}
&\bigg{|}\sum_{n \ne 0} L(k + \frac{2\pi n \lambda}{\Delta x_0}) \Phi(k + \frac{2\pi n \lambda}{\Delta x_0}) | B(\frac{k}{\lambda} + \frac{2\pi n }{\Delta x_0})|^2 
\bigg{|}\nonumber \\
&\le C^2 \sum_{n \ne 0}\bigg{|}  L(k + \frac{2\pi n \lambda}{\Delta x_0}) \Phi(k + \frac{2\pi n \lambda}{\Delta x_0})   \bigg{|}\nonumber \\
&\le C^2\sum_{n \ne 0}\bigg{|}  \frac{2\pi n \lambda}{\Delta x_0}   \bigg{|}^{\beta - \alpha} = \lambda^{\beta - \alpha}C^2\sum_{n \ne 0}\bigg{|}  \frac{2\pi n}{\Delta x_0}   \bigg{|}^{\beta - \alpha}
\end{align}

The term inside the sum is independent of the scaling. Therefore we have an object which scales as $\mathcal{O}(\lambda^{\beta-\alpha})$, which we identify with $\mathcal{O}(\Delta x^{\alpha-\beta})$, since $\Delta x = \Delta x_0/\lambda$. We repeat the argument with the denominator, and get a term proportional to $\mathcal{O}(\Delta x^{\alpha})$. We use the Taylor expansion for $1/(1-\epsilon) \approx 1+\epsilon+ \epsilon^2+\cdots$ to expand the denominator in eqn.~\ref{errorscaling} into something more useful:

\begin{equation}
\frac{1}{ \Phi(k) | B(\frac{k}{\lambda})|^2 + \mathcal{O}(\Delta x^{\alpha})} =\frac{1}{ \Phi(k) | B(\frac{k}{\lambda})|^2}+ \mathcal{O}(\Delta x^{\alpha})
\end{equation}

Thus the whole error expression scales as:
\begin{align}
&\bigg{(}L(k)\Phi(k)| B(\frac{k}{\lambda})|^2 + \mathcal{O}(\Delta x^{\alpha-\beta}) \bigg{)}\bigg{(}\frac{1}{ \Phi(k) | B(\frac{k}{\lambda})|^2}+ \mathcal{O}(\Delta^{\alpha}) \bigg{)} -L(k) \\
&=L(k) + \mathcal{O}(\Delta x^{\alpha})+ \mathcal{O}(\Delta x^{\alpha-\beta})+ \mathcal{O}(\Delta x^{2\alpha-\beta}) - L(k) = \boxed{ \mathcal{O}(\Delta x^{\alpha-\beta})} \nonumber
\end{align}
 
The other $\mathcal{O}$ terms cancel because only the term with the worst scaling (lowest power) matters. For a term of order $\Delta t^p$ in eqn.~\ref{errorscaling}, we repeat the argument and get 
\begin{equation}
\boxed{E_d \propto \mathcal{O}(\Delta t^p \Delta x^{\alpha-p\beta})}. 
\end{equation}
The total error scaling is determined by the worst scaling of any of the individual terms.\newline
%The term with the worst scaling will obviously form the limiting case, which is why the $\mathcal{O}(\lambda^{\beta-2\alpha})$ term was thrown out. For $L(k)$'s with growth according to a positive polynomial, the limiting factor will be 
%$\mathcal{O}(\lambda^{\beta-\alpha})$, which we identify with the scaling in spatial resolution, so we write it as $\mathcal{O}(\Delta x^{\alpha-\beta})$,
%thus giving an error scaling for the first order expansion of $\mathcal{O}(\Delta t, \Delta x^{\alpha-\beta})$. For a higher order expansion of $\Delta t^n$, we repeat the argument and get $\mathcal{O}(\Delta t^n, \Delta x^{\alpha-n\beta})$.\newline

We see from this formula that going to higher orders in $\Delta t$ decreases the spatial order. This is fine for $L =\partial_x$ because the total order remains the same, but for higher derivatives, the spatial order decreases faster in $p$ than the time order increases, decreasing the total order and making the scaling worse.
This can be thought of in the following way: if the prior drops off as some power $\alpha$, then it is only $\alpha$ times differentiable, so it is not smooth. In the limit of high resolutions, the bins approximate something non-smooth, and thus there is a maximum possible order in the expansion. This doesn't necessarily imply that higher orders in time give worse IFD simulations. We know this is not true for the case of straight advection, where second order codes are stable, and first order codes are not. Furthermore, we already know that for normal finite-difference derivatives, higher orders aren't better when the fields are rough. \newline

This formula can be immediately used to get an error estimate on the toy model. Note that the boxes in this model are not smooth, and the Fourier transforms have a $1/k$ scaling independent of the box width, which shows up in the denominator of the transport operator (eqn.~\ref{kernel}) in addition to the $1/k^4$ from the prior.
Thus, for the second order implementation, the denominator on the transport operator scales as $k^{6}$ and $L=ik$, so 

\begin{equation}
E_d = \mathcal{O}(\Delta t \Delta x^{-5})+\mathcal{O}(\Delta t^2 \Delta x^{-4})
\end{equation}

\begin{figure}[h]%{0.4\textwidth}
\includegraphics[width=\textwidth]{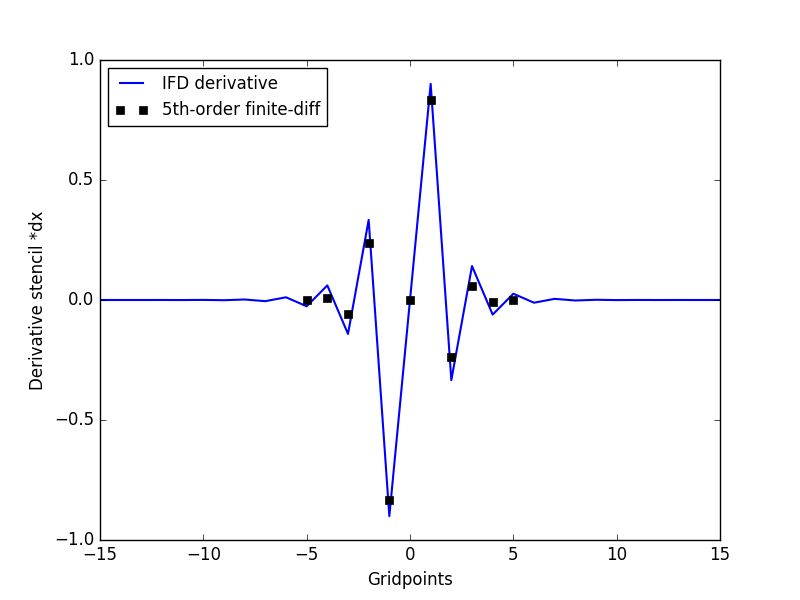}
\caption{Comparison of the numerical approximation to the first derivative produced by IFD, and the standard numerical fifth-order derivative approximation. 
The two are quite similar, though the IFD derivative extends over a slightly larger distance. The IFD derivative was calculated on a periodic interval of 50 gridpoints, using the toy model, which has a $1/k^4$ prior.}
\label{stencil}
\end{figure}

We now know that the derivative that we have been implementing is fifth-order, and this confirms something that experience has shown in the codes; they appear to be implementing something analogous to higher order derivatives. When we compare the first derivative operator for the toy code, vs the standard fifth order derivative, we see that they are quite similar, as shown in figure \ref{stencil}. This isn't to say that they are the same. The error scaling we just derived only kicks in in the limit in which the shape of the bins is lost. At lower resolutions, the structure of the bins will provide some additional subgrid structure, which should be advantageous, if the bins are intelligently chosen. Important to note also is that the operators produced in IFD still have nonlocal effects, i.e. it is not just the resolution, but the resolution compared to the size of the domain which has an effect on the resulting operators.

\section{Extension to nonconstant velocities}

The code was extended to cover nonconstant velocity fields, and nonperiodic boundary conditions. This means that the transport operator had to be computed numerically.
Adding a diffusion term to the system would have been a possible extension, but because diffusion simply destroys fine structure, the addition of diffusion actually makes numerical simulations \textit{easier}. In fact, adding artificial diffusion is a common way to stabilize misbehaving codes \cite{Godunov}. \newline 

We picked an approximation to signal space whose resolution was higher than that of data space by a factor of 100. A position space representation of the prior and responses was then chosen, and the $(R\Phi R^\dagger)_{ij}$ matrix was computed according to the convolutional action of the prior:
\begin{equation}
(R\Phi R^\dagger)_{ij}=\int B_i(x) \bigg{(}\int \Phi(x-y)B_j(y) dy\bigg{)}dx 
\end{equation}
The domain of definition of the convolution depends on the boundary conditions. Thus the inverse matrix $(R\Phi R^\dagger)^{-1}$ contains information about the geometry. To compute the second matrix $RL\Phi R^\dagger$, we constructed a matrix representation of $L=\partial_x v(x)$ on signal space by taking a standard central difference numerical derivative, i.e. $L f(x) = (f_{i+1}v_{i+1} - f_{i-1}v_{i-1})/(100\Delta x)$. The rest of the computation was done with normal matrix algebra. For stability, a second order time expansion was used, $\bar{U} = \mathbbm{1} + \Delta t L + \Delta t^2 L^2/2$. \newline

\begin{figure}
  \begin{minipage}[b]{0.5\linewidth}
    \includegraphics[width=\textwidth]{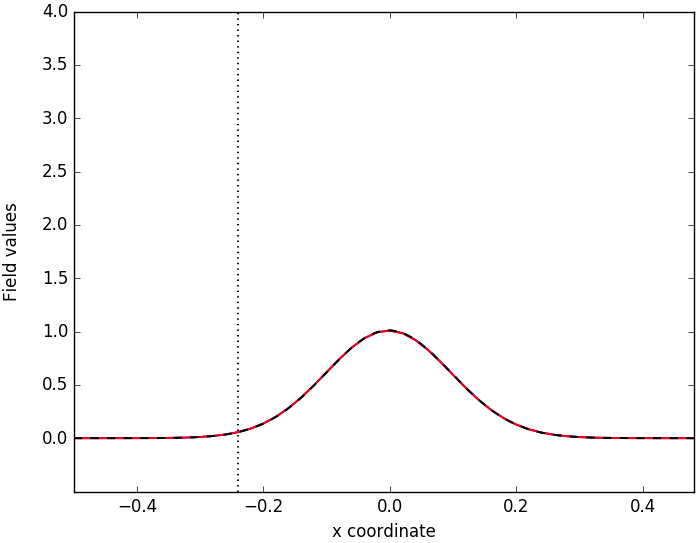}
    %\caption{Picture 1}
    %\label{fig:1}
  \end{minipage}
  \begin{minipage}[b]{0.5\linewidth}
    \includegraphics[width=\textwidth]{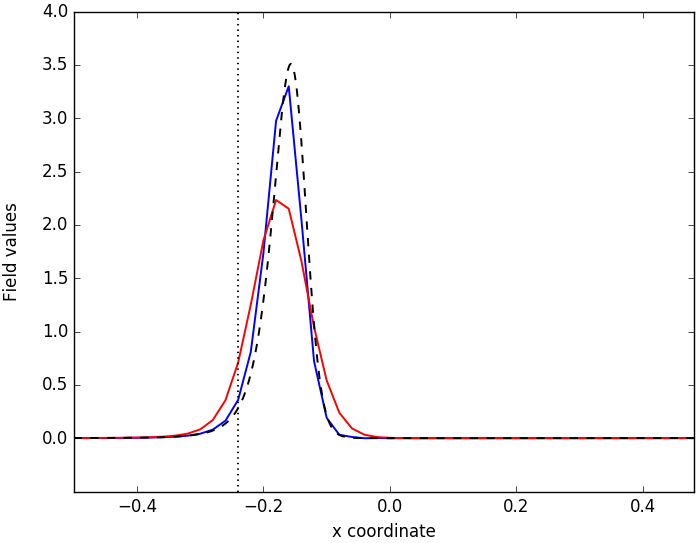}
    %\caption{Picture 2}
    %\label{fig:2}
  \end{minipage}
\begin{minipage}[b]{0.5\linewidth}
    \includegraphics[width=\textwidth]{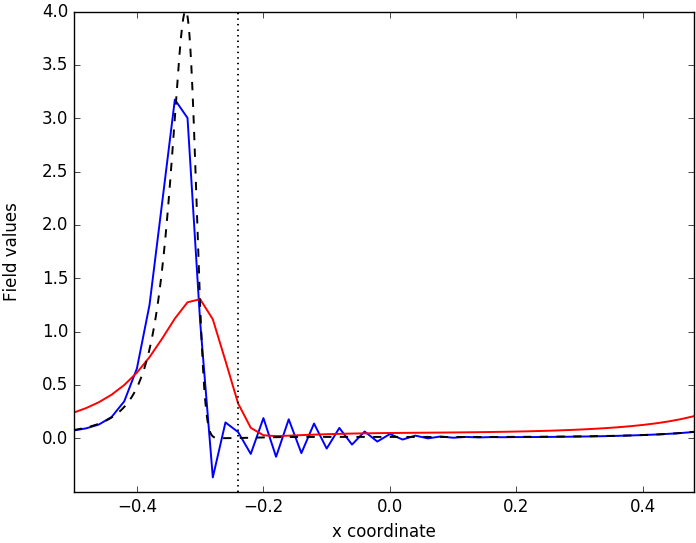}
    %\caption{Picture 2}
    %\label{fig:3}
  \end{minipage}
	\begin{minipage}[b]{0.5\linewidth}
    \includegraphics[width=\textwidth]{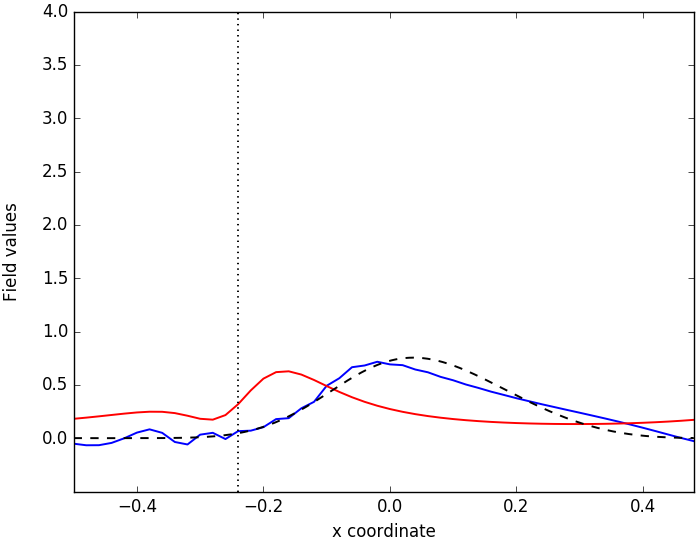}
    %\caption{Picture 2}
    %\label{fig:4}
  \end{minipage}
\caption{Simulated advection of an exponential pulse for a nonconstant leftward directed velocity field, with periodic boundary conditions. The simulation space has 50 grid points and 1000 timesteps, with a length and velocity of 1 and runtime of 2, in arbitrary units. The minimum in the velocity field is denoted by a dotted vertical black line. The first panel shows the initial conditions. The second panel shows the beginning of compression. The third panel shows the pulses after passing through the compression region; The IFD code has kept its shape better than the traditional solver, but has developed spurious oscillations due to the poor handling of shocks. The final panel shows the pulses after one full cycle; the IFD code still displays artefacts, but otherwise performs well.}
\label{sinev}
\end{figure}

\begin{figure}
  \begin{minipage}[b]{0.5\linewidth}
    \includegraphics[width=\textwidth]{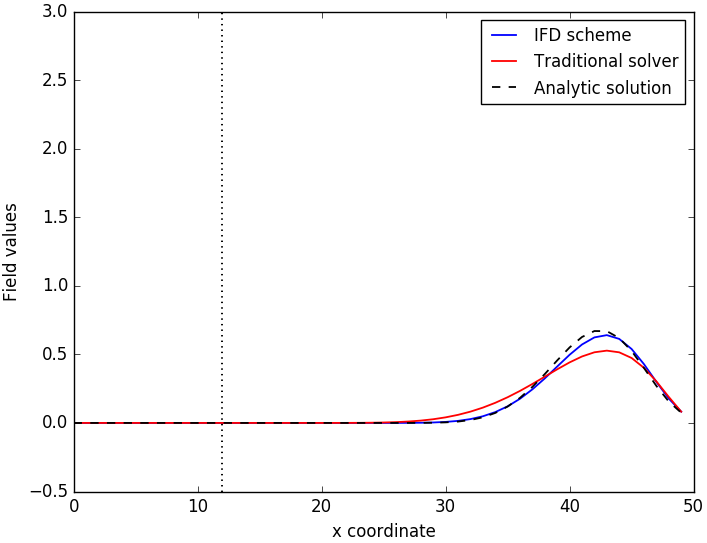}
    %\caption{Picture 1}
    %\label{fig:1}
  \end{minipage}
  \begin{minipage}[b]{0.5\linewidth}
    \includegraphics[width=\textwidth]{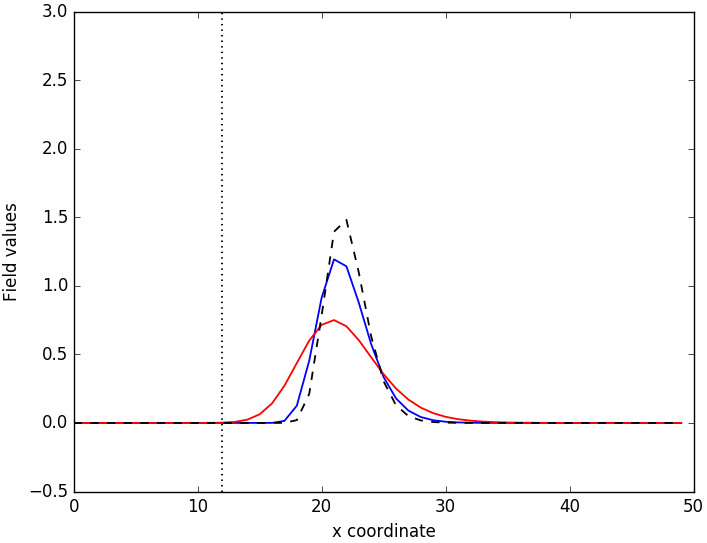}
    %\caption{Picture 2}
    %\label{fig:2}
  \end{minipage}
\begin{minipage}[b]{0.5\linewidth}
    \includegraphics[width=\textwidth]{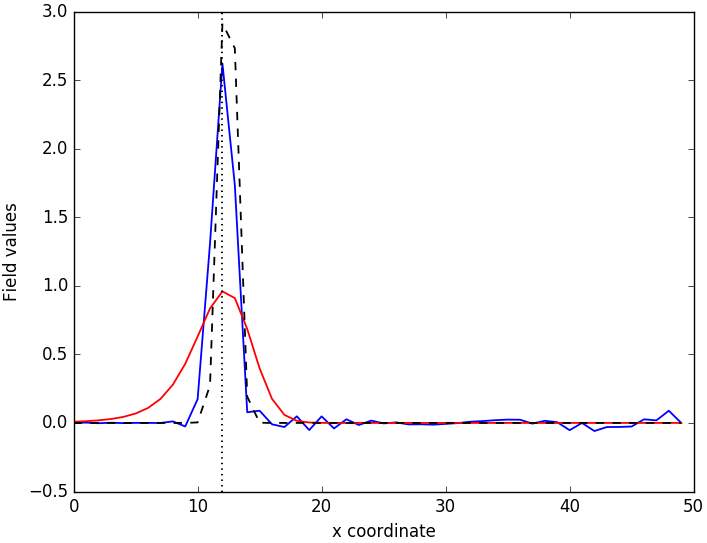}
    %\caption{Picture 2}
    %\label{fig:3}
  \end{minipage}
	\begin{minipage}[b]{0.5\linewidth}
    \includegraphics[width=\textwidth]{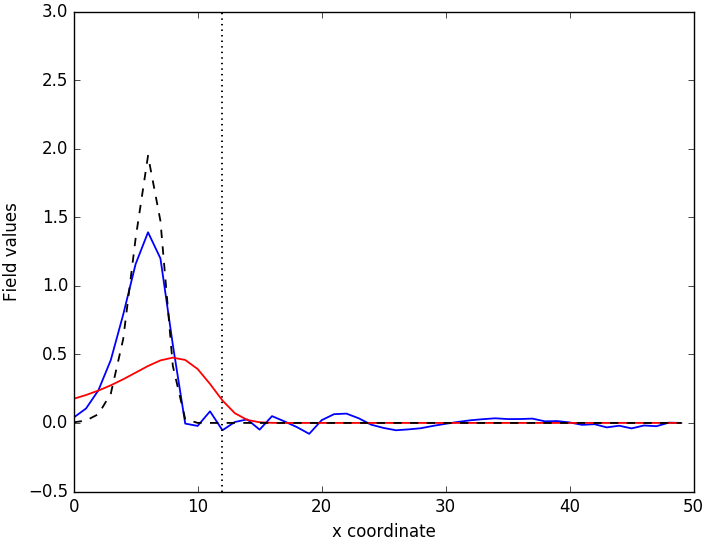}
    %\caption{Picture 2}
    %\label{fig:4}
  \end{minipage}
\caption{Simulated advection of an exponential pulse for a nonconstant leftward directed velocity field, with nonperiodic boundary conditions. The simulation space has 50 grid points and 1000 timesteps, with a length and velocity of 1 and runtime of 2, in arbitrary units. The minimum in the velocity field is denoted by a dotted vertical black line. The first panel shows the pulse being injected from the right hand side. The second panel shows the beginning of compression. The third panel shows the pulse at the minimum of the velocity field. As with the previous figure, the IFD has retained its shape better than the traditional solver, but has developed oscillations, as well as a new oscillation which collects on the boundary. The final panel shows the pulse leaving the compression region. Overall, the IFD solver performs better than the traditional solver, but its performance is much worse compared to the constant-velocity case.}
\label{sinebc}
\end{figure}

We present two pairs of example outputs, one with periodic boundary conditions (fig.\ref{sinev}), and one without (fig.\ref{sinebc}). The rest of the simulation parameters were kept as similar as possible to aid comparison. The prior was chosen to be a position-space representation of $1/(k^4 + m^4)$, with the width adjusted to that $\Phi(x)$ effectively drops to zero within the length of the simulation domain, so that convolution over $\Phi(x)$ was valid in both spaces. The mass term $m$ was kept very small and was only used to regularize the integrals. The response bins were the square boxes from the previous toy model. A velocity field valid on both spaces was also chosen, $1 + c \sin(2 \pi x/l)$ for a tuneable constant $c$, whose value was kept $<1$ to ensure the velocity field has a consistent direction. The field flows from right to left, and the sinusoidal character means that there is a velocity minimum in the left-hand corner of the box. This compresses the fields and causes shocks. \newline

For the case with boundary conditions, the numerical convolution was computed by allowing $\Phi(x-y)$ to run off the edge of the simulation domain. Thus the $R\Phi R^\dagger$ matrix produced is not translation-invariant in data space, and shows boundary effects, though explicit boundary conditions have not yet been applied. This matrix is still invertible, and $RL\Phi R^\dagger$ was computed similarly. In this scheme, the matrix inversions carry information about the geometry, which means the transport operator automatically accounts for the existence of a boundary. The boundary conditions themselves were then imposed by fixing the values of the gridpoints on each end, as with a normal simulation. These boundary conditions injected an exponential pulse at one end, and let it pass through at the other. As stated earlier, this approach to boundary conditions is somewhat ad-hoc. Various other approaches were attempted, but this method functioned the best. \newline

With both examples, there is no analytic solution to compare to, so they are compared to an ordinary finite-difference scheme carried out at 100 times resolution in time and space. This is regarded as the `true' solution. \newline % The simulation grid was 50 points long with length 1, and a runtime of 2 (arbitrary units) with 1000 timesteps. \newline

Both examples show the same weakness; a vulnerability to shocks as predicted by the analytic analysis. When the incoming pulses become compressed, the solutions develop extra maxima and minima. This can be expected from Godunov's theorem, now that the previous chapter has established that these IFD schemes are of very high order. Aside from the spurious oscillations, both examples retain the shape of the pulse better than the ordinary forward-difference code. Note that the periodic example was run with a factor of $c=1/1.2$ (arbitrary units), whereas the nonperiodic code was run with $c=1/1.5$. The second code is not able to handle compression as well, as edge effects coming from the $(R\Phi R^\dagger)^{-1}$ matrix in the transport operator induce an additional oscillation which originates on the boundary. \newline

This poor handling of compression can also be seen in light of the Bayesian interpretation. We have picked a prior such that the field is correlated over some length scale that is on the order of multiple bins. During compression, any initial variation in the field is squeezed to within the scale of a few bins, and thus there is sharp variation over a length which the prior says should be smooth. This violates the prior, and delivers poor results. We have again touched on the theme encountered at the end of subsection \ref{phasesec}. \newline

\chapter{SPH-like schemes}\label{ChapSPH}

\section{A brief introduction to Smooth Particle Hydrodynamics}
During the course of this project, a second type of model was attempted. Despite being unsuccessful, it is presented here for two reasons. The first is that it serves to highlight the importance of the theme discovered with the previous models: if the equations of motion contradict the prior, then the simulations begin to show errors. The second reason is that we believe it shows more promise than the translation-invariant schemes. For this problem, we once again attempt to solve the basic problem of 1D advection, but now with an added diffusion term:  $\partial_t f(x,t) = \partial_x(v(x)f(x,t))+ K \partial_x^2f(x,t)$, where $K$ is some constant diffusion coefficient. \newline

This model takes inspiration from so-called ``smoothed particle hydrodynamics'' (SPH) methods\cite{SPH}\cite{SPH2}\cite{SPH3} \cite{Springel}.In such models, rather than solving the equations of motion on a fixed set of $N$ gridpoints, the field in question (typically a fluid) is instead modelled by a set of $N$ ``particles'' which represent a sampling of the true field, each posessing a location and velocity. At each timestep, an estimate of the continuous field is reconstructed via some \textit{smoothing kernel} denoted by $W(x)$ which is positive, has a peak at zero, and drops off to 0 with increasing $x$. For a collection of particles at locations $x_i$ with mass $m$, the reconstructed density is given by 

\begin{equation}\label{sph}
\rho(x) = \frac{m}{N}\sum_i W(x-x_i)
\end{equation}

Typical choices for the smoothing kernel are Gaussians, or polynomial splines which fall to zero inside some compact region \cite{Springel}. The density reconstruction in an SPH code converges to the true density in the limit of $N \to \infty$ provided the width of the smoothing kernel $W(x)$ is scaled with $1/N$ \cite{SPH}. From this reconstructed density field, using the equations of motion of the field, $\rho(x)$, and the particle velocities $u_i$, the force on the $ith$ particle $F_i$ is then calculated, which is used to update $u_i \to u_i +\Delta t F_i$. This new velocity is then used to update the positions $x_i \to x_i + \Delta t u_i$, and the cycle then repeats.\newline

There are many advantages to this type of model, the most relevant of which is that it can be thought of as a model with an adaptive simulation grid. The variable particle locations automatically increase the spatial resolution of the code in regions where there is most activity. This makes them extremely useful for cosmological problems in which there are large density gradients, such as cosmic ray simulations.

\section{The IFD approach}

The analogy between IFD and SPH is immediately apparent. Both frameworks take a sampling from the field, from which a reconstruction is derived, which is then fed into the equations of motion. Typically the smoothness prior will take the form of a convolution over some integration kernel, so we can immediately make the comparison $\Phi(x-y) \Leftrightarrow W(x-y)$. The ``particles'' would then obviously take the form of a set of delta functions, representing point-measurements of the field i.e. $R_i = \int \delta(x-x_i)$. The locations of these delta functions would then be shifted in time according to the equations of motion of the field.\newline

We call our new model the \textit{SPH-like} code. According to the data/response equivalence as discussed in section \ref{redundancy},
the responses can be updated in such a way that it is equivalent to updating the data. The model we are about to define is a 'hybrid' IFD model, in which the computation is split between the data and the response. The particles represent point measurements of the field, thus their mass can vary depending on the value of the field at that location. Advection is modelled by moving the locations of the response delta functions, and diffusion is modelled by redistributing mass between the data points (particles). This is in contrast to an SPH code, where the particles have fixed masses, and diffusive effects are modelled by adding a repulsive force between them.\newline

The SPH-like code needs to be expressed in equations. The responses will be dependent on time and space, so we write $R_i^j$ with lower indices denoting spatial coordinates, and upper indices denoting time coordinates. The responses are then given by $R_i^j =\int \delta(x-x_i^j)$ for some finite set of locations $\{x_i^j\}$ at timestep $t^j$. The prior $\Phi$ will be constant in time, and will be given by a convolution over some kernel $\Phi(x)$. The time evolution operator is expanded to first order $\bar{U} =\mathbbm{1}+\Delta t \partial_x v(x) +\Delta t K \partial_x^2$. Because the terms will be broken up, we denote $\partial_x v(x)$ by $L_A$ for the advection part, and $K \partial_x^2$ by $L_D$ for the diffusion part, with associated $U_A$ and $U_D$ time evolution operators. For dynamically changing responses, we write down the transport operator:

\begin{equation}
T^j=R^{j+1} (\underbrace{1 + \Delta t v(x) \partial_x}_{U_A} + \underbrace{\Delta t K \partial_x^2}_{L_D }) W^j %\Phi R^{j\dagger} (R^j \Phi R^{j\dagger})^{-1}
\end{equation}

Note that $W^j$ is the Wiener filter at time $t^j$, $W^j=\Phi R^{j\dagger} (R^j \Phi R^{j\dagger})^{-1}$, and \textit{not} the SPH smoothing kernel. We define ${R^{j+1} = R^j U_A^{-1}=R^j (1 + \Delta t v(x)\partial_x)^{-1}}$, so that the advection part is absorbed into the evolving responses. Now we do a little bit of lazy mathematics: the action of $R^j U_A^{-1}$ on a field corresponds to taking the field, advecting it backward in time, and then measuring it with the delta functions located at the positions $x^j_i$. This is the same as taking the locations of the delta functions, and advecting them forward in time: $x_i^{j+1} = x_i^j +\Delta t v(x_i^j)$. Now that we have defined the updating rule for the responses, we can write the transport operator as:

\begin{align}
&T^j =\underbrace{R^{j+1} (1 + \Delta t v(x) \partial_x}_{=R^j U_A^{-1}U_A=R^j} + \Delta t K \partial_x^2) W^j \nonumber \\
&=\big{(}R^{j}  + R^{j+1} \Delta t L_D\big{)} W^j= \mathbbm{1} + \Delta t R^{j+1}L_D W^j
\end{align}

Where the last equality follows from the fact that $R^j W^j =\mathbbm{1}$.
One full cycle of the scheme is then as follows:

\begin{enumerate}
\item Construct the Wiener filter using the response locations at timestep $j$,  $W^j = \Phi R^{j\dagger} (R^j \Phi R^{j\dagger})^{-1}$. 
\item Reconstruct the field at timestep $j$ based on the Wiener filter and the data: $\phi(x)=W^j d^j$.
\item Apply the diffusion operator $L_D= K \partial_x^2$ to the reconstructed field.
\item Advect the response locations ${x^{j+1} = x_i^j+\Delta t v(x_i^j)}$, such that $R_i^{j+1} = \int \delta(x-x_i^{j+1})$.
\item Apply the new response to the diffused field: $ R^{j+1}\phi(x)= R^{j+1} L_D W^j d^j$.
\item Construct the new data: $d^{j+1}=T^j d^j=(\mathbbm{1} + \Delta t R^{j+1} L_D W^j) d^j$.
\end{enumerate}

This method should hopefully give an IFD scheme that has all the advantages of an SPH code. This scheme has a changing response at each timestep, 
which means the Wiener filter will need to be continually recomputed. 
This involves performing operations on objects in signal space, which should in general be computationally expensive; however the choice of responses and some clever computational tricks render this quite easy. Since the prior is static in time, we can store a single representation of $\Phi(x)$ at a very high resolution. We can exploit a handy property of convolution here: for two functions $f$ and $g$, $\partial_x (f * g) = (\partial_x f)*g =f* (\partial_x g)$. So, to know the action of $\partial_x^2$ on $\Phi * \phi$, we only need to know $\partial_x^2 \Phi(x)$, which we also store a representation of. The $\Delta t$ part of the transport operator is then computed in two parts, 
\begin{equation}
R^{j+1} L_D W^j = \underbrace{R^{j+1}L_D \Phi R^{j\dagger}}_{\text{part one}} \underbrace{(R^j \Phi R^{j\dagger})^{-1}}_{\text{part two}}
\end{equation}

Since the responses are delta functions, these two matrices can be computed extremely easily, using the fact that $(R \Phi R)_{ik}^\dagger = \Phi(x_i-x_k)$ and similarly, $(R \partial_x^2 \Phi R^\dagger)_{ik} = K\partial_x^2 \Phi|_{x_i-x_k}$. In the actual algorithm, this means that the two matrices are computed by just indexing into the stored copies of 
$\Phi$ and $\partial_x^2 \Phi(x)$ respectively, which is quite cheap in terms of computation time. This is however hard on memory, given that any approximation to signal space must have much higher resolution than that of data space in order to get good results. \newline

For a cutting-edge simulation, we obviously would like that the actual data simulation space is at the limits of our available memory. This code uses a 50$\times$ factor of resolution, and a translationally-invariant prior, which means the stored prior is 50 times larger than the actual data being simulated. This is at least better than the case with a general position-space prior, of the form $\int \Phi(x,y) dy$, which would be larger by a factor of $50^2$. There are two ways around this scaling problem: the first is to simply use a prior whose analytic form, and that of its derivatives, is known. If that is not possible, we can anticipate eventually going to three dimensions, where we can pick an isotropic prior $\Phi(\vec{x}) =\Phi(|x|)$, and then wait until the simulation gets large enough such that $50N< N^3$. Since the code in this report is just a proof of concept, we store the full prior and simply tolerate the terrible memory performance.\newline

We avoid the costly process of inverting $R \Phi R^\dagger$ at each timestep, as matrix inversion scales as $\mathcal{O}(N^3)$, and instead compute the cheaper (and more stable) problem of solving the linear system of equations $(R \Phi R^\dagger)v=d^j$ for the specific $d^j$ at each timestep, and some unknown vector $v$. This only scales as $\mathcal{O}(N^2)$. We then apply $R_{i+1}L_D\Phi R_i^\dagger$ to this vector $v$, thus computing $R^{j+1} L_D W^j d^j$ without having to compute the full transport operator.\newline

There are some additional differences between this code and an SPH code. For an SPH code, observation of eqn.~\ref{sph} shows that the reconstruction is roughly analogous to $\Phi R^\dagger$ in the IFD framework, and is thus local. Our code differs by the factor of $(R \Phi R^\dagger)^{-1}$, thus rendering the reconstructions slightly nonlocal. The nonlocality is supposed to be advantageous, as it exploits the correlation structure of the field to give better reconstructions. Furthermore, in our code, the particles do not have their own velocities. This is due to the fact that we are modelling advection, so the particle velocity is entirely determined by the vector field which pushes it. If this code were to be extended to model a more complex fluid-dynamics problem, where the velocity of the fluid itself comes into the equations, then the inclusion of a particle velocity would become necessary. \newline

\section{Results and post-mortem}

\begin{figure}
  \begin{subfigure}[b]{0.5\textwidth}
    \includegraphics[width=\textwidth]{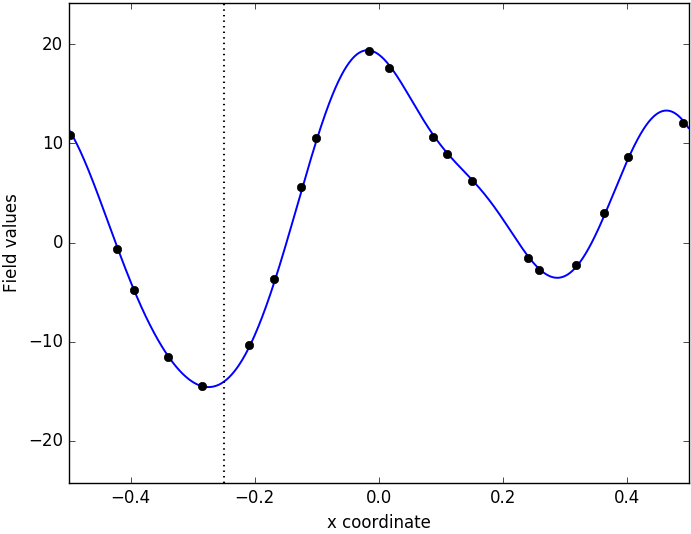}
    %\caption{Picture 1}
    %\label{fig:1}
  \end{subfigure}
  \begin{subfigure}[b]{0.5\textwidth}
    \includegraphics[width=\textwidth]{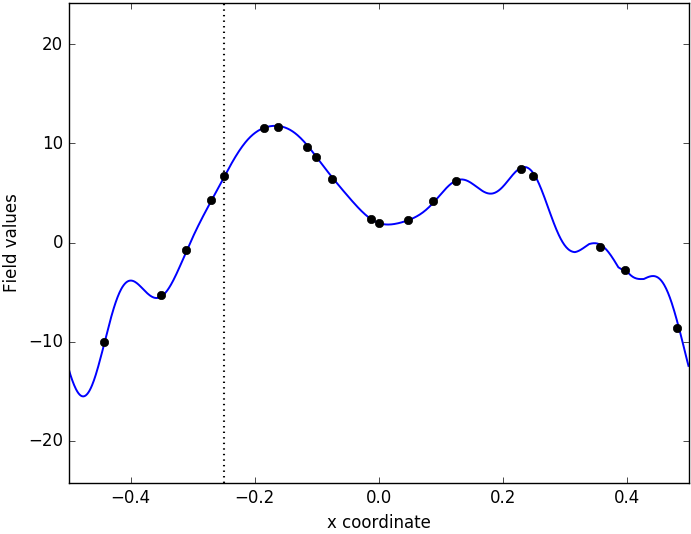}
    %\caption{Picture 2}
    %\label{fig:2}
  \end{subfigure}
\begin{subfigure}[b]{0.5\textwidth}
    \includegraphics[width=\textwidth]{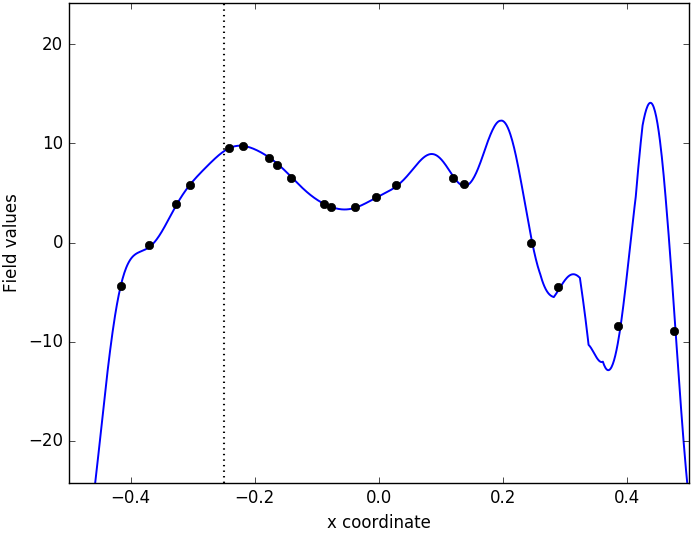}
    %\caption{Picture 2}
    %\label{fig:3}
  \end{subfigure}
	\begin{subfigure}[b]{0.5\textwidth}
    \includegraphics[width=\textwidth]{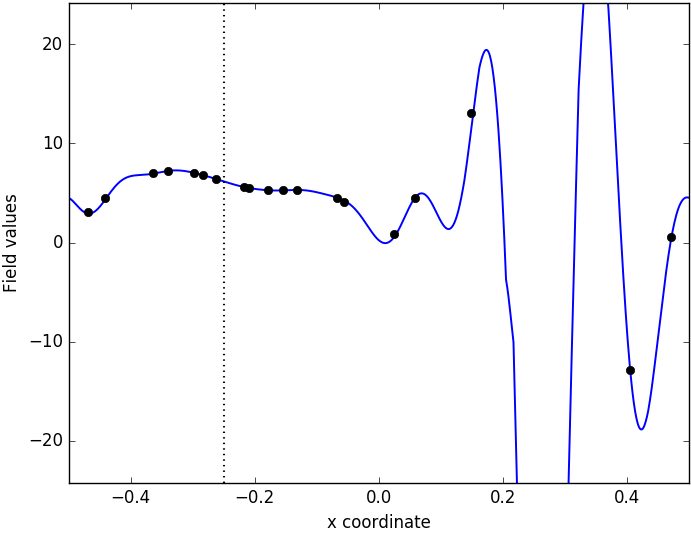}
    %\caption{Picture 2}
    %\label{fig:4}
  \end{subfigure}
\caption{Output of the SPH-like code on box of length 1 and a runtime of 1 in arbitrary units, with periodic boundary conditions. The velocity field is sinusoidal and leftward-directed, with the minima denoted by a dotted vertical black line. Outputs shown are for timeteps 0, 106, 148, and 238. The figure shows the development of instabilities in the reconstruction as the initially well-spaced population of particles gets compressed in the left hand side of the box. Note that a violation of mass conservation can clearly be seen.}
\label{pp}
\end{figure}

The SPH-like code was implemented, and is was seen to be extremely unstable. It diverges sharply when compressed, which was exactly what it was designed not to do.  As with the previous examples, we took a simulation domain with periodic boundary conditions, and a sinusoidally-varying velocity field $1 +c \sin(2\pi x)$ for some tunable constant $c$, and arbitrary units. This model of course included a constant diffusion term. An example output is shown in figure \ref{pp}, which was run with $c= 1/2.9$ and a diffusion coefficient of $K=1/60$. The simulation used 20 test particles, and the initial conditions were a randomly sampled smooth field, whose correlation structure was the same as the prior. The prior was an exponential which drops to zero within the length of the simulation space. Various other priors were attempted, all with uniformly bad performance.\newline 

This code is stable as long as the particles remain evenly spread out, but as soon as multiple particles are compressed into a small space, the reconstruction diverges, and the particle values themselves then diverge. This behaviour is surprising, given that the no-noise Wiener filter always has the property that $RW= \mathbbm{1}$, which means that the reconstruction always perfectly agrees with the data at the exact locations of the particles. However it can be seen that the reconstruction begins to oscillate wildly in the space \textit{between} particles, and thus the gradients of the reconstruction at the particle locations become large, which bleeds into the data in the next timestep under the action of $\partial_x^2$. \newline

\begin{figure}\centering
  %\begin{subfigure}[b]{0.55\textwidth}
    \includegraphics[width=0.8\textwidth]{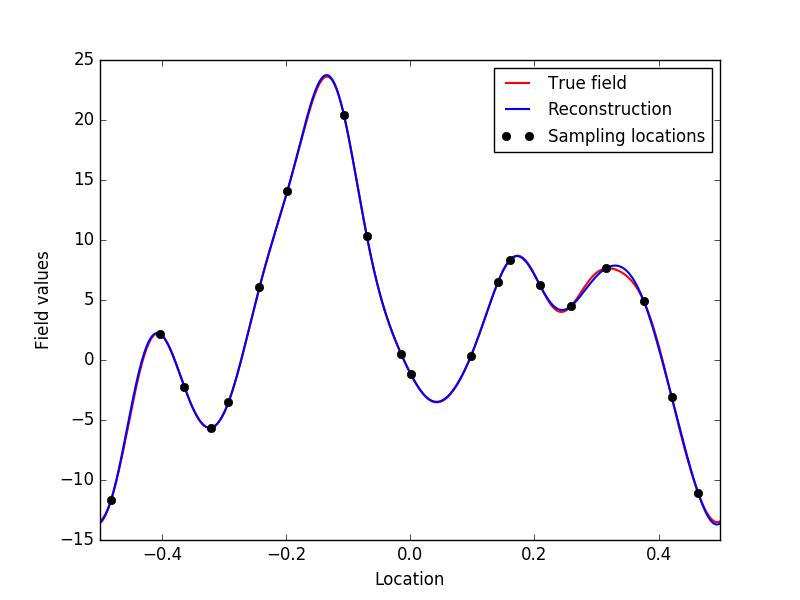}
    %\caption{Picture 1}
    %\label{fig:1}
  %\end{subfigure}
  %
  %\begin{subfigure}[b]{0.55\textwidth}
    \includegraphics[width=0.8\textwidth]{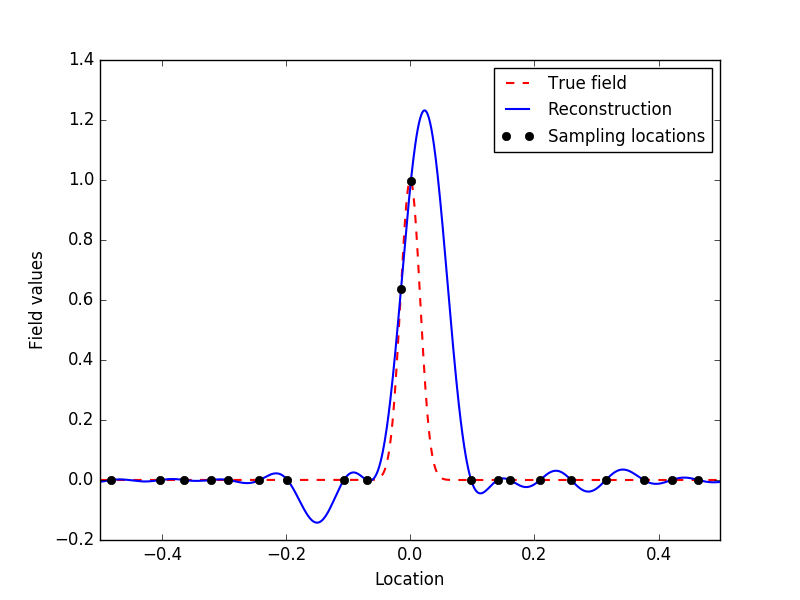}
    %\caption{Picture 2}
    %\label{fig:2}
 % \end{subfigure}
\caption{Comparison of reconstructions for two different fields given a set of irregular point measurements, reconstructed with the same prior. The prior has a Gaussian correlation structure. The plot on the left shows a randomly sampled field which also has Gaussian spatial correlations of width slightly smaller than that of the prior; the reconstruction is scarcely distinguishable from the original field. The plot on the right shows the reconstruction of a single narrow bump of width much smaller than the prior. This field is considered unlikely according to the prior, and displays numerous artefacts such as overshooting, and extra maxima and minima.}
\label{reconstruction}
\end{figure}

In human terms, the failure of this code can be described as follows: In the IFT framework, the reconstruction of a field given some data typically gives very accurate results if the data is considered to be likely according to the prior. However, if the data is considered unlikely, then the reconstructions can  become very, very bad. An example of this is shown in figure \ref{reconstruction}, which compares the reconstructions of a field given a set of point samples, for both a ``likely'' and ``unlikely'' data set. The reconstruction was performed using a basic Gaussian smoothness prior. The first result was random noise sampled with a very similar power spectrum to the prior, whereas the second was a localised bump of width much narrower than the correlation length of the prior.  \newline

We can now interpret this observed divergence under strong compression. The simulation takes an initial thermally agitated field, with variations on the scale of the correlation length of the prior. This field is then compressed so that there is significant variation within the supposed correlation length. Thus the data is considered unlikely, and the reconstruction diverges. We have yet again touched on our common theme that the equations of motion must be consistent with the prior. In this case, we have supposed translation-invariance, which supposes that no locations are special, despite the fact that the compression region is clearly special.\newline

In mathematical terms, the problem lies in the inversion of the $R \Phi R^\dagger$ matrix. Indeed, given almost any static prior, we can show that there will always be a scale at which the matrix $(R \Phi R^\dagger)^{-1}$, and thus the Wiener filter, diverges.

\begin{theorem}
Given any prior which takes the form of an integration over a continuous, symmetric kernel $\Phi \phi(x) = \int \Phi(x,y) \phi(y) dy$ over some metric space, 
and a time-varying response of the form $R_i^j =\delta(x-x_i^j)$ for some finite set of locations $\{x_i^j\}$ at a set of timesteps $\{ t_j \}$, then there will always be some compression scale for which the determinant of $(R\Phi R^\dagger)^{-1}$ becomes arbitrarily large.

\begin{proof}
Suppose hypothetically that one had a pair of response locations that were identical, so that $x_1=x_2$, then $(R \Phi R^\dagger)_{ik} = \Phi(x_i, x_k)$ will have two identical columns $(R \Phi R^\dagger)_{i1} = \Phi(x_i, x_1)  = \Phi(x_i, x_2) =(R \Phi R^\dagger)_{i2}$, and thus its determinant will be zero. \newline

The rest follows as a simple consequence of continuity. Suppose there is any sequence of response locations $\{ x_i^j \}$ in time for $j \to \infty$ which converge to some locations $x_i$, such that one pair approaches each other: $||x_1^j -x_2^j||\to 0$ as $j \to \infty$. Continuity of the kernel $\Phi(x,y)$ then implies that  the sequence of matrices $\Phi(x_i^j,x_k^j)$ approaches some matrix $M_{ij}$ in the euclidean matrix norm, which is defined as $||M||=\sqrt{\sum_{ik} |M_{ik}|^2}$. That is to say 
$\sqrt{\sum_{ik} (\Phi(x_i^j,x_k^j) -M_{ij} )} \to 0$. This is true because each entry of $\Phi(x_i^j,x_k^j)$ converges by continuity of $\Phi(x,y)$. $M_{ij}$ has two equal columns and thus it's determinant is zero. Since the determinant function is also continuous in the euclidean norm,  $\det(\Phi(x_i^j,x_k^j)) \to \det(M) = 0$, and thus $\det((\Phi(x_i^j,x_k^j)^{-1}) \to \infty$. Hence, the determinant of $(R\Phi R^\dagger)^{-1}$ can be made arbitrarily large by choosing the response locations close enough.

%The determinant function $\det(M_{ik})$ is continuous in the euclidean entry norm: $||M||=\sqrt{\sum_{ik} |M_{ik}|^2}$. 
%an integration kernel, and the responses are delta functions, if, for any time there is a $x(t_i)_j = x(t_i)_k$ then $\det(R\Phi R^\dagger)=0$. Now suppose I have any (time) sequence of $\{ x(t_i)_j \}$'s such that there is one pair for which $|x(t_i)_j -x(t_i)_k|\to 0$. Continuity of the kernel $K(x,y)$ implies sequential continuity so that $ K(x(t_i)_j,x(t_i)_k)$ sequentially approaches some matrix $\widehat K_{ij}$ in the euclidean matrix norm (because all it's entries do). Since $\det(\widehat K_{ij}) = 0$, $\det(K(t_i)_{jk}) \to 0$. Thus the determinant of the inverse blows up. This shows that for any static prior, the $(R \Phi R^\dagger)^{-1}$ matrix can be made arbitrarily large.  

\end{proof} 

\end{theorem}

Note that in this limit of high compression, the matrix elements themselves do not become smaller. This shows that the matrix inversion is blowing up due to a true degeneracy, rather than a cosmetic scaling which could potentially be cancelled by the $\Phi R^\dagger$ term in the Wiener filter. This instability caused by the divergence of the Wiener filter can be reduced by introducing enough diffusion into the model such that structure is eliminated faster than it is compressed. However the amount of diffusion required is extremely large, and also does not counter the inaccuracy introduced into the equations by the unstable Wiener filter.\newline

There is another flaw with the SPH-like code, caused by the use of delta functions. Observe that, for diffusion on a convolutional prior, the matrix elements on the main diagonal of $RL\Phi R^\dagger$ are given by $\partial_x^2 \Phi(x)|_{x=0}$. This means that the matrix entries depend entirely on the derivative of the prior at a single point, which is in contrast to box-responses or bump functions, where the matrix elements become averages over some region, and so are less dependent on the exact form of the prior. This did indeed have a noticeable effect on the simulations, where we experimented with various different reasonable-looking priors, and obtained very different diffusion speeds. This is in contrast to the toy code in the previous section, which was much more robust to small changes in the prior. Robustness is considered desirable, because often the correlation structure will be nothing more than an educated guess. \newline

A mathematical justification for this is available, if we imagine a hypothetical case where the particles are evenly spaced out and the velocity field is constant. Then the system is translation-invariant and we can apply eqn.~\ref{main}. Delta functions have a flat Fourier spectrum, so the transport operator becomes: 

\begin{equation} 
T(k)=  
\frac{ \sum^{}_{b \in 2f_N \mathbb{Z}} \bar{U}(k+b) \Phi(k+b)}
{\sum^{}_{\hat b \in 2f_N \mathbb{Z}} \Phi(k+\hat b)}
\end{equation}

Making small changes to the prior $\Phi(x)$ in a small region about the origin, by the uncertainty principle, corresponds to modifying the frequencies at infinity. Given that the derivative operators form polynomials in $k$, this drastically changes the convergence properties of the above sum. This is not the case 
when the bins are smooth and compactly supported, as they decay faster than any polynomial by the Paley-Wiener theorem. This gives the other codes their robustness in response to a changing prior. An extreme example of this flaw is when one inserts a seemingly reasonable momentum space prior:
$\Phi(k) = 1/ (k^2+m^2)$ which has a position-space representation of $\Phi(x)=e^{-m|x|}$; this is not differentiable and thus the code is unuseable. Once again, we expect the insight we gain in the translation-invariant case to hold in the non-invariant case. \newline

%For a model with dynamically updating delta-function responses, and prior which is convolution over a function $K$, the $L_1$ norm of the Wiener filter reconstruction is not conserved:
%\begin{proof}

%\begin{align*}
%&\int |\sum_{ij} \Phi(x-x_i) (R\Phi R^\dagger)^{-1}_ij d_j| dx = \big{(} \int |\Phi(x)| dx \big{)} \cdot \big{|} \sum_{ij}  (R\Phi R^\dagger)^{-1}_{ij} d_j  \big{|} \\
%& = ||\Phi||_1 \big{|} \sum_{ij}  (R\Phi R^\dagger)^{-1}_{ij} d_j  \big{|}
%\end{align*}
%We know that since $(R\Phi R^\dagger)^{-1}$ is a self-adjoint operator on a finite space, there exists a basis of eigenvectors. Since we know from the previous lemma that the determinant (product of eigenvalues, all positive) can be made arbitrarily large, the sum term can also be made arbitrarily large for a suitable choice of response locations and data. TRY DOING THIS FOR THE L2 NORM?
%\end{proof}
%\end{theorem}

%So, our reconstructions do not conserve mass (L1 norm), this is in contrast to particle-pusher codes which do. Also, we see that these models aren't guaranteed positivity either, simply because we require the space to be a vector space, so for a field in our space with positive mass, there also exists one with negative mass too.\newline

The last problem with this code is that the no-noise Wiener filter on delta functions doesn't preserve positivity of the fields. This makes the reconstructions unsuitable for simulating everywhere positive quantities, as was the case for the original goal of simulating cosmic ray streaming. Suppose the  prior $\Phi(x)$ is everywhere greater than zero (i.e. not just positive in the operator sense). If $R\Phi R^\dagger$ has any elements off the main diagonal, which it must in order to be useful, then these elements are all positive. Thus, the inverse matrix must have some negative entries in order to cancel the off-diagonal terms and yield the identity matrix. It is these negative elements which result in the fields not being everywhere positive. This of course is not a proof, rather an explanation of the observed behaviour. On a deeper level, this comes from the fact that the Gaussian regime of IFD requires signal space to be a vector field. Therefore for every possible field, it's negative must also be possible and is in fact equally likely to occur. A common approach in IFT is to enforce positivity by taking the logarithm of the fields, and supposing the statistics of the logged fields are Gaussian. However if we want to extend to more complex systems such as fluid dynamics, log fields become very inconvenient.

\section{Suggested improvements}

Despite the fact that the SPH-like code was the least successful of the two presented, it is the author's opinion that the second model shows the most promise. 
We know that the translation-invariant codes are implementing something analogous to very high-order finite difference schemes. The field of fixed-grid linear finite-difference solvers has been around for a very long time, and it is not suspected that there is much room for anything novel. SPH solvers are in contrast rather new, and already fit well with the IFD formalism. If the IFD reconstructions can be stabilized, then they could easily be inserted into existing SPH solvers, such as GADGET-2 \cite{Springel}, replacing the typical SPH reconstruction with a Bayesian one, and then proceeding as normal with the simulation. \newline

We previously saw that in the SPH-like model, a static prior will always give a Wiener filter that diverges under enough compression. In regions of high compression, very often SPH codes will also have to dynamically resize the width of the smoothing kernel, so that the density estimates remain accurate \cite{Springel}. Thus we know it is at least feasible to dynamically update basic features of the prior, such as its correlation width. This could be a potential fix for this problem. Alternatively, diffusive effects could be modelled by by a repulsive force between particles, thus preventing them from approaching each other too closely. The model would then need to be applied to a more complex fluid dynamics problem, where there are more forces than just advection and diffusion, so that the reconstructions have a purpose.\newline

A more elegant alternative to dynamically changing the prior width, would be to derive a more sensible starting prior. Taking a translation-invariant prior on a system which is not translationally invariant is in hindsight a terrible idea. This project has shown that prior selection is a very important part of IFD, and the sensitivity of the delta grid models to changes in the prior indicates that priors should not be chosen arbitrarily, but rather after careful consideration of the system at hand. This is of course much harder, and it is not known what a properly-selected prior for a system with nonconstant velocity fields should look like. \newline

Due to the delta-function responses, even with a more fitting prior, the simulations will still be very vulnerable to small changes in said prior. It is possible however to have responses which move under advection, but still have a volume. This would be something similar to a moving-mesh code.
Instead of having delta function responses, one could take the moving set of point locations from before, and define a set of box responses $R =\int B_i(x) dx$ around the points via a Voroni cell decomposition. A similar process is carried out in  \cite{mesh}, and we therefore know that recomputing the responses in this fashion at every timestep is at least computationally feasible. \newline

However, even with better responses and priors, we still have to solve the problem that the reconstructions do not preserve positivity, which will forever cause headaches whenever we try and model an everywhere-positive field. Under the assumption that going to log-fields is impractical, we may be required to leave the Gaussian regime of IFT, and consider more exotic reconstructions. \newline

\chapter{Conclusions}\label{summary}
\section{Summary}
This project did not produce a prototype cosmic ray simulation as intended. Many initial attempts at producing workable IFD codes were found to be unsuccessful. The source of the persistent problems with the attempted algorithms was not tracked down until a theoretical analysis of error and stability was undertaken. It took many months to realize that all the various unsuccessful codes could be united under a single unified framework of translation-invariant problems. This realization allowed us to develop the general form of the transport operator in momentum space, eqn.~\ref{main}, from which all the major conclusions of this work followed. Along the way, a number of smaller results were proven, such as the noise-invariance, as well as implementing minor fixes, such as taking the KL divergence in the right direction. \newline

Crucially, we showed that translation-invariant schemes in IFD are all consistent under a broad set of assumptions, which proves the basic validity of the IFD approach. A working prototype was then developed, which functions similarly to a higher-order linear finite difference scheme. The prototype model does indeed perform much better than basic finite-difference schemes. However the translationally-invariant schemes will generally perform poorly when simulating shocks and other numerically-interesting problems. Furthermore, given their similarity to existing linear codes, which are already extensively studied, it is not expected that this class of schemes will deliver particularly novel results. \newline

It is believed that the most promising code, despite its present flaws, is the SPH-like code. It can easily be integrated into current SPH algorithms, which are currently a very active field. It is also believed that there are techniques already used to stabilize SPH codes (adaptive smoothing lengths) which can be used to stabilize the prototype algorithm. It is hoped that the Bayesian nature of the reconstructions will provide an edge over the traditional field reconstructions in SPH codes. \newline

Though both types of algorithms explored in this project displayed the same general flaw; if the behaviour of the system contradicts the prior, then the results are bad. In the case of the SPH-like code, they can be catastrophically bad. This topic warrants its own discussion.

\section{Prior selection}

Every simulation scheme relies on a set of assumptions. The difference between traditional simulation schemes and IFD is that our assumptions are hard-coded into the simulation by our choice of prior. The experience gained in this project shows that this is a double-edged sword; if the data to be simulated is considered likely given the prior, then the simulations perform well; if the equations of motion  are inconsistent with the prior, then the results are bad, or even divergent.\newline

We have learned that correct prior selection is probably the primary consideration when developing a simulation scheme using IFD. There is still an unresolved question as to whether the prior is simply a tunable parameter which can be arbitrarily adjusted to give good simulations, or if it is an honest Bayesian belief about the system under study. The challenge going forward, is how to select a sensible prior in advance of performing the simulation. \newline

Given any initial prior, one can naively evolve it in time such that it is consistent with the equations of motion, by picking $\Phi(t) = U(t) \Phi_0 U(t)^\dagger$. This ensures that for any initial field $\phi_0$, the evolved field has the same probability of occurring according to the prior probability distribution. Strictly speaking, this is the logical thing to do anyway for a dynamic system. However this naive approach is not computationally feasible. For any interesting system, the exact time evolution is unknown, so the prior evolution must be simulated. Any approximation to signal space must be of much higher dimension than that of data space in order to be useful. If $m$ is the dimension of the approximation to signal space, and $n$ the dimension of data space, then evolving the prior correctly in time would involve simulating an $m \times m$ dimensional object, for the sole purpose of getting better simulations of an $n$-dimensional object. \newline

Cheap fixes for this problem may be found. As a suggested improvement for the SPH-like code, the correlation with of the prior could be scaled by some factor $\lambda$ such that $\Phi(x) \to \Phi(\lambda x)$ in regions where the particle density is high. This is already known to be technically feasible. This fix is somewhat 'ad hoc' though, and means that the prior becomes more of a tunable parameter rather than expressing our beliefs about the system.\newline

It appears that the best way forward is to derive a prior based on considerations of the equations of motion of the system itself. Philosophically speaking, whenever we perform a simulation, information theoretic or not, we assume that the mathematical equations of motion accurately describe the system under consideration, otherwise we wouldn't be doing the simulation. So, up to some factors such as boundary conditions, and the influence of random noise/thermal agitation, the prior should be able to be deduced from the equations of motion themselves. \newline

For example, if we can associate an energy to a field configuration, $E[\phi]$, then we can describe our initial ignorance about the system by a thermodynamic prior:
$\mathcal{P}(\phi) = \exp(-\beta E[\phi])$. For a typical Klein-Gordon field with mass, then $E[\phi] =\int \phi^\dagger ( m^2 + \nabla) \phi dx$, yielding the familiar 
$\Phi(k) = 1/(k^2 + m^2)$ prior.\newline

There is a danger here, we are taking the equations of motion and feeding them back into a simulation of the equations of motion, meaning that this process may simply be introducing redundant information. As an extreme example, for a linear system, the Green's functions give the correlation structure of the field (in time and space), but finding the Green's functions for a given system is equivalent to solving the system itself. So it appears that for nontrivial systems we should not expect to get the true correlation structure, but rather a very rough guess obtained by a consideration of symmetries, boundary conditions, and the physical scales of the relevant formulae. \newline

\section{Final remarks}

Information field dynamics is an extremely young theory; this thesis marks only the fourth publication on the topic. As with all new inventions, a significant amount of blood, sweat and tears will be required before it finally starts to generate novel results. In hindsight, attempting to develop a prototype simulation scheme for the challenging problem of cosmic ray transport was probably over-ambitious. Even on an extremely reduced problem, implementing a scheme that worked at all proved to be a significant challenge. We did not arrive at a scheme which worked better than an ordinary finite-difference scheme until a few months before the due date. However, future work can proceed armed with a set of theoretical results showing what works and what does not, and a general framework for analysing whatever problems may emerge. We also have a roadmap for the future, identifying a promising solution which doesn't work at present, but has a clearly prescribed set of fixes.
The issue of prior selection has turned out to be both crucial and rather tricky. The limited time requirements of a masters thesis means that the job of solving this issue must fall on the shoulders of some other (un)fortunate student.\newline

\medskip

%\begin{appendices}
%  \chapter{KL divergence derivation}
%blahblkasdjf;lksdkjf;alslkjf;asdlkfja
%\end{appendices}

\bibliographystyle{unsrt}
\bibliography{mastersbib_arxiv}

\begin{thebibliography}{10}

\bibitem{Hydro}
C.P. Dullemond and H.H. Wang.
\newblock Lecture numerical fluid dynamics.
\newblock Course given at Heidelberg university, retrieved from
  \url{http://www.ita.uni-heidelberg.de/~dullemond/lectures/num_fluid_2009/Intro.pdf},
  2009.

\bibitem{IFD}
Torsten~A. En\ss{}lin.
\newblock Information field dynamics for simulation scheme construction.
\newblock {\em Phys. Rev. E}, 87:013308, Jan 2013.

\bibitem{IFT}
Torsten~A. En\ss{}lin, Mona Frommert, and Francisco~S. Kitaura.
\newblock Information field theory for cosmological perturbation reconstruction
  and nonlinear signal analysis.
\newblock {\em Phys. Rev. D}, 80:105005, Nov 2009.

\bibitem{IFDmath}
Christian M{\"u}nch.
\newblock {Mathematical foundation of Information Field dynamics}.
\newblock Master's thesis, Technische Universit{\"a}t M{\"u}nchen, Germany,
  2014.
\newblock Retrieved from:
  \url{http://wwwmpa.mpa-garching.mpg.de/~ensslin/research/research_IFT.html#Publications}.

\bibitem{Jaynes}
E.T. Jaynes and G.L. Bretthorst.
\newblock {\em Probability Theory: The Logic of Science}.
\newblock Cambridge University Press, 2003.

\bibitem{Wiener}
N.~Wiener.
\newblock {\em Extrapolation, Interpolation, and Smoothing of Stationary Time
  Series: With Engineering Applications}.
\newblock Massachusetts Institute of Technology : Paperback series. M.I.T.
  Press, 1964.

\bibitem{KLref}
Anderson~D.R. Burnham~K.P.
\newblock {\em Model Selection and Multimodel Inference}.
\newblock Springer-Verlag, New York, 2002.

\bibitem{Reimar}
R.H. Leike and T.A. En\ss{}lin.
\newblock Optimal belief approximation.
\newblock Unpublished paper, retrieved from
  \url{https://arxiv.org/abs/1610.09018}, 2017.

\bibitem{MultiKL}
John Duchi.
\newblock Lecture notes for statistics 311/electrical engineering 377.
\newblock Stanford University, retrieved from
  \url{https://stanford.edu/class/stats311/Lectures/full_notes.pdf}, 2016.

\bibitem{Numerics}
R.~Corless and N.~Fillion.
\newblock {\em A Graduate Introduction to Numerical Methods: From the Viewpoint
  of Backward Error Analysis}.
\newblock SpringerLink : B{\"u}cher. Springer New York, 2013.

\bibitem{Lax}
P.~D. Lax and R.~D. Richtmyer.
\newblock Survey of the stability of linear finite difference equations.
\newblock {\em Communications on Pure and Applied Mathematics}, 9(2):267--293,
  1956.

\bibitem{Godunov}
C.~Hirsch.
\newblock {\em Numerical Computation of Internal and External Flows:
  Fundamentals of Computational Fluid Dynamics}.
\newblock Butterworth-Heinemann. Elsevier/Butterworth-Heinemann, 2007.

\bibitem{Skilling}
J.~Skilling.
\newblock Cosmic ray streaming—i effect of alfvén waves on particles.
\newblock {\em Monthly Notices of the Royal Astronomical Society},
  172(3):557--566, 1975.

\bibitem{CRmain}
{Enßlin, T.}, {Pfrommer, C.}, {Miniati, F.}, and {Subramanian, K.}
\newblock Cosmic ray transport in galaxy clusters: implications for radio
  halos, gamma-ray signatures, and cool core heating.
\newblock {\em A \& A}, 527:A99, 2011.

\bibitem{CRsecond}
{Enßlin, T. A.}, {Pfrommer, C.}, {Springel, V.}, and {Jubelgas, M.}
\newblock Cosmic ray physics in calculations of cosmological structure
  formation.
\newblock {\em A \& A}, 473(1):41--57, 2007.

\bibitem{Miniati}
Francesco Miniati.
\newblock Cosmocr: A numerical code for cosmic ray studies in computational
  cosmology.
\newblock {\em Computer Physics Communications}, 141(1):17 -- 38, 2001.

\bibitem{Miniati-new}
Francesco Miniati, Dongsu Ryu, Hyesung Kang, and T.~W. Jones.
\newblock Cosmic-ray protons accelerated at cosmological shocks and their
  impact on groups and clusters of galaxies.
\newblock {\em The Astrophysical Journal}, 559(1):59, 2001.

\bibitem{Reedsimon}
M.~Reed and B.~Simon.
\newblock {\em Methods of Modern Mathematical Physics}.
\newblock Number v. 2 in Methods of Modern Mathematical Physics. Academic
  Press, 1975.

\bibitem{SPH}
Monaghan~J.J. Gingold~R.A.
\newblock Smoothed particle hydrodynamics - theory and application to
  non-spherical stars.
\newblock {\em Monthly Notices of the Royal Astronomical Society},
  181(11):375--389, 1977.

\bibitem{SPH2}
Lucy L.B.
\newblock A numerical approach to the testing of the fission hypothesis.
\newblock {\em Astronomical Journal}, 82(12):1013--1024, 1977.

\bibitem{SPH3}
Volker Springel and Lars Hernquist.
\newblock Cosmological smoothed particle hydrodynamics simulations: the entropy
  equation.
\newblock {\em Monthly Notices of the Royal Astronomical Society},
  333(3):649--664, 2002.

\bibitem{Springel}
Volker Springel.
\newblock The cosmological simulation code gadget-2.
\newblock {\em Monthly Notices of the Royal Astronomical Society},
  364(4):1105--1134, 2005.

\bibitem{mesh}
Philip Mocz, Rüdiger Pakmor, Volker Springel, Mark Vogelsberger, Federico
  Marinacci, and Lars Hernquist.
\newblock A moving mesh unstaggered constrained transport scheme for
  magnetohydrodynamics.
\newblock {\em Monthly Notices of the Royal Astronomical Society},
  463(1):477--488, 2016.

\end{thebibliography}

\appendix
\chapter{Derivation of the Gaussian KL divergence}
%\begin{equation}
\begin{lemma*}
For two Gaussians $\mathcal{G}(\phi-a,A)$ and $\mathcal{G}(\phi-b,B)$, the KL divergence between the two $\mathrm{KL}(\mathcal{G}(\phi-a,A)||\mathcal{G}(\phi-b,B))$is given by:
\begin{equation}
\frac{1}{2}\bigg{(} \Tr [\ln(BA^{-1}) -\mathbbm{1} + B^{-1}A ]+\bra{b-a} B^{-1} \ket{b-a} \bigg{)}
%\frac{1}{2} \big[ \Tr(BA^{-1}) + (b-a)^\dagger B (b-a) - \Tr{\mathbbm{1}} - \Tr\ln(BA^{-1}) \big]
\end{equation}
\begin{proof}

\begin{multline}
\mathrm{KL}(\mathcal{G}(\phi-a,A)||\mathcal{G}(\phi-b,B)) = 
\int \mathcal{D}\phi \mathcal{G}(\phi-a,A) \bigg{[} \ln(\mathcal{G}(\phi-a,A)) - \ln \big( \mathcal{G}(\phi-b,B) \big) \bigg{]}\\
= \frac{1}{|2\pi A|^{1/2}}  \int \mathcal{D}\phi \exp \bigg{(}-\frac{1}{2}\bra{\phi-a} A^{-1} \ket{\phi -a} \bigg{)} \bigg{[} \bigg{(}  
-\frac{1}{2}\bra{\phi-a} A^{-1} \ket{\phi -a} - \frac{1}{2}\ln(2\pi \det A) \bigg{)} \\
-  \bigg{(}  -\frac{1}{2}\bra{\phi-b}^\dagger B^{-1} \ket{\phi -b} - \frac{1}{2}\ln(2\pi \det B) \bigg{)} \bigg{]}%\\
\end{multline}
The $\ln(2\pi \det A)$ terms may be factored out of the integral as they are independent of the fields. This just leaves a multiple of the integral of the Gaussian $\mathcal{G}(\phi-a,A)$, which equals one. We also exploit the identity that for a positive matrix $\ln (\det A)= \Tr(\ln A)$, which simplifies the equation to:

\begin{align}
=\frac{1}{2}[\Tr (\ln(B) - \ln(A))] +  &\\
\frac{1}{2|2\pi A|^{1/2}}  \int \mathcal{D}\phi \exp &\big{(}-\bra{\phi-a}  A^{-1} \ket{\phi -a} \big{)} \bigg{[} 
\bigg{(} \bra{\phi-b} B^{-1} \ket{\phi -b}  \bigg{)}  \nonumber\\
& \qquad \qquad \qquad- \bigg{(} \bra{\phi-a} A^{-1} \ket{\phi -a}  \bigg{)} \bigg{]} \nonumber 
%\bigg{(} -(\phi-a)^\dagger A^{-1} (\phi -a)  \bigg{)} \\
%-  \bigg{(} (\phi-b)^\dagger B^{-1} (\phi -b))  \bigg{)} \bigg{]} \\
%-\Tr(A A^{-1}) \\+ \int \mathcal{D}\phi \frac{1}{|2\pi A|^{1/2}}\exp [-\frac{1}{2}(\phi-a)^\dagger A^{-1} (\phi -a)](\phi-b)^\dagger B^{-1} (\phi -b)) ]\\
\end{align}
For the $\bra{\phi-a} A^{-1} \ket{\phi -a}$ term, we can shift to a new variable $\phi'=\phi-a$, which brings the integral into the form an expectation value:

\begin{align}
&\mathbb{E}\big{[} \bra{\phi'} A^{-1} \ket{\phi'} \big{]}_{\mathcal{G}(\phi',A)} = \mathbb{E}\big{[} \Tr( A^{-1} \ket{\phi'} \bra{\phi'})\big{]} \nonumber \\
&=\Tr ( A^{-1}\mathbb{E}\big{[} 
\ket{\phi'} \bra{\phi'} \big{]}) =\Tr (A^{-1} A) =\Tr(\mathbbm{1})
\end{align}

Whereas for the $\bra{\phi-b} B^{-1} \ket{\phi -b}$ term, we can perform a similar trick by decomposing it into 
$ \bra{\phi-a-(b-a)} B^{-1} \ket{\phi -a-(b-a)}$ and calling $(b-a)=y$, so we can rewrite it as $\bra{\phi'-y} B^{-1} \ket{\phi' -y}$ with $\phi'$ as before. 
Expanding this quadratic out gives us:
\begin{align}
&\mathbb{E}\big{[} \bra{\phi'-y} B^{-1} \ket{\phi' -y} \big{]}_{\mathcal{G}(\phi-a,A)} \\
&= \mathbb{E}\big{[} \bra{\phi'} B^{-1} \ket{\phi'} \big{]}
+\mathbb{E}\big{[} -\bra{\phi'} B^{-1} \ket{y} \big{]}+\mathbb{E}\big{[} -\bra{y} B^{-1} \ket{\phi'} \big{]}+\mathbb{E}\big{[} \bra{y} B^{-1} \ket{y} \big{]} \nonumber \\
&=\Tr \bigg{(} B^{-1} \underbrace{\mathbb{E}\big{[} \ket{\phi'} \bra{\phi'} \big{]}}_{=A} \bigg{)}
-\underbrace{\mathbb{E}\big{[}\bra{\phi'} \big{]}}_{=0}B^{-1} \ket{y} -\bra{y} B^{-1} \underbrace{\mathbb{E}\big{[}\ket{\phi'}\big{]}}_{=0}+
\underbrace{\mathbb{E}\big{[} \bra{y} B^{-1} \ket{y} \big{]}}_{=\bra{y} B^{-1} \ket{y}} \nonumber
\end{align}

%\begin{multline}
%=\frac{1}{2}[\Tr (\ln(BA^{-1})) -\Tr(A A^{-1}) + \langle (\phi-b)^\dagger B^{-1} (\phi -b))\rangle \\
%\end{multline}

%\begin{multline}
%\frac{1}{|2\pi A|^{1/2}}\int \mathcal{D}\phi \bigg[(\phi-a)^\dagger B^{-1} (\phi -a))-(\phi-a)^\dagger B^{-1} y-y^\dagger B^{-1} (\phi -a)+y^\dagger B^{-1} y \bigg]%\exp [-\frac{1}{2}(\phi-a)^\dagger A^{-1} (\phi -a)] \\
%=\Tr(B^{-1}A) +(b-a)^\dagger B^{-1} (b-a)
%\end{multline}
Combining all the parts together yields
\begin{equation}\label{KL}
\frac{1}{2}\bigg{(} \Tr [\ln(BA^{-1}) -\mathbbm{1} + B^{-1}A ]+\bra{b-a} B^{-1} \ket{b-a} \bigg{)}
\end{equation}
as desired.
\end{proof}
\end{lemma*}

\newpage
\section*{Selbst\"andigkeitserkl\"arung}
\normalsize
Hiermit erkläre ich, Martin Dupont, die beiliegende Arbeit
``On the Practical Applications of Information Field Dynamics''
selbständig und ohne unerlaubte fremde Hilfe angefertigt zu haben.
Keine anderen als die von mir angegebenen Schriften und Hilfsmittel wurden benutzt.
Die den benutzten Werken wörtlich und inhaltlich entnommenen Stellen sind kenntlich
gemacht.

\vspace{2cm}
Datum, Unterschrift

\end{document}